\documentclass[aps,pre,reprint,onecolumn,groupedaddress]{revtex4-2}

\usepackage{graphicx}
\usepackage{amsmath}
\usepackage{color,soul}
\usepackage{caption}
\usepackage{subcaption}
\usepackage{float}

\usepackage{amssymb}
\usepackage{pifont}

\newcommand{\xmark}{\ding{53}}
\begin{document}


\title{A fugacity-based Lattice Boltzmann method for multicomponent multiphase systems}


\author{Muzammil Soomro}
\email{msoomro@psu.edu}
\author{Luis F. Ayala}
\affiliation{Department of Energy and Mineral Engineering, The Pennsylvania State University, University Park, PA 16802, USA}

\author{Cheng Peng}
\affiliation{Key Laboratory of High Efficiency and Clean Mechanical Manufacture, Ministry of Education, School of Mechanical Engineering, Shandong University, Jinan 250061, China}

\author{Orlando M. Ayala}
\affiliation{Department of Engineering Technology, Old Dominion University, Norfolk, VA 23529, USA}


\date{\today}

\begin{abstract}
The free energy model can extend the Lattice Boltzmann method to multiphase systems. However, there is a lack of models capable of simulating multicomponent multiphase fluids with partial miscibility. In addition, existing models cannot be generalized to honor thermodynamic information provided by any multicomponent equation of state of choice. In this paper, we introduce a free energy Lattice Boltzmann model where the forcing term is determined by the fugacity of the species, the thermodynamic property that connects species partial pressure to chemical potential calculations. By doing so, we are able to carry out multicomponent multiphase simulations of partially miscible fluids and generalize the methodology for use with any multicomponent equation of state of interest. We test this fugacity-based Lattice Boltzmann method for the cases of vapor-liquid equilibrium for two and three-component mixtures in various temperature and pressure conditions. We demonstrate that the model is able to reliably reproduce phase densities and compositions as predicted by multicomponent thermodynamics and can reproduce different characteristic pressure-composition and temperature-composition envelopes with a high degree of accuracy. We also demonstrate that the model can offer accurate predictions under dynamic conditions.
\end{abstract}


\maketitle

\section{Introduction}

\begin{figure}[h]
    \centering
    \includegraphics[scale=0.45]{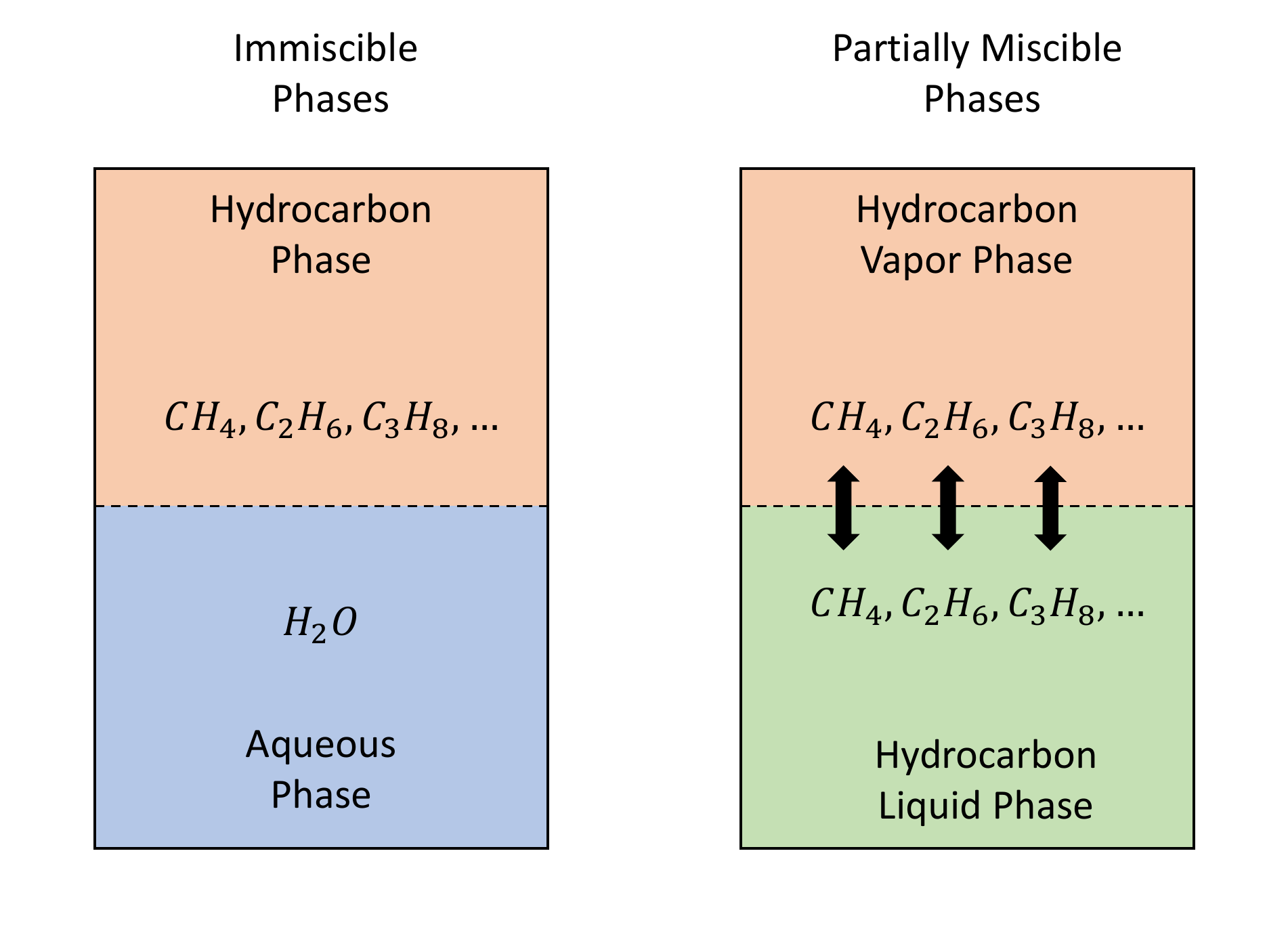}
    \vspace{-0.35cm}
    \caption{Left: An immiscible system of a hydrocarbon phase composed of the components $CH_4$, $C_2H_6$, $C_3H_8$ amongst others, and an aqueous phase composed of the component $H_2O$. The components are restricted to their phases and there is no interfacial mass transfer. Right: A partially miscible system of hydrocarbons with a light vapor phase and dense liquid phase. Both phases are composed of the same components ($CH_4$, $C_2H_6$, $C_3H_8$ amongst others) and there is interfacial mass transfer.}
    \label{figMPSchematic}
\end{figure}

Multiphase flows of fluids with multiple chemical components occur in many important settings like hydrocarbon reservoirs, carbon dioxide sequestration in aquifers, and chemical reactors. To achieve reliable modeling of these systems, it is essential to develop tools that can accurately model multicomponent multiphase (MCMP) flow, which would involve fluid dynamics coupling with thermodynamics. Under a thermodynamic framework, MCMP systems can be divided into two categories: immiscible and partially miscible. In immiscible phases, some components in the system are confined to a single phase, and there is no interfacial mass transfer. In partially miscible phases, however, components can be exchanged between phases. An example of immiscible and partially miscible phases is shown in Figure \ref{figMPSchematic}. It is worth noting that the terms immiscible and partially miscible are not applicable to single component systems as a single component multiphase system would always have to allow for interfacial mass transfer. Although modeling immiscible phases is quite popular, immiscibility is only a useful idealization meant to simplify modeling. There is always some degree of interfacial mass transfer in all multiphase systems; therefore, an accurate model for multiphase systems needs to account for partial miscibility.

In the study of partially miscible phases an equation of state (EOS) can play an important role. This is because in addition to relating the pressure of a system to its density, temperature and composition, an EOS also contains information about the equilibrium state of multiphase systems. Therefore, the choice of EOS is crucial to the modeling of partially miscible fluids. Cubic equations of state (EOSs), such as the Peng-Robinson (PR) \cite{Peng1976} and the Soave-Redlich-Kwong (SRK) \cite{Soave1972}, are widely regarded as the most accurate ones to describe the thermodynamics of fluids. When multiple components are present, an EOS needs to account for interactions between ``like" and ``unlike" component pairs captured through mixing rules \cite{Kwak1986}. A realistic model of partially miscible systems, and hence MCMP systems, must incorporate an accurate EOS and the multicomponent mixing rules.
The chemical potential governs the transport of a component within and across phases in MCMP systems. At equilibrium, the chemical potential of each component is constant throughout the multiphase system. In order to simplify phase equilibrium calculations, and to connect chemical potential and partial pressure calculations, the concept of fugacity was introduced \cite{Lewis1901}. At phase equilibrium, the equality of the fugacity of each component is mathematically equivalent to the equality of the chemical potential of each component. Although fugacity has primarily been used to simplify the phase equilibrium problem, it can also be a powerful quantity for simulating multiphase flow because it can be tied directly to an EOS. Thus, fugacity can be used as a bridge to incorporate accurate multicomponent EOSs into flow simulations.

Over the last thirty years, the Lattice Boltzmann Method (LBM) has emerged as a powerful fluid dynamics simulation tool with extensions to simulate multiphase flow. While early multiphase versions of LBM were restricted to studying the flow of immiscible fluids (e.g., \cite{Rothman1988}), models eventually emerged that allowed interfacial mass transfer, albeit only for single component systems. Two of these models that have gained popularity are the pseudopotential model \cite{Shan1993} and the free energy model \cite{Swift1995}. The pseudopotential model achieves phase separation by introducing a force to replicate intermolecular interactions, and it has been extended to incorporate any EOS \cite{Yuan2006}. It has also been extended to multicomponent systems of immiscible phases \cite{Martys1996}, and attempts have been made to consider multicomponent systems of partially miscible phases \cite{Bao2013}. However, a well-known drawback of this method is its inconsistency with thermodynamics when using an EOS, which comes in two forms. The first inconsistency has to do with multicomponent versions of the pseudopotential LBM where a single component EOS is used to calculate a pressure for each component \cite{Bao2013}. The pressure is a property of a phase, not a component, and any multicomponent formulation should have one thermodynamic pressure at each point which comes from a multicomponent EOS with mixing rules. The second inconsistency applies to all pseudo potential models, and it is that at equilibrium, the fugacity of each component (as predicted by the EOS) is not constant throughout the system at equilibrium conditions \cite{He2002}. In a single component system, this is equivalent to violating the Maxwell equal area rule. While the first form of inconsistency can be addressed \cite{Peng2021}, the second form has only been fully resolved for single component systems \cite{PengAyala2020,Czelusniak2022}.

The free energy model introduces macroscopic thermodynamics directly into Lattice Boltzmann through a functional of the Helmholtz free energy \cite{Luo1998}, which makes it a good candidate for achieving full consistency with thermodynamics. While the free energy model can be shown to be theoretically consistent with thermodynamics, it still does not lead to fully consistent simulations that can achieve full equality of chemical potentials and fugacities for phases in equilibrium \cite{Guo2011}. Guo \cite{Guo2021} showed that this inconsistency resulted from errors produced due to discretization, and the ``well-balanced" LBM could eliminate these errors, making the free energy model consistent with thermodynamics. Multiphase free energy LBM has been used extensively with free energy functionals that can be generalized to any EOS like the PR or SRK EOS but only for single-component (pure) systems \cite{Mazloomi2015PRLSC,Mazloomi2015PRESC,Siebert2014SC,Wen2020SC,Zhang2022WB,Qiao2019}. Its extension to MCMP systems has mostly relied on the use of simpler free energy functionals which can only qualitatively predict phase separation and are largely limited to immiscible MCMP systems \cite{Liu20162C,Scarbolo20132C,Liang20163C,Lamura19993C,Yuan2022MC,Zheng2020MC,Semprebon2016MC}. In general, there remains a significant lack of models capable of performing LB simulations that fully honor multicomponent thermodynamics as predicted by multicomponent EOSs--including cases of systems of partial miscibility.

In this paper, we introduce a new free energy LBM model which relies on the fugacity of each component, a property readily available through any EOS, to honor multicomponent thermodynamics fully. Through this new model, we can simulate flows of MCMP fluids with partial miscibility, using any multicomponent EOS. Additionally, by extending Guo's well-balanced LBM to multicomponent systems, we can show that our model fully complies with macroscopic thermodynamics and predicts phase densities and compositions in agreement with flash calculations done with the multicomponent EOS at a variety of different pressure and temperature conditions. Our paper is structured as follows. Section \ref{secReview} will review the free energy LBM and the limitations of previous models. In Section \ref{secFug}, the fugacity-based LBM will be introduced, and the procedure to implement it will be shown. In Section \ref{secResults}, the fugacity-based LBM will be tested in various cases of vapor-liquid equilibrium, including tests for two-component and three-component mixtures and the generation of pressure-composition and temperature-composition envelopes. Additionally, the cases of spinodal decomposition and oscillating droplet will also be shown in this section. Finally, the main conclusions of the paper and discussions will be presented in Section \ref{secConclusion}.

\section{A review of current Free Energy LBM models}
\label{secReview}

LBM relies on the Lattice Boltzmann Equation (LBE) and for the case of a single component fluid, this is given by Equation \ref{eqLBE} \cite{He1997}.
\begin{subequations}
    \label{eqLBE}
    \begin{equation}
            g_{\alpha}\left(\mathbf{r}+\mathbf{e}_{\alpha}\delta t,t+\delta t\right)-g_{\alpha}\left(\mathbf{r},t\right)=-\frac{1}{\tau}\left[g_{\alpha}\left(\mathbf{r},t\right)-g_{\alpha}^{(eq)}\left(\mathbf{r},t\right)\right]+\left[1-\frac{1}{2\tau}\right]F_{\alpha}\left(\mathbf{r},t\right)\delta t
    \end{equation}
    \begin{equation}
             g_{\alpha}^{(eq)}=\rho w_{\alpha}\left[1+\frac{\mathbf{u}\cdot \mathbf{e}_{\alpha}}{c_s^2}+\frac{\left(\mathbf{u}\cdot \mathbf{e}_{\alpha}\right)^2}{2c_s^4}-\frac{\mathbf{u}\cdot \mathbf{u}}{2c_s^2}\right]
    \end{equation}
    \begin{equation}
    \label{eqLBEfinal3}
            F_{\alpha}=\mathbf{F}\cdot w_{\alpha}\left[\frac{\mathbf{e}_{\alpha}-\mathbf{u}}{c_s^2}+\frac{\left(\mathbf{u}\cdot \mathbf{e}_{\alpha}\right)\mathbf{e}_{\alpha}}{c_s^4}\right]
    \end{equation}
\end{subequations}

Here $\mathbf{r}$ is the position vector, $\mathbf{u}$ is the macroscopic velocity, $\mathbf{F}$ is the body force, $c_s$ is the speed of sound, $t$ is the time, $\tau$ is the relaxation time, $\rho$ is the macroscopic mass density, and  $\mathbf{e}_\alpha$, $w_\alpha$, $g_\alpha$, and $g_\alpha^{eq}$ are the lattice velocity, weighing parameter, distribution function and equilibrium distribution function for direction $\alpha$ respectively. $F_\alpha$ is the forcing term given in Ref. \cite{Guo2002Forcing}. It can be shown that at the macroscopic scale, LBE can replicate the Navier-Stokes equation (Equation \ref{eqNSE}) \cite{Guo2002Forcing}.

\begin{equation}
    \label{eqNSE}
    \frac{\partial}{\partial t}\left(\rho\mathbf{u}\right)+\nabla\cdot\left(\rho\mathbf{u}\mathbf{u}\right)=-\nabla \left(c_s^2\rho\right)+\nu\nabla\cdot\left[\rho\left(\nabla\mathbf{u}+\left(\nabla\mathbf{u}\right)^T\right)\right]+\mathbf{F}
\end{equation}
Here $\nu$ is the kinematic viscosity, and the pressure is given by $c_s^2\rho$. The Free Energy model introduces multiphase thermodynamics into LBM by replacing the gradient of the standard LBM pressure, $c_s^2\rho$, with the divergence of an updated pressure tensor, $\overline{\overline{P}}$, in the Navier-Stokes equation. This pressure tensor contains the contributions from the bulk fluid, denoted by subscript `$B$', and the interface, denoted by subscript `$I$'. To derive the pressure tensor, we need a functional of the Helmholtz Free Energy, referred to as the free energy functional, and it is shown in Equation \ref{eqFreeEn}.
\begin{equation}
    \label{eqFreeEn}
    \Psi(T,V,n)=\int{\psi(T,\tilde{\rho})}dV=\int{\left(\psi_B+\psi_I\right)}dV
\end{equation}
Here $\Psi$ is the extensive free energy, which is a function of temperature ($T$), volume ($V$), and amount in moles ($n$) and $\psi$ is the intensive free energy per unit volume which is a function of temperature and molar density ($\tilde{\rho}$). The chemical potential, $\mu$, is then defined as the functional derivative of the extensive free energy with respect to moles (at constant $T$ and $V$) or intensive free energy with respect to molar density (at constant $T$), as shown in Equation \ref{eqchempot1}.

\begin{equation}
    \label{eqchempot1}
    \mu=\frac{\delta\Psi}{\delta n}=\frac{\delta\psi}{\delta \tilde{\rho}}
\end{equation}

Next, the divergence of the pressure tensor is calculated as a function of the gradient of the chemical potential through the application of the Gibbs-Duhem equation (Equation \ref{eqGibbsDuhem}) for isothermal conditions.
\begin{equation}
    \label{eqGibbsDuhem}
    \nabla\cdot \overline{\overline{P}}=\tilde{\rho}\nabla\mu
\end{equation}
The divergence of the pressure tensor is introduced into the LBE as a body force and this can either be done by the force in the ``pressure form" (Equation \ref{eqPressform}) or the ``potential form" (Equation \ref{eqPotform}) \cite{Kruger2017}. This body force will be input through Equation \ref{eqLBEfinal3} and will appear in the Navier-Stokes equation (Equation \ref{eqNSE}).
\begin{equation}
\label{eqPressform}
    \mathbf{F}=-\nabla\cdot \overline{\overline{P}}+\nabla\left(c_s^2\rho\right)
\end{equation}
\begin{equation}
\label{eqPotform}
    \mathbf{F}=-\tilde{\rho}\nabla\mu+\nabla\left(c_s^2\rho\right)
\end{equation}
The term $\nabla\left(c_s^2\rho\right)$ in both forms of the force is an artificial term needed to cancel out the standard pressure gradient produced from the LBE  at the Navier-Stokes level, leaving it only in terms of the real thermodynamic pressure. The interface free energy represents an energy penalty for having an interface and is expressed as $\psi_I=\frac{\kappa}{2}\left(\nabla\tilde{\rho}\right)^2$ \cite{Cahn1958}. Through this expression, the ``pressure form" and ``potential form" of the force can be written as Equation \ref{eqPressform2} and Equation \ref{eqPotform2} respectively \cite{Kruger2017,Guo2021}. 
\begin{equation}
\label{eqPressform2}
    \mathbf{F}=-\nabla p_B + \kappa\tilde{\rho}\nabla\left(\nabla^2\tilde{\rho}\right) +\nabla\left(c_s^2\rho\right)
\end{equation}
\begin{equation}
\label{eqPotform2}
    \mathbf{F}=-\tilde{\rho}\nabla\mu_B+\tilde{\rho}\nabla\left(\kappa\nabla^2\tilde{\rho}\right)+\nabla\left(c_s^2\rho\right)
\end{equation}
Here $\kappa$ is a parameter controlling interfacial tension strength, and $p_B$ and $\mu_B$ are the bulk (without contributions from the interface) pressure and chemical potential, respectively. Theoretically, the free energy model is supposed to be consistent with thermodynamics (it should fulfill the iso-fugacity criterion at equilibrium); however, Guo \cite{Guo2021} showed that due to the different discretizations involved in the LBE and the body force, the $\nabla\left(c_s^2\rho\right)$ term does not entirely cancel out. To solve this, the ``Well-balanced" LBM was introduced with the equilibrium distribution function and forcing term given by Equation \ref{eqfeqWB} and Equation \ref{eqFWB} respectively.
\begin{equation}
\label{eqfeqWB}
    g_{\alpha}^{(eq)}=  \begin{cases}
    \rho-\left(1-w_0\right)\rho_0+w_0\rho \left[\frac{\mathbf{u}\cdot \mathbf{e}_{\alpha}}{c_s^2}+\frac{\left(\mathbf{u}\cdot \mathbf{e}_{\alpha}\right)^2}{2c_s^4}-\frac{\mathbf{u}\cdot \mathbf{u}}{2c_s^2}\right] & \text{if $\alpha=0$ }\\
    w_\alpha\rho_0+w_\alpha\rho\left[\frac{\mathbf{u}\cdot \mathbf{e}_{\alpha}}{c_s^2}+\frac{\left(\mathbf{u}\cdot \mathbf{e}_{\alpha}\right)^2}{2c_s^4}-\frac{\mathbf{u}\cdot \mathbf{u}}{2c_s^2}\right] & \text{if $\alpha\ne0$}\\
    \end{cases}
\end{equation}
\begin{equation}
\label{eqFWB}
        F_{\alpha}=\mathbf{F}\cdot w_{\alpha}\left[\frac{\mathbf{e}_{\alpha}-\mathbf{u}}{c_s^2}+\frac{\left(\mathbf{u}\cdot \mathbf{e}_{\alpha}\right)\mathbf{e}_{\alpha}}{c_s^4}\right]+\nabla\rho\cdot w_{\alpha} \left[-\mathbf{u}+ \frac{\left(\mathbf{u}\cdot \mathbf{e}_{\alpha}\right)\mathbf{e}_{\alpha}}{c_s^2} +\frac{1}{2}\left(\frac{\mathbf{e}_\alpha^2}{c_s^2}-D\right)\mathbf{u} \right]
\end{equation}
Here $\rho_0$ is a numerical constant set to 0 as suggested by Guo \cite{Guo2021} and $D$ is the spacial dimension of the problem. With this new formulation, the artificial $\nabla\left(c_s^2\rho\right)$ term does not need to be included in Equations \ref{eqPressform2} and \ref{eqPotform2}. 

Single-component free energy models allow for interfacial mass transfer and have been able to successfully incorporate EOSs through both the pressure form \cite{Mazloomi2015PRLSC,Mazloomi2015PRESC,Siebert2014SC} and potential form \cite{Wen2020SC,Zhang2022WB,Qiao2019} of the force. For the pressure form, this is straightforward as the pressure $p_B$, in Equation \ref{eqPressform2}, is simply provided through the EOS. For the potential form, the intensive free energy per unit volume, $\psi(T,\tilde{\rho})$, is first obtained for an isothermal case through an EOS by solving the Ordinary Differential Equation given by Equation \ref{eqODE} to obtain the general solution given by Equation \ref{eqODEsol} \cite{Wen2020SC}.
\begin{equation}
\label{eqODE}
    \psi_B-\tilde{\rho}\frac{\tilde{d\psi_B}}{\tilde{d\rho}}-p_B=0
\end{equation}
\begin{equation}
\label{eqODEsol}
    \psi_B=\tilde{\rho}\left[\int\frac{p_B}{\tilde{\rho}^2}d\tilde{\rho}+C\right]
\end{equation}
Here $C$ is a temperature dependent constant of integration. Then the chemical potential is found, again for an isothermal case, as the derivative of the intensive free energy with respect to the molar density \cite{Wen2020SC}.

However, neither the approach for incorporating an EOS through the pressure form nor the potential form can be extended to multicomponent fluids. For a multicomponent fluid, the LBM requires a forcing term for each component, as shown later in Section \ref{secFug}, and therefore a body force must be derived for each component. Equation \ref{eqPressform2}, used to calculate the force in the pressure form, cannot be written for each component as dividing the pressure gradient between components would not be physically meaningful since pressure is a property of the phase and not the component. Equation \ref{eqPotform2} for the potential form, on the other hand, can be written for each component as the chemical potential is a property of the component and is the natural variable that drives mass transfer for each component. However, for a multicomponent case, the intensive free energy cannot be derived the same way as for a single component case. This is because the intensive free energy for a multicomponent system is a function of temperature and molar density of each component, $\tilde{\rho}_i$ ($\tilde{\rho}_i=n_i/V$, where $n_i$ is the moles of component $i$). Integrating an EOS will produce a constant of integration that depends not only on the temperature but also on the fluid composition, which is not constant throughout the system. Detailed analysis and discussion of this argument are given in Appendix \ref{appendixA}.

Multicomponent extensions of the free energy model do exist and have primarily been restricted to two-component systems \cite{Liu20162C,Scarbolo20132C}, with some extensions being to three component systems \cite{Liang20163C,Lamura19993C}, and some being applicable to any number of components \cite{Yuan2022MC,Zheng2020MC,Semprebon2016MC}. While majority of these models are for immiscible phases, there have been extensions to partially miscible phases in the context of phase field modelling \cite{Fu2016,Fu2017}. However, these models rely on simple free energy models with the ``multi-well" geometry needed to trigger phase separation. They may qualitatively predict phase separation but they will not offer quantitatively accurate predictions of phase density and composition. Recently, Ridl and Wagner \cite{Ridl2018} introduced a multicomponent model for partially miscible systems based on the multicomponent vdW EOS. vdW fluids have a theoretical basis in statistical mechanics, through which their free energy functional can be derived \cite{Hill2012}. However, this approach does not apply to any EOS in general due to their empirical nature, as demonstrated in Appendix \ref{appendixB}. W\"ohrwag et al. \cite{Wohrwag2018} introduced a ternary fluid model applicable to any single component EOS. They introduced a free energy model using a combination of single component EOS and qualitative ``double-well" potentials. However, single component EOSs do not apply to multicomponent mixtures. For a multicomponent mixture, interactions between like and unlike component pairs are captured through mixing rules, and these are not accounted for in a single component EOS. Common EOSs and multicomponent mixing rules are given in Appendix \ref{secAppendixEOS}. There is a lack of LBM models equipped to deal with partially miscible MCMP fluids, which can additionally incorporate any multicomponent EOS, and that is what we aim to address with the proposed fugacity-based LBM. Some important studies discussed in this section and their limitations have been summarized in Table \ref{tabLitReview}, along with the aim of the proposed fugacity-based LBM model, which will be introduced in Section \ref{secFug}.

\begin{table}[h]
\caption{Selected studies from Section \ref{secReview} categorized by relevant features and the aim of the proposed fugacity-based LBM.}
\label{tabLitReview}
\centering
\begin{tabular}{|c|c|c|c|c|}
\hline
Model &  \begin{tabular}[c]{@{}c@{}}Can handle multiple \\ components\end{tabular} & \begin{tabular}[c]{@{}c@{}}Can incorporate any \\ equation of state\end{tabular} & \begin{tabular}[c]{@{}c@{}}Allows partial miscibility\\ (for multicomponent models)\end{tabular} & \begin{tabular}[c]{@{}c@{}}Applies mixing rules (for \\ multicomponent models) \end{tabular} \\
\hline
\citet{Mazloomi2015PRESC}     & \xmark     &  \checkmark  & -    &  -   \\
\citet{Siebert2014SC}         &  \xmark     &   \checkmark    &   -  &      - \\
\citet{Wen2020SC}    &   \xmark      &   \checkmark    &   -   & -   \\
\citet{Liang20163C}        &  \checkmark  &  \xmark  &    \xmark    & \xmark \\
\citet{Yuan2022MC}         &  \checkmark  & \xmark     &    \xmark    & \xmark \\
\citet{Zheng2020MC}        &  \checkmark  & \xmark     &   \xmark     & \xmark \\
\citet{Ridl2018}         &  \checkmark   &  \xmark    &    \checkmark    &  \checkmark \\
\citet{Wohrwag2018}         & \checkmark    &\checkmark     &\xmark       & \xmark \\
\textbf{Proposed model}         & \pmb{\checkmark}     & \pmb{\checkmark}      & \pmb{\checkmark}        & \pmb{\checkmark}  \\
\hline
\end{tabular}
\end{table}

\section{The fugacity-based lattice Boltzmann method}
\label{secFug}
\subsection{Theory}
To solve problems involving MCMP mixtures with a greater degree of accuracy, we introduce a new approach to the free energy model, with a body force for each component `$i$' ($\mathbf{F}_i$) based on the fugacity of component `$i$' ($f_i$). For the well-balanced LBM formulation, this force is shown in Equation \ref{eqforceFugLBM}.
\begin{equation}
\label{eqforceFugLBM}
    \mathbf{F}_i=-\tilde{\rho}_i RT\nabla\ln{f_i}-\tilde{\rho}_i\nabla\mu_{I,i}
\end{equation}
Here $R$ is the universal gas constant, and $\mu_{I, i}$ is the interface chemical potential for component `$i$'. Since this force is to be used with the well-balanced LBM, there will be no artificial term to cancel out the standard pressure from the LBE. However, if the standard LBM formulation is used, Equation \ref{eqforceFugLBM} should have an additional $c_s^2\nabla\rho_i$ term added to it. 
In order to arrive at Equation \ref{eqforceFugLBM}, we start by extending the well-balanced LBM formulation for a multicomponent system; the well-balanced LBE for a component `$i$' is given by Equation \ref{eqLBEMC}.
\begin{subequations}
    \label{eqLBEMC}
    \begin{equation}
    \label{eqLBEMCa}
            g_{\alpha,i}\left(\mathbf{r}+\mathbf{e}_{\alpha}\delta t,t+\delta t\right)-g_{\alpha,i}\left(\mathbf{r},t\right)=-\frac{1}{\tau}\left[g_{\alpha,i}\left(\mathbf{r},t\right)-g_{\alpha,i}^{(eq)}\left(\mathbf{r},t\right)\right]+\left[1-\frac{1}{2\tau}\right]F_{\alpha,i}\left(\mathbf{r},t\right)\delta t
    \end{equation}
    \begin{equation}
    \label{eqLBEMCb}
          g_{\alpha,i}^{(eq)}=  \begin{cases}
    \rho_i-\left(1-w_0\right)\rho_0+w_0\rho_i \left[\frac{\mathbf{u}\cdot \mathbf{e}_{\alpha}}{c_s^2}+\frac{\left(\mathbf{u}\cdot \mathbf{e}_{\alpha}\right)^2}{2c_s^4}-\frac{\mathbf{u}\cdot \mathbf{u}}{2c_s^2}\right] & \text{if $\alpha=0$ }\\
    w_\alpha\rho_{0,i}+w_\alpha\rho_i\left[\frac{\mathbf{u}\cdot \mathbf{e}_{\alpha}}{c_s^2}+\frac{\left(\mathbf{u}\cdot \mathbf{e}_{\alpha}\right)^2}{2c_s^4}-\frac{\mathbf{u}\cdot \mathbf{u}}{2c_s^2}\right] & \text{if $\alpha\ne0$}\\
    \end{cases}
    \end{equation}
    \begin{equation}
    \label{eqLBEMCfinal3}
            F_{\alpha,i}=\mathbf{F}_i\cdot w_{\alpha}\left[\frac{\mathbf{e}_{\alpha}-\mathbf{u}}{c_s^2}+\frac{\left(\mathbf{u}\cdot \mathbf{e}_{\alpha}\right)\mathbf{e}_{\alpha}}{c_s^4}\right]+\nabla\rho_i\cdot w_{\alpha} \left[-\mathbf{u}+ \frac{\left(\mathbf{u}\cdot \mathbf{e}_{\alpha}\right)\mathbf{e}_{\alpha}}{c_s^2} +\frac{1}{2}\left(\frac{\mathbf{e}_\alpha^2}{c_s^2}-D\right)\mathbf{u} \right]
    \end{equation}
\end{subequations}
Here $\rho_{0, i}$ will be set to 0 as in the single component formulation. The forcing term for each component, given by Equation \ref{eqLBEMCfinal3}, requires a body force for each component. For this, we write the potential form of the body force for each component \cite{Zhang2009,Ridl2018} as shown in Equation \ref{eqForceComp}. Again, the force will not have the artificial $c_s^2\nabla\rho_i$ term because of the use of the well-balanced LBM.
\begin{equation}
    \label{eqForceComp}
    \mathbf{F}_i=-\tilde{\rho}_i\nabla\mu_{i}
\end{equation}
The chemical potential for component `$i$' is defined as the functional derivative of the extensive free energy with respect to moles of `$i$' (at constant $T$, $V$, and $n_{j\ne i}$) or intensive free energy with respect to molar density of `$i$' (at constant $T$ and $\tilde{\rho}_{j\ne i}$), as shown in Equation \ref{eqchempotMC}. Writing the free energy in terms of its contribution from the bulk fluid and the interface, as shown in Equation \ref{eqFreeEn}, the chemical potential for component `$i$' can also be split into its bulk and interface contributions as in Equation \ref{eqChemPot}.

\begin{equation}
    \label{eqchempotMC}
    \mu_i=\frac{\delta \Psi}{\delta n_i}=\frac{\delta\psi}{\delta \tilde{\rho_i}}
\end{equation}
\begin{subequations}
    \label{eqChemPot}
    \begin{equation}
    \label{eqChemPotTot}
       \mu_i=\mu_{B,i}+\mu_{I,i} 
    \end{equation}
    
    \begin{equation}
    \label{eqChemPotBulk}
        \mu_{B,i}=\frac{\partial\psi_B}{\partial \tilde{\rho_i}}
    \end{equation}
        \begin{equation}
        \label{eqChemPotI}
        \mu_{I,i}=\frac{\delta\psi_I}{\delta \tilde{\rho_i}}
    \end{equation}
    
\end{subequations}

As seen in Equation \ref{eqChemPotBulk}, the bulk chemical potential would require a multicomponent free energy functional, which is not available for all EOSs (e.g. the PR and SRK EOS). However, this free energy functional is no longer needed if we use a formulation based on the fugacity, a property readily available for any EOS. The fugacity of component `$i$' is defined using the differential given by Equation \ref{eqfugdef1} and the reference state given by Equation \ref{eqfugdef2} \cite{Lewis1901}.
\begin{subequations}
\label{eqfugdef}
\begin{equation}
\label{eqfugdef1}
    d \ln{f_i}\equiv \frac{d \mu_{B,i}}{RT}\ \ at\ const.\ T
\end{equation}
\begin{equation}
\label{eqfugdef2}
    \lim_{p \to 0} \frac{f_i}{x_ip}\equiv 1
\end{equation}
\end{subequations}
Using Equation \ref{eqfugdef1}, the gradient of the bulk chemical potential for an isothermal system can be written as:
\begin{equation}
\label{eqgradmuB}
    \nabla\mu_{B,i}=RT\nabla \ln{f_i}
\end{equation}
Splitting Equation \ref{eqForceComp} into its bulk and interface contributions and using Equation \ref{eqgradmuB}, we can arrive at our proposed form of the force, Equation \ref{eqforceFugLBM}. An expression for the fugacity of component `$i$' can be obtained by integrating Equation \ref{eqfugdef1} and using the reference state given by Equation \ref{eqfugdef2} and then this expression can be evaluated using an EOS as shown in Appendix \ref{secAppendixEOS}.

To perform MCMP simulations, we still require the interface chemical potential. The interface chemical potential should represent an energy penalty for having an interface. Several multicomponent models exist for the interface free energy from which the interface chemical potential can be derived \cite{Semprebon2016MC,Wohrwag2018,Kim2007}. However, this paper will use the interface free energy given by Ridl and Wagner \cite{Ridl2018} since it considers the interactions between component pairs, which is crucial for partially miscible phases. According to this interface free energy, the interface chemical potential is given by Equation \ref{eqChemPotI2}.
\begin{equation}
    \label{eqChemPotI2}
    \mu_{I,i}=-\sum_{j=1}^{N_c}\left(\kappa_{ij}\nabla^2\tilde{\rho}_j\right),
\end{equation}
where $\kappa_{ij}$ controls the strength of the interfacial tension and comes from the molecular interactions between component `$i$' and component `$j$'. Since $\kappa_{ij}$ comes from intermolecular interactions, we can calculate its value through van der Waals mixing rules (similar to Equation \ref{eqmixingrulesavdW2} in Appendix \ref{secAppendixEOS}) using pure component strength of interfacial tension, $\kappa_i$, as shown in Equation \ref{eqkappa}.
\begin{equation}
    \label{eqkappa}
    \kappa_{ij}=\sqrt{\kappa_i\kappa_j}
\end{equation}
The pure component interfacial tension strengths can be adjusted to achieve the desired interfacial tension in the system. In general, the less volatile a component is, the greater the intermolecular forces and hence the greater the surface tension. Therefore, for a pure component, the lower the volatility, the greater the value of $\kappa_i$. With this, the final form of the force can be written as shown in Equation \ref{eqforcefinal}.

\begin{equation}
\label{eqforcefinal}
    \mathbf{F}_i=-\tilde{\rho}_i RT\nabla\ln{f_i}+\tilde{\rho}_i\sum_{j=1}^{N_c}\nabla\left(\sqrt{\kappa_i\kappa_j}\ \nabla^2\tilde{\rho}_j\right)
\end{equation}
\subsection{Implementation}

The proposed model will require the calculation of the gradient and Laplacian for certain properties, which will require approximating the first and second-order spatial derivatives of those properties. For this study, the following finite difference schemes will be used to evaluate the derivatives for any generic property $h$ and the direction $x$:
\begin{equation}
\label{eq1stder}
    \frac{dh}{dx}\approx \frac{h\left(x+\delta x\right)-h\left(x-\delta x\right)}{2\delta x}
\end{equation}
\begin{equation}
\label{eq2ndder}
    \frac{d^2h}{dx^2}\approx \frac{h\left(x+\delta x\right)-2h\left(x\right)+h\left(x-\delta x\right)}{\delta x^2}
\end{equation}
The procedure to implement the fugacity-based LBM is straightforward and outlined as follows. 
\begin{itemize}
    \item[1.] Given the distribution functions and body forces of each component at a particular lattice node from the previous time step, compute the component mass density $\rho_i$, total mass density $\rho$, and velocity $\mathbf{u}$ at that lattice node, using the following equations: 
    \begin{equation*}
    \label{eqMaccompden}
        \rho_i=\sum_\alpha g_{\alpha,i}\ \ ; \ \ \rho=\sum_i \rho_i\ \ ; \ \ \mathbf{u}=\frac{1}{\rho} \sum_i\left[\sum_\alpha g_{\alpha,i} \mathbf{e_\alpha}+\frac{\mathbf{F}_i\delta t}{2}\right]
    \end{equation*}
    \item[2.] Calculate the component molar density $\tilde{\rho}_i$, the total molar density $\tilde{\rho}$, and component mole fraction (or composition) $x_i$, using the following relations:  
    \begin{equation*}
        \tilde{\rho}_i=\frac{\rho_i}{M_i}\ \ ; \ \ \tilde{\rho}=\sum_i\tilde{\rho}_i\ \ ; \ \ x_i=\frac{\tilde{\rho}_i}{\tilde{\rho}}
    \end{equation*}
    where $M_i$ is the molar mass of component `$i$'.
    \item[3.] Apply the mixing rules to obtain the attraction parameter for the mixture ($a_m$ for vdW or $a\alpha_m$ for SRK and PR EOS) and co-volume for the mixture ($b_m$). This process is outlined in Appendix \ref{secAppendixEOS}.
    \item[4.] Obtain the pressure through an EOS (the vdW, SRK and PR EOS are given in Appendix \ref{secAppendixEOS} by Equation \ref{eqvdwEOS}, \ref{eqSRKEOS}, and \ref{eqPREOS} respectively).
    \item[5.]  Obtain the natural log of fugacity for each component, $\ln{f_i}$,  using a fugacity expression relevant to the EOS used (the fugacity expressions for vdW, SRK and PR EOS are given in Appendix \ref{secAppendixEOS} by Equations \ref{eqfugvdW}, \ref{eqfugSRK} and \ref{eqfugPR} respectively).
    \item[6.] Obtain the interface chemical potential using Equation \ref{eqChemPotI2}. Use Equation \ref{eq2ndder} when evaluating the second-order spatial derivatives to calculate the Laplacian of the component molar density.
    \item[7.] Calculate the body force for each component as given by Equation \ref{eqforceFugLBM}. Use Equation \ref{eq1stder} to evaluate the first-order space derivatives when calculating the gradient of relevant properties.
    \item[8.] Calculate the equilibrium distribution function and forcing term for each component using Equations \ref{eqLBEMCb} and \ref{eqLBEMCfinal3} respectively.
    \item[9.] Perform collision for each component followed by propagation for each component to obtain the new distribution functions. Repeat steps 1-9 until the simulation is run for the desired time.
    
\end{itemize}

\section{Results}
\label{secResults}
To test the fugacity-based LBM, we perform several simulations on a two-phase vapor-liquid equilibrium system with the phases separated by a flat interface and compare the simulation results to those predicted by a flash calculation. We start by testing whether the model works for different EOSs by simulating a two-component mixture using the PR and SRK EOS. Next, the model is tested for a two-component mixture under various pressure and temperature conditions, and the pressure-composition and the temperature-composition envelopes are generated. Finally, a three-component mixture is simulated to show that the model can be extended to any number of components. After testing the cases of vapor liquid equilibrium, we simulate the case of a uniform system undergoing spinodal decomposition to form two separate phases and validate our results by testing the equality of component fugacity in each phase. Lastly, we test our model under dynamic conditions by simulating the case of an oscillating droplet, and comparing the obtained oscillation period with its theoretical value. Different mixtures were created for each case from the components listed in Table \ref{tabCompProp}.

\begin{table}[h]
\caption{The properties of relevant components used in the LBM simulations.}
\label{tabCompProp}
\centering
\begin{tabular}{|c|c|c|c|c|}
\hline
Component & Critical Pressure (bar) & Critical Temperature (K) & Accentric factor & Molar Mass (g/mol) \\
\hline
Methane (C1)            & 45.947        & 190.74       & 0.0104       & 16.043              \\
Ethane (C2)             & 48.711        & 305.51       & 0.0979       & 30.070              \\
Propane (C3)            & 42.472        & 370.03       & 0.1522       & 44.097              \\

n-Pentane (C5)         & 33.688        & 469.89       & 0.2514       & 72.150
\\
\hline
\end{tabular}
\end{table}

For all the simulations, a D2Q9 lattice is used in a periodic computational domain, and the relevant conversions between lattice units and physical units are established by fixing the universal gas constant and the attraction parameter, co-volume, and molar mass for the most volatile component in each mixture to the following values: $R=1$, $a_i=2/49$, $b_i=2/21$, and $M_i=1$. In Table \ref{tabCompProp}, the volatility decreases from Methane to n-Pentane. The binary interaction parameter between each component pair is 0.

\subsection{Two component mixture using PR and SRK equations of state}
\label{seccase1}
\begin{figure}[H]
    \centering
        \begin{subfigure}{0.48\textwidth}
            \centering
            \caption{}
            \includegraphics[width=\textwidth]{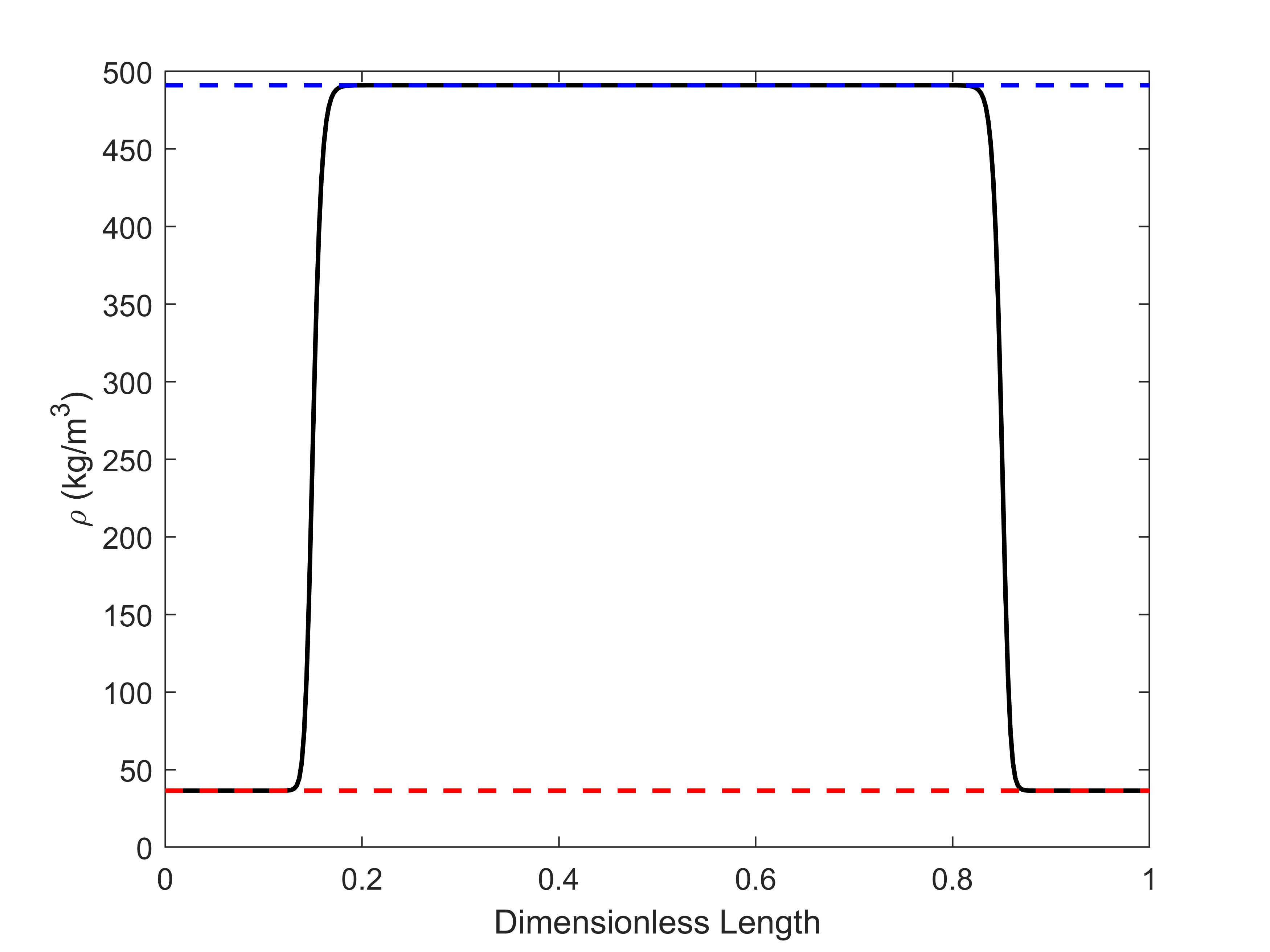}
        \end{subfigure}%
        \hfill
        \begin{subfigure}{0.48\textwidth}
            \centering
            \caption{}
            \includegraphics[width=\textwidth]{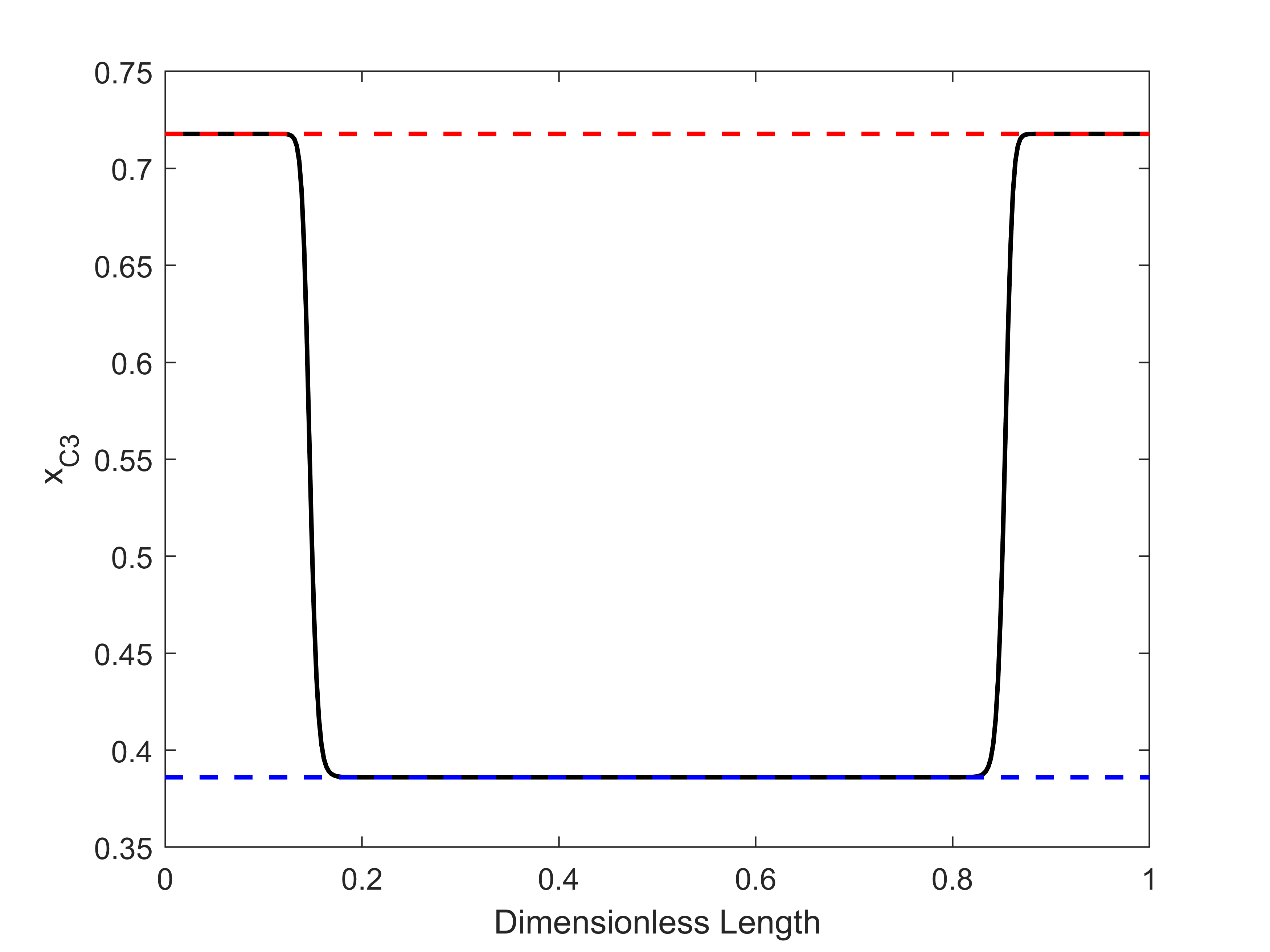}
        \end{subfigure}
        \hfill        
        \begin{subfigure}{0.48\textwidth}
            \centering
            \caption{}
            \includegraphics[width=\textwidth]{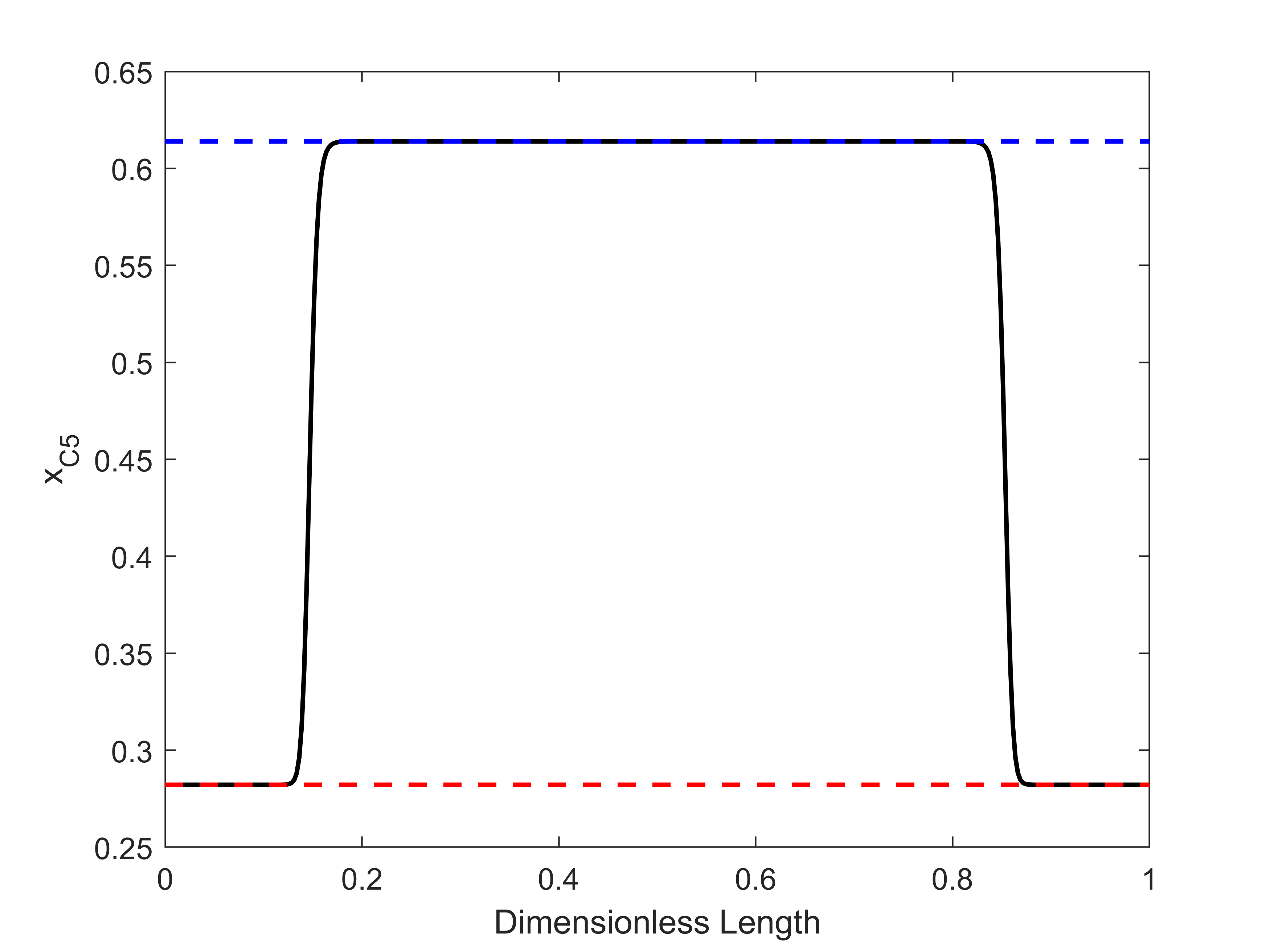}
        \end{subfigure}
        \hfill      
\caption{The equilibrium profiles for (a) density, (b) composition of C3 and (c) composition of C5 using the PR EOS. The solid black line shows the results from LBM and the dashed blue and red lines show the theoretical values from a flash calculation for the liquid and vapor phase respectively.}
\label{figcase1PR}
\end{figure}

\begin{figure}[H]
    \centering
        \begin{subfigure}{0.48\textwidth}
            \centering
            \caption{}
            \includegraphics[width=\textwidth]{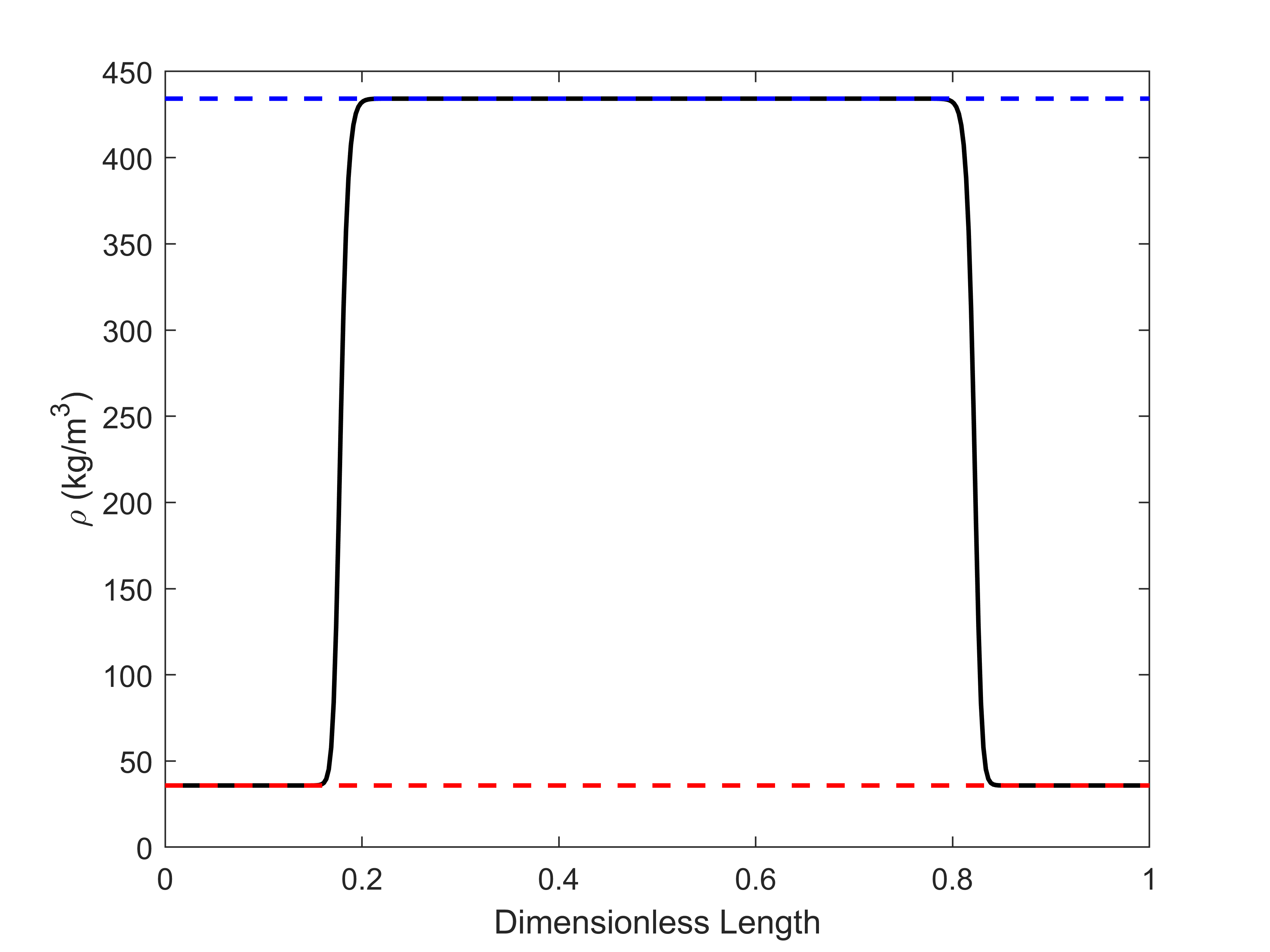}
        \end{subfigure}%
        \hfill
        \begin{subfigure}{0.48\textwidth}
            \centering
            \caption{}
            \includegraphics[width=\textwidth]{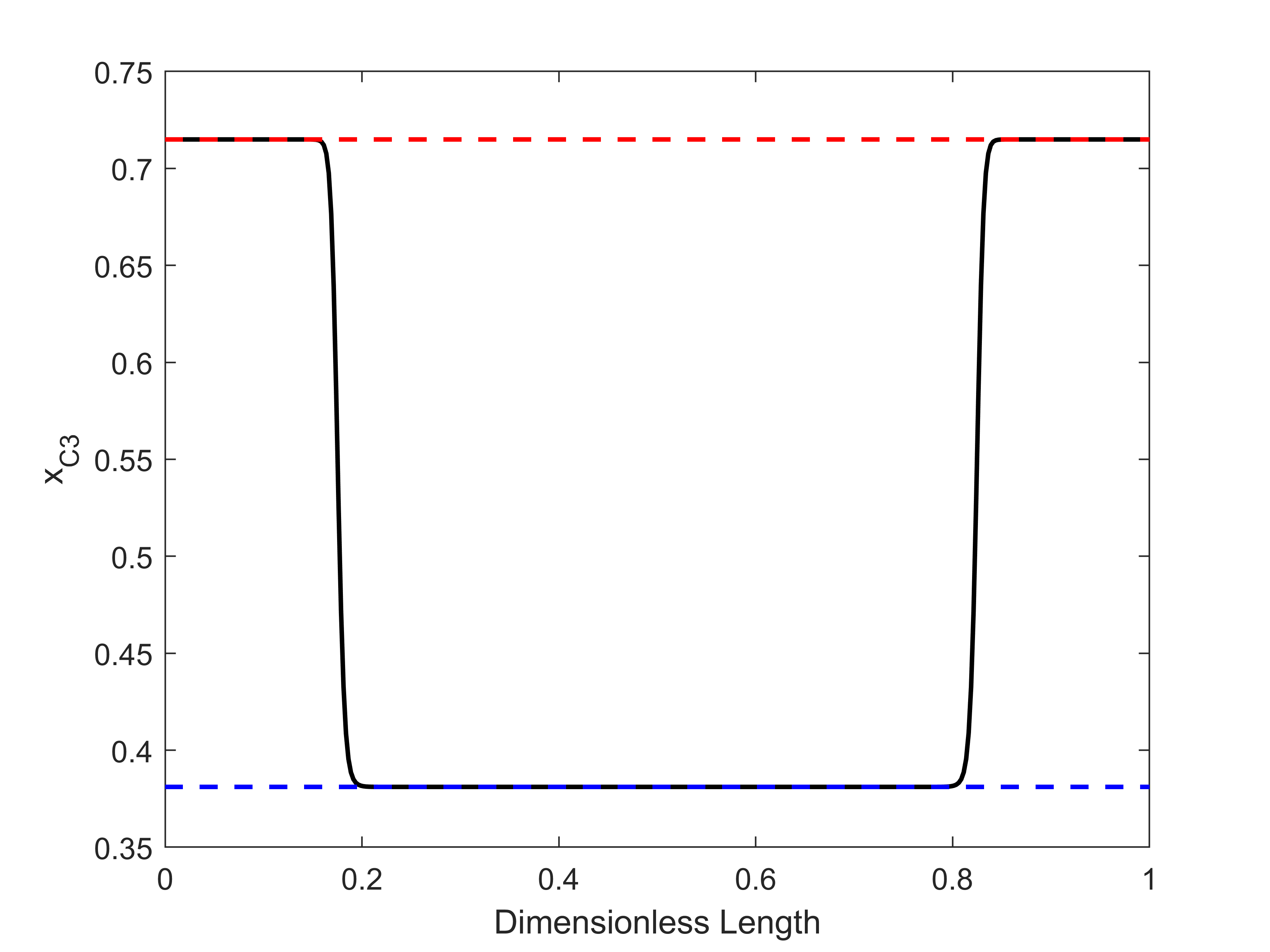}
        \end{subfigure}
        \hfill        
        \begin{subfigure}{0.48\textwidth}
            \centering
            \caption{}
            \includegraphics[width=\textwidth]{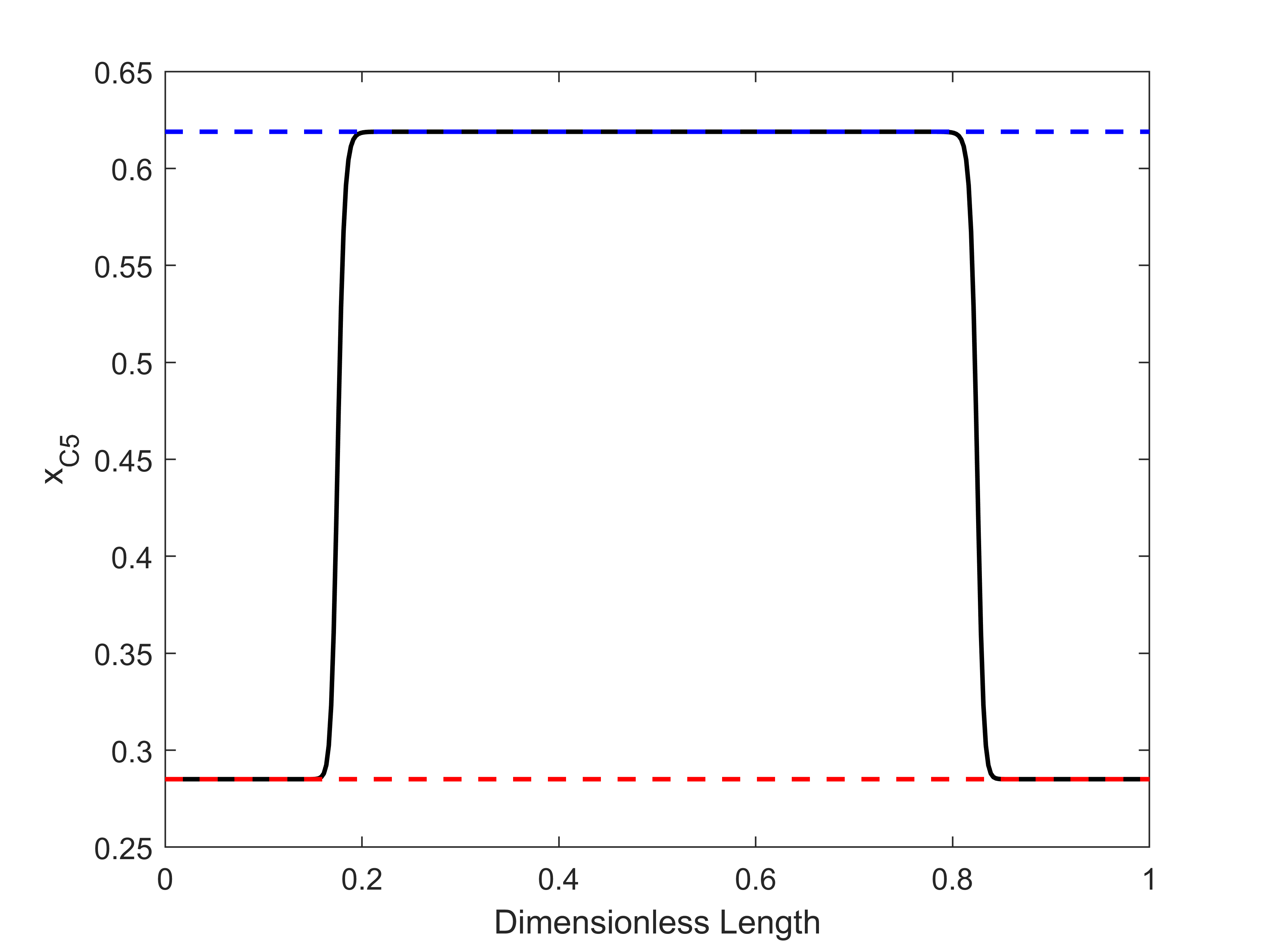}
        \end{subfigure}
        \hfill      
\caption{The equilibrium profiles for (a) density, (b) composition of C3 and (c) composition of C5 using the SRK EOS. The solid black line shows the results from LBM and the dashed blue and red lines show the theoretical values from a flash calculation for the liquid and vapor phase respectively.}
\label{figcase1SRK}
\end{figure}

In this case, we will simulate a system of Propane (C3) and n-Pentane (C5) at equilibrium using the PR and SRK EOS, with the mole fraction of C3 and C5 in the system being 0.4 and 0.6, respectively. For reference, a flat interface, vapor-liquid equilibrium case using these two components has been carried out in Ref. \cite{Peng2021} with the PR EOS  (the results are shown in Figures 5c and 5d of the Ref.). However, in Ref. \cite{Peng2021}, two additional parameters were needed to achieve the iso-fugacity criterion: a  force split coefficient ($\gamma$) depending on the pressure and temperature, and a tuning parameter ($\beta$). We will show that the fugacity-based LBM can achieve similar results under similar conditions without needing any tuning or empirical parameters. The simulations are initialized at a pressure of 16.547 bar (240 psia) and a temperature of 370.03 K (666.06 $^oR$). The size of the computational domain is $400\times 2$ ($n_x\times n_y$), the relaxation time $\tau=0.8$ and the values interfacial tension strength are $\kappa_{C1}=0.10$ and $\kappa_{C5}=0.15$. The density of each component in the x direction is initialized as shown in Equation \ref{eqInit} (the domain will be symmetric in the y direction).
\begin{equation}
    \label{eqInit}
    \rho_i(x,t=0)=\rho_{i,V}+\frac{\rho_{i,L}-\rho_{i,V}}{2}\left[\tanh{\left(\frac{2\left(x-\frac{S_V}{2}n_x\right)}{W}\right)}-\tanh{\left(\frac{2\left(x-\left(1-\frac{S_V}{2}\right)n_x\right)}{W}\right)}\right]
\end{equation}
Here $W$ is the initial interface width set to be 8, $\rho_{i, V}$, $\rho_{i, L}$, and $S_V$ are the density of component `$i$' in the vapor phase, density of component `$i$' in the liquid phase and saturation (volume fraction) of the vapor phase, respectively, which are calculated by performing a flash calculation for the given mixture at $p=16.547\ bar$ and $T=370.03\ K$ using the relevant EOS. The simulations are run for 1,000,000 time steps to achieve equilibrium, and the density and composition (mole fractions) versus the dimensionless length ($x/n_x$) are shown in Figure \ref{figcase1PR} and Figure \ref{figcase1SRK} for the PR and SRK EOS, respectively. The results predicted from a flash calculation using the relevant EOS are also included in the figures. It should be noted that when initializing the densities using Equation \ref{eqInit}, the equilibrium values of densities at the given pressure and temperature conditions are used in the bulk vapor and liquid regions. However, the densities at the interface are only approximations of the equilibrium state, as the exact equilibrium profile at the interface is not given by the ``tanh" profile. As a result, a new equilibrium will be reached where the pressure will slightly deviate from 13.79 bar. The LBM simulations are compared with flash calculations at the updated pressures.

The deviation of the fugacity-based LBM results from the flash calculation results is quantified with a relative error percentage: 
\begin{equation}
\label{eqerror}
    Relative\ Error\ (\%) = \frac{|z^{LBM}-z^{Theory}|}{z^{Theory}}\times 100\%
\end{equation}
where $z$ is the property to be measured (density or composition). The errors are tabulated in Table \ref{tabcase1error}. As can be seen from the error values, a near exact match between the LBM results and theoretical results is obtained.
\begin{table}[h]
\caption{The relative error values in the density, C3 composition and C5 composition in the different phases and with different EOSs.}
\label{tabcase1error}
\centering
\begin{tabular}{|p{2cm}|p{3cm}|p{3cm}|p{3cm}|p{3cm}|}
\hline
Property         & Liquid Region Error using PR (\%) & Vapor Region Error using PR (\%) & Liquid Region Error using SRK (\%) & Vapor Region Error using SRK (\%) \\
\hline
Density          & $1.70\times 10^{-5}$      & $1.40\times 10^{-4}$     & $4.27\times 10^{-5}$       & $3.00\times 10^{-4}$      \\
$x_{C3}$         & $7.94\times 10^{-7}$      & $4.69\times 10^{-6}$       & $2.62\times 10^{-6}$     & $1.27\times 10^{-6}$    \\
$x_{C5}$         & $4.99\times 10^{-7}$      & $1.19\times 10^{-5}$       & $1.61\times 10^{-6}$     & $3.17\times 10^{-6}$  \\
\hline
\end{tabular}
\end{table}

\subsection{Pressure-composition and temperature-composition envelopes}
\label{seccase2}
\begin{figure}[H]
    \centering
        \begin{subfigure}{0.48\textwidth}
            \centering
            \caption{}
            \includegraphics[width=\textwidth]{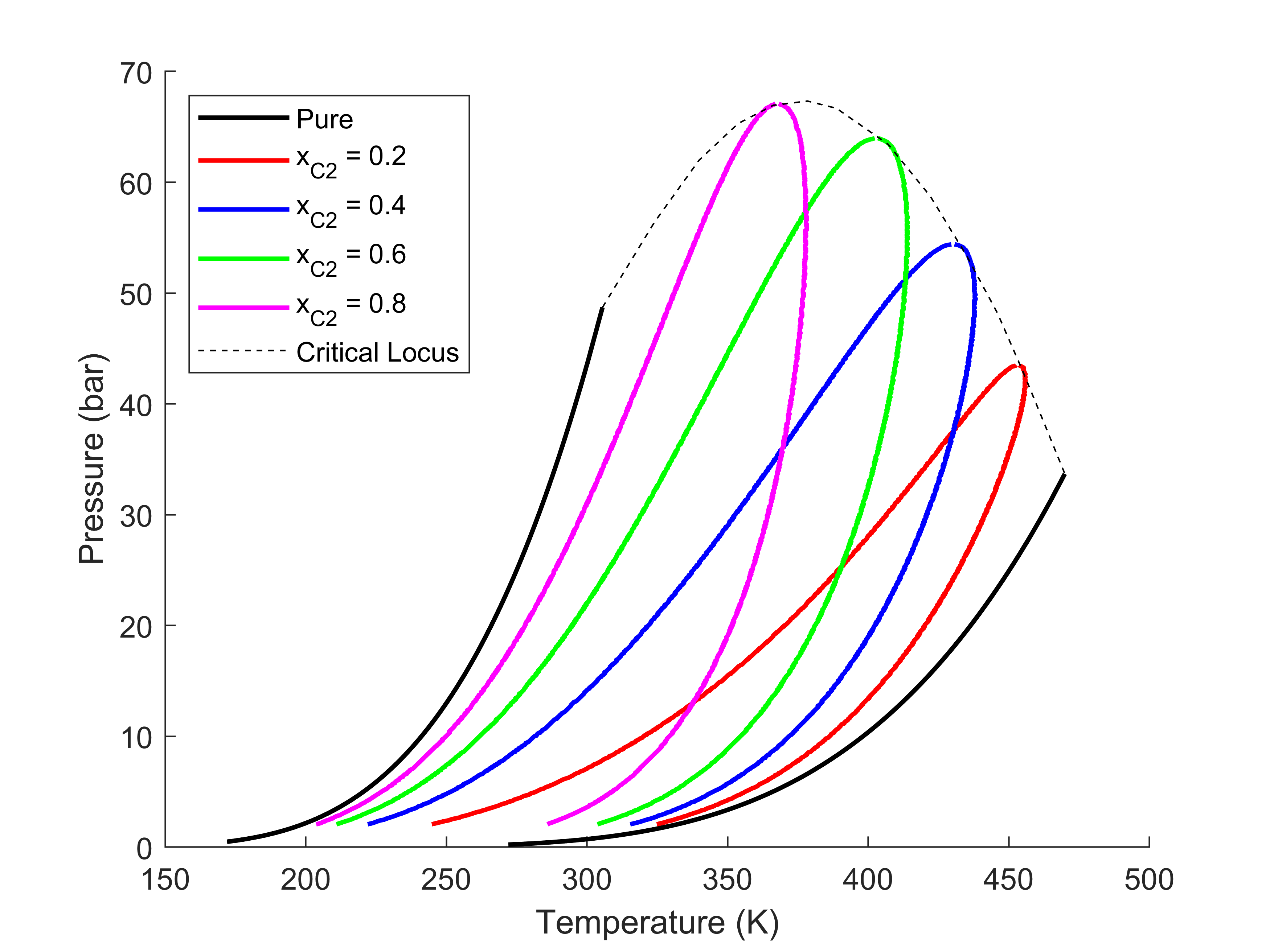}
            \label{figPTmaina}
        \end{subfigure}%
        \hfill
        \begin{subfigure}{0.48\textwidth}
            \centering
            \caption{}
            \includegraphics[width=\textwidth]{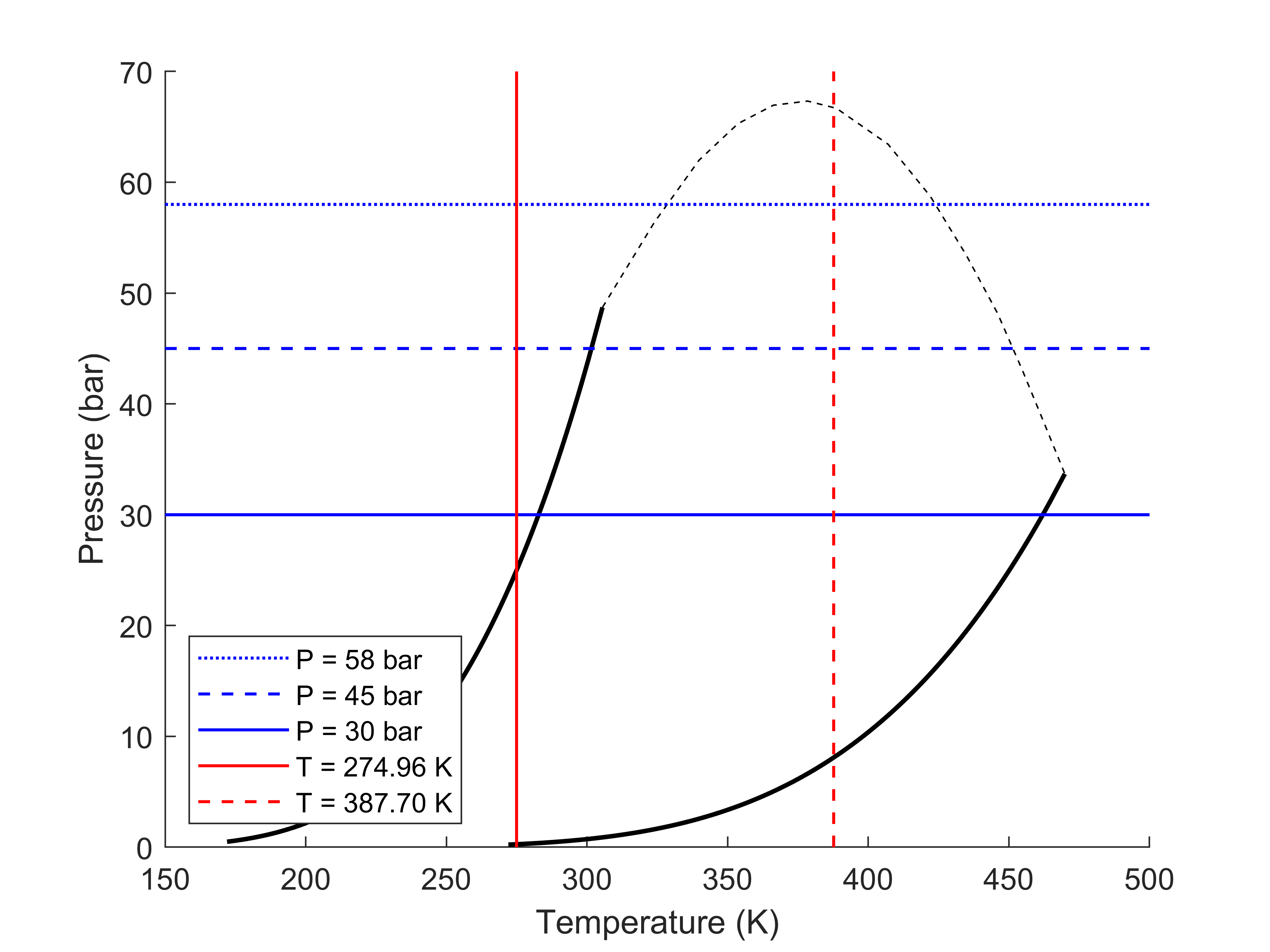}
            \label{figPTmainb}
        \end{subfigure}
        \hfill        

\caption{(a) The pressure-temperature envelopes for the C2-C5 mixture at different compositions of C2. The solid black lines represent the pure species with the line on the right representing pure C5 and the line on the left representing pure C2. (b) The black lines are the pure component curves and critical locus from (a), the blue lines represent the relevant isobars and the red lines represent the relevant isotherms.}
\label{figPTmain}
\end{figure}

For this case, we will investigate the phase behavior of a two-component system at different pressure and temperature conditions and create the pressure-composition (p-x) and temperature-composition (T-x) envelopes. We consider an Ethane (C2) and Pentane (C5) system and use the PR EOS. The pressure-temperature (p-T) envelopes of this system at different compositions (mole fractions) of C2 generated using the PR EOS are shown in Figure \ref{figPTmaina}. The critical locus, which is the curve connecting all the critical points of the mixtures at different compositions, along with the saturation curves for pure C2 and pure C5, bound the possible two-phase region on the p-T plane. For a p-x envelope, flash calculations are performed for different pressures at a constant temperature and the composition of C2 in the vapor phase ($x_{C2,V}$) and liquid phase ($x_{C2,L}$) are recorded. A plot of pressure versus $x_{C2,V}$ will make the dew point curve, whereas a plot of pressure versus $x_{C2,L}$ will make the bubble point curve. For a T-x envelope, this process is repeated with the pressure held constant, and the vapor and liquid compositions plotted with temperature. Different characteristic plots can be produced depending on the choice of the isotherm for a p-x plot or the choice of isobar for the T-x plot. Several different isotherms and isobars of interest which will be tested are shown in Figure \ref{figPTmainb}.

\begin{figure}[H]
    \centering
        \begin{subfigure}{0.46\textwidth}
            \centering
            \caption{}
            \includegraphics[width=\textwidth]{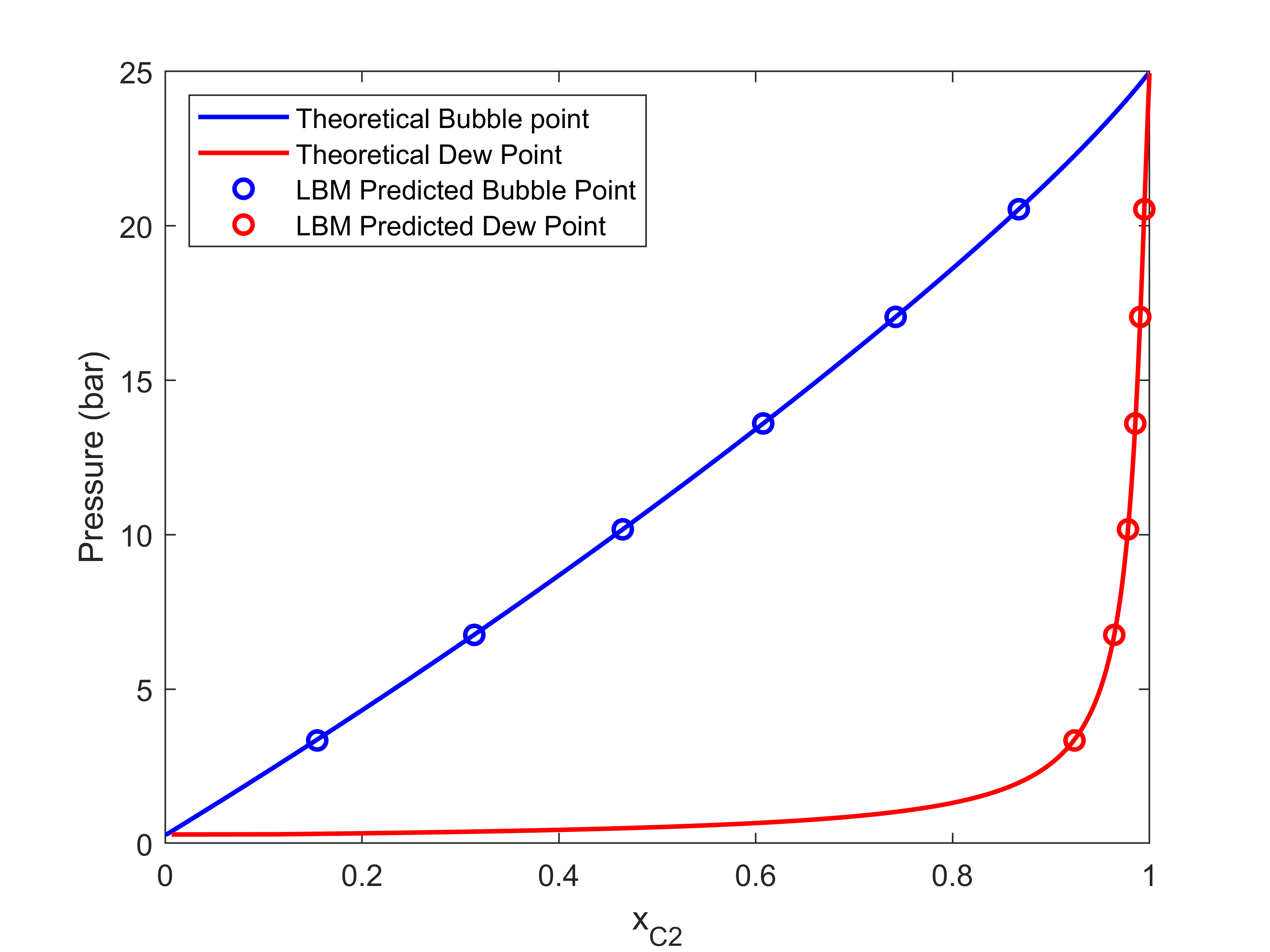}
            \label{figPxEnvelopeA}
        \end{subfigure}%
        \hfill
        \begin{subfigure}{0.46\textwidth}
            \centering
            \caption{}
            \includegraphics[width=\textwidth]{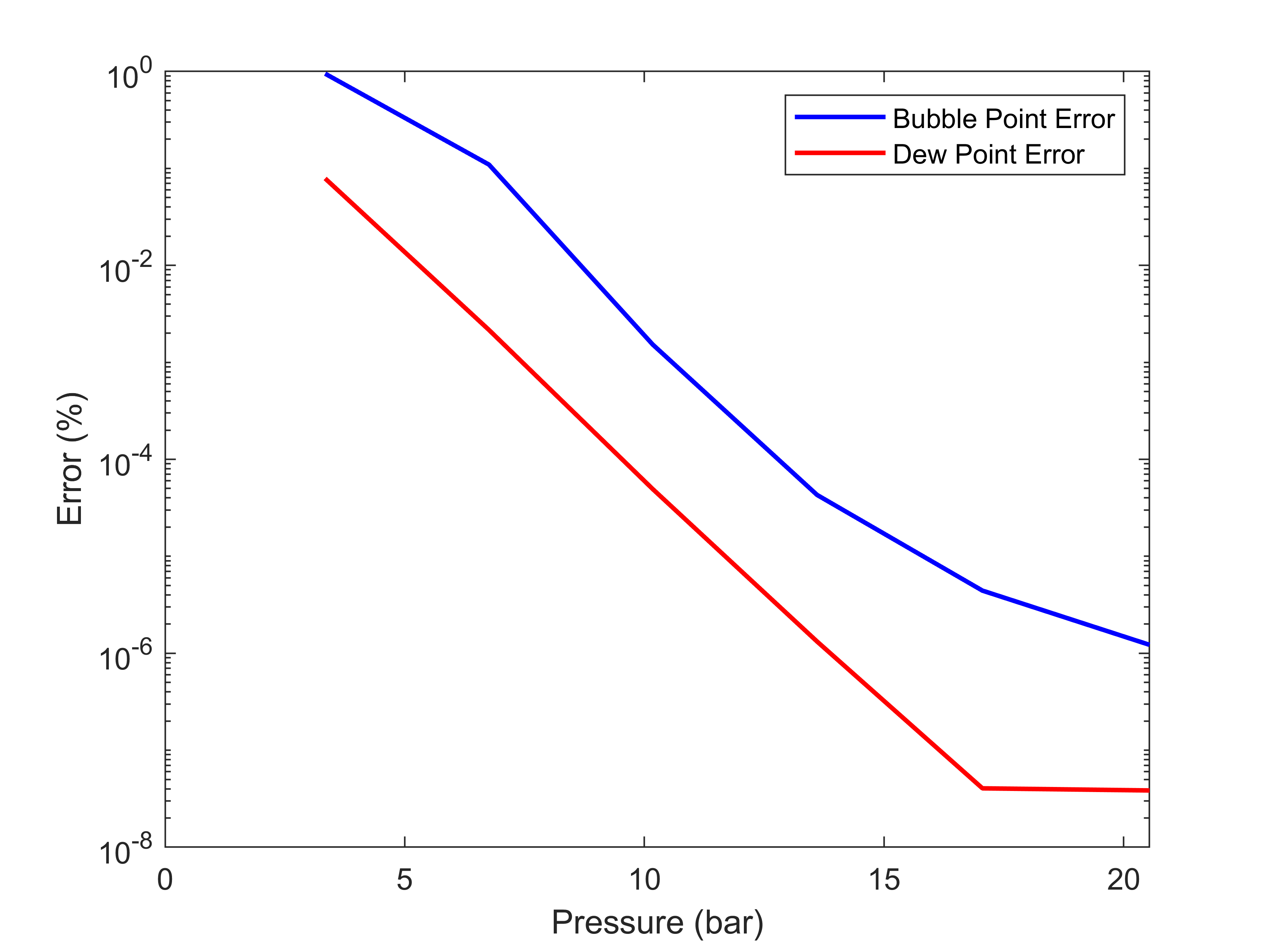}
            \label{figPxEnvelopeB}
        \end{subfigure}%
        \hfill
        \begin{subfigure}{0.46\textwidth}
            \centering
            \caption{}
            \includegraphics[width=\textwidth]{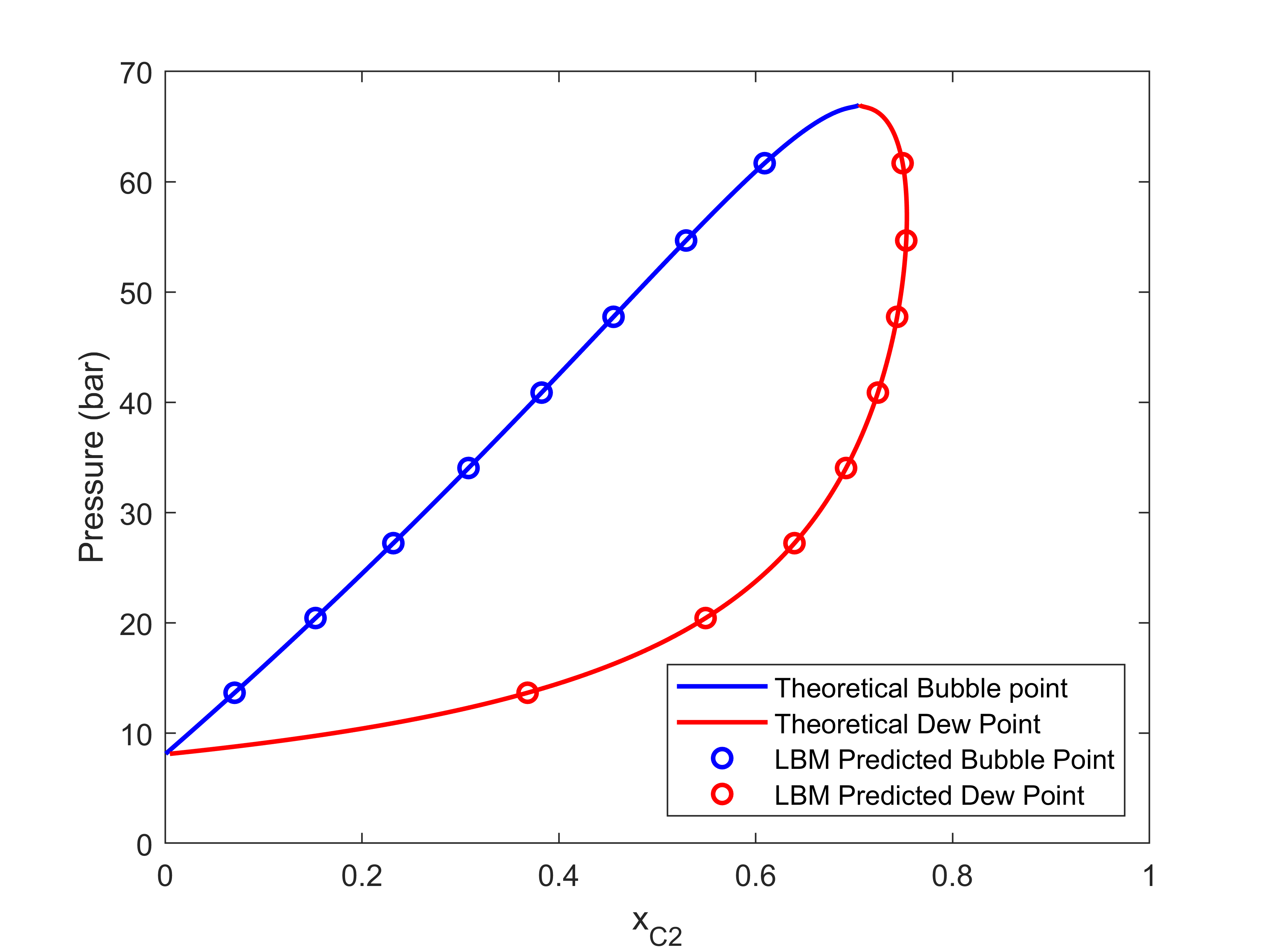}
            \label{figPxEnvelopeC}
        \end{subfigure}
        \hfill        
        \begin{subfigure}{0.46\textwidth}
            \centering
            \caption{}
            \includegraphics[width=\textwidth]{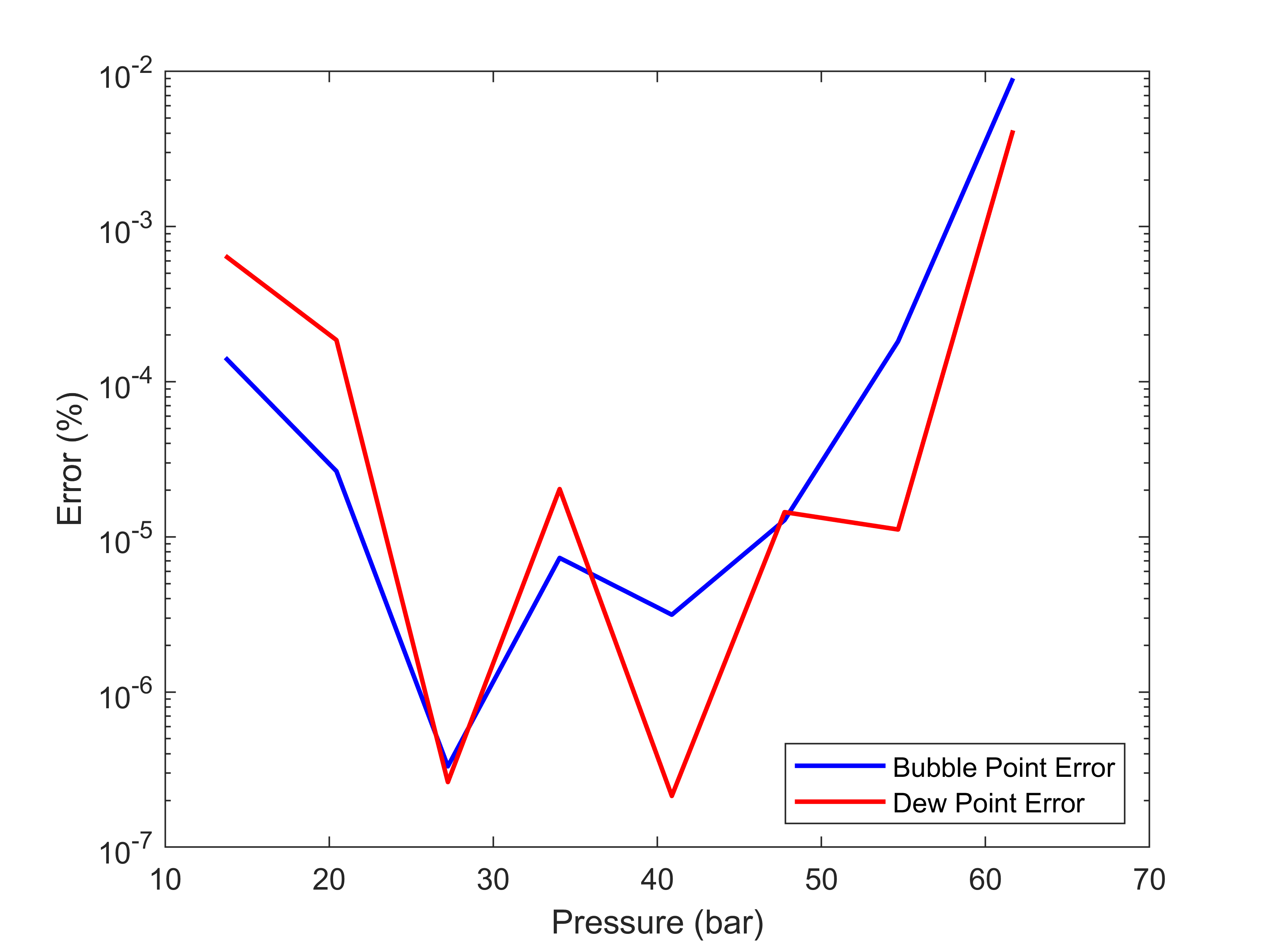}
            \label{figPxEnvelopeD}
        \end{subfigure}
        \hfill 

\caption{Left: The p-x envelopes at (a) T = 274.96 K and (c) T = 387.70 K, where the solid lines represent the theoretical results and the dots represent the results predicted by LBM. Right: The relative error (\%) in the predicted values at (b) T = 274.96 K and (d) T = 387.70 K.}
\label{figPxEnvelope}
\end{figure}

In Figure \ref{figPTmainb}, we see two isotherms $T_1=274.96\ K$ and $T_2=387.70\ K$ such that: $T_1<T_{c,C2}<T_{c,C5}$ and $T_{c,C2}<T_2<T_{c,C5}$. At $T_1$ the p-x envelope will begin from $x_{C2}=0$ and end at $x_{C2}=1$. However, at $T_2$, the p-x envelope will begin from $x_{C2}=0$ and end before reaching $x_{C2}=1$. We perform several simulations at different pressures and a fixed temperature to replicate this behavior with the fugacity-based LBM. A $200\times 2\ (n_x\times n_y)$ computational domain is used, with the relaxation time $\tau=1.25$ and the values interfacial tension strength being $\kappa_{C2}=0.1$ and $\kappa_{C5}=0.2$. The density is initialized using Equation \ref{eqInit} with $W=4$ and $\rho_{i,V}$ and $\rho_{i,L}$ calculated from a flash calculation at the respective initial pressure and temperature. The overall composition in the system is chosen such that $S_V=0.5$ for all simulations. For a binary system, this will not impact the values of phase compositions. Each simulation is run for 500,000 time steps to ensure equilibrium is reached. The theoretical p-x envelope generated by performing flash calculations using the PR EOS and the p-x envelope generated by plotting the equilibrium values of pressure and phase composition from the fugacity-based LBM are shown in Figures \ref{figPxEnvelopeA} and \ref{figPxEnvelopeC}. In order to show the deviation of the values of compositions predicted by the fugacity-based LBM at a certain pressure from the theoretical values, the relative errors are calculated using Equation \ref{eqerror}. The relative error for versus pressure plots are shown in Figures \ref{figPxEnvelopeB} and \ref{figPxEnvelopeD}. It can be seen that the error lies between the order of magnitude of $1\ \%$ to $10^{-8}\ \%$ indicating excellent agreement between the theoretical and predicted values. Again, this excellent agreement is achieved without any type of user intervention, tuning, or empirical manipulation.

\begin{figure}[H]
    \centering
        \begin{subfigure}{0.48\textwidth}
            \centering
            \caption{}
            \includegraphics[width=\textwidth]{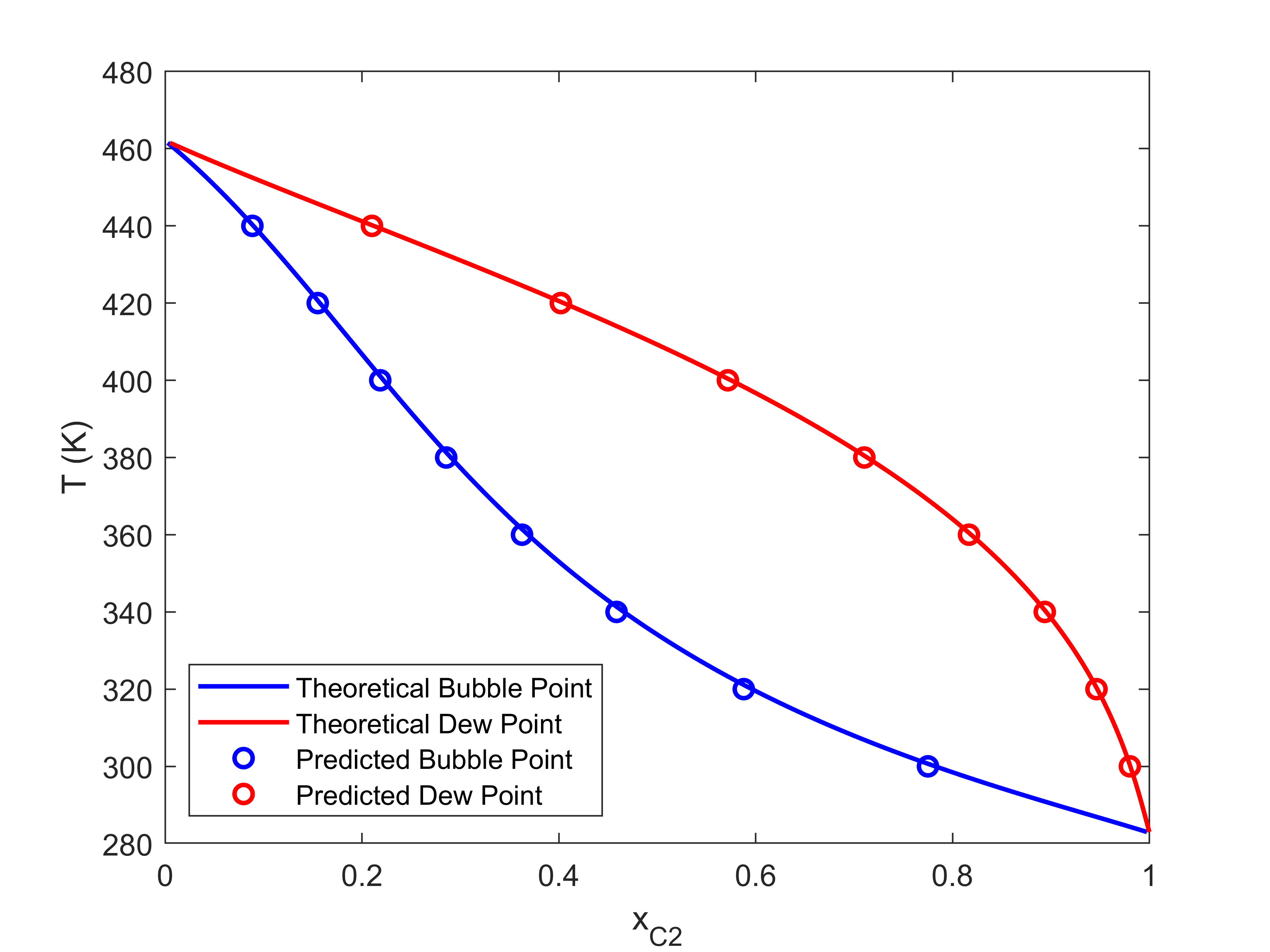}
            \label{figTxEnvelopesA}
        \end{subfigure}%
        \hfill
        \begin{subfigure}{0.48\textwidth}
            \centering
            \caption{}
            \includegraphics[width=\textwidth]{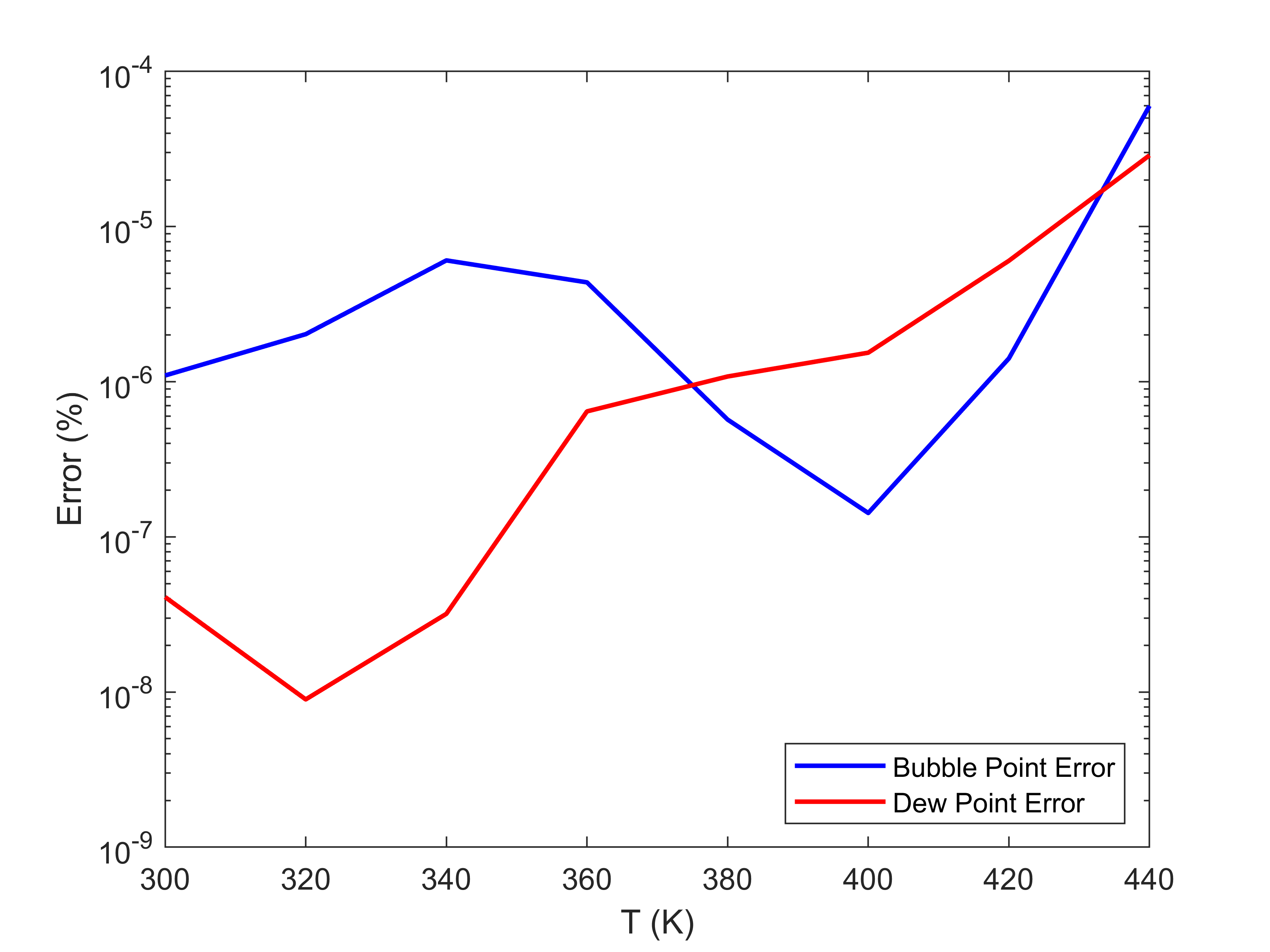}
            \label{figTxEnvelopesB}
        \end{subfigure}%
        \hfill
        \begin{subfigure}{0.48\textwidth}
            \centering
            \caption{}
            \includegraphics[width=\textwidth]{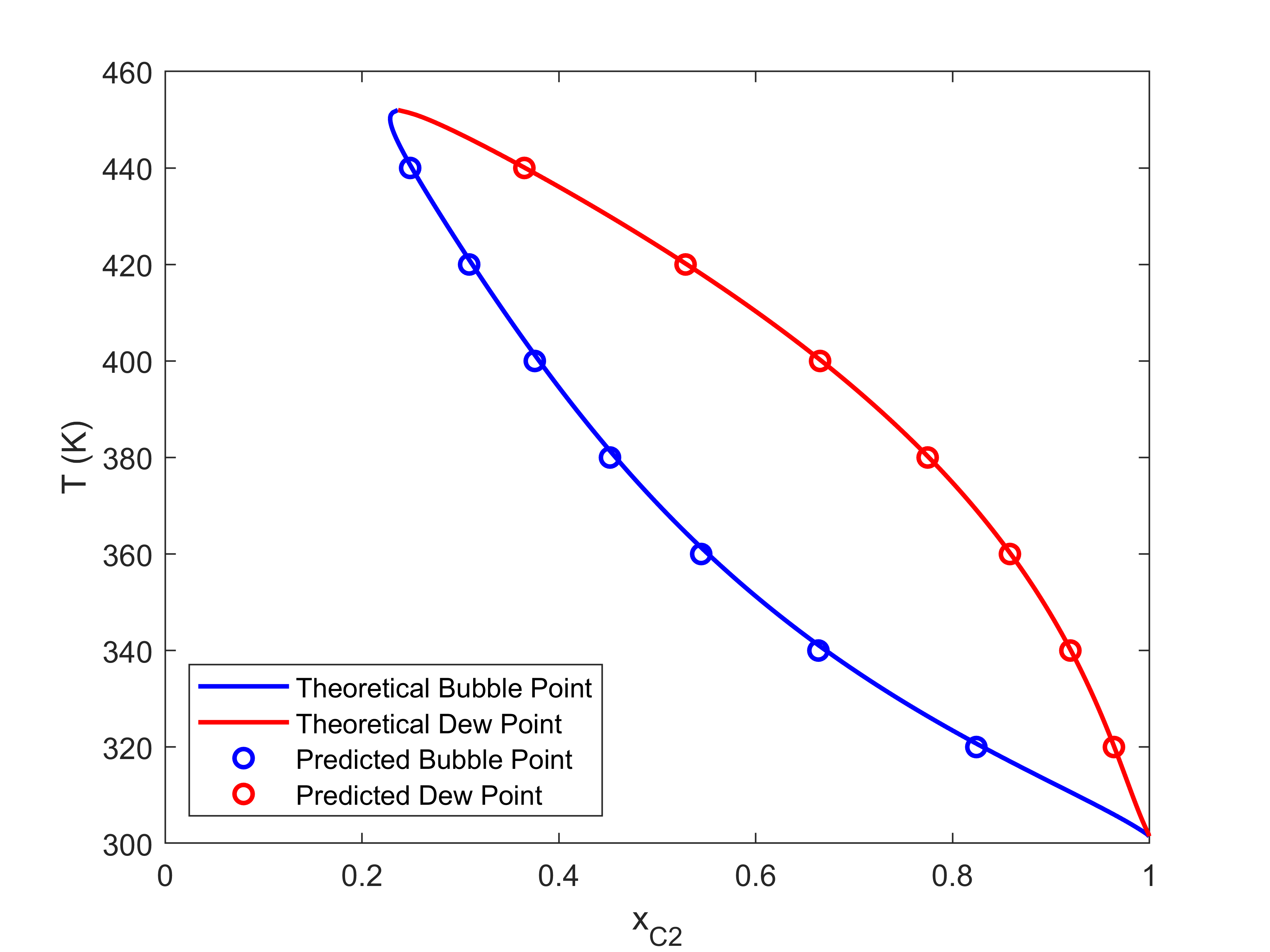}
            \label{figTxEnvelopesC}
        \end{subfigure}
        \hfill    
        \begin{subfigure}{0.48\textwidth}
            \centering
            \caption{}
            \includegraphics[width=\textwidth]{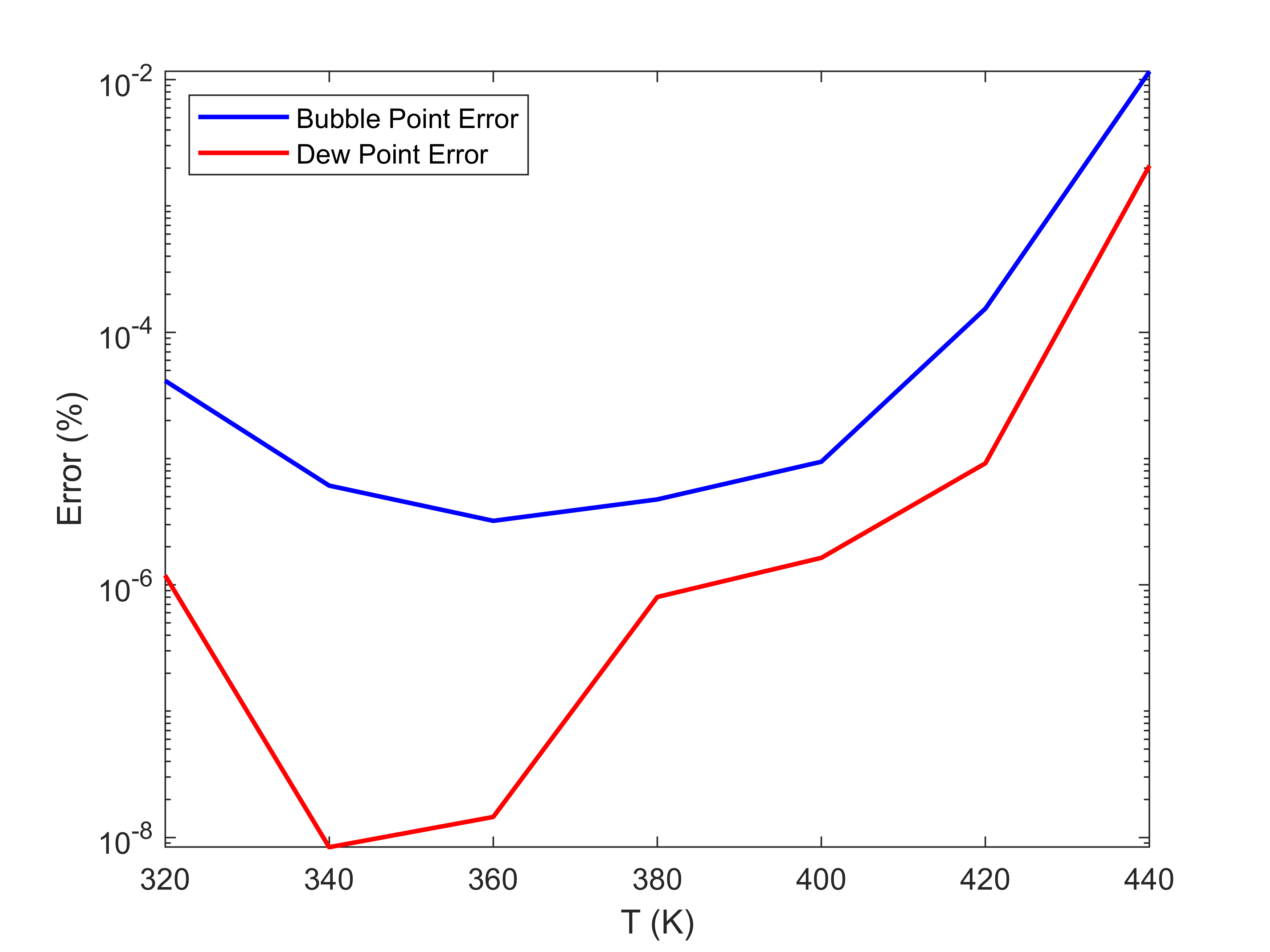}
            \label{figTxEnvelopesD}
        \end{subfigure}
        \hfill 
        \begin{subfigure}{0.48\textwidth}
            \centering
            \caption{}
            \includegraphics[width=\textwidth]{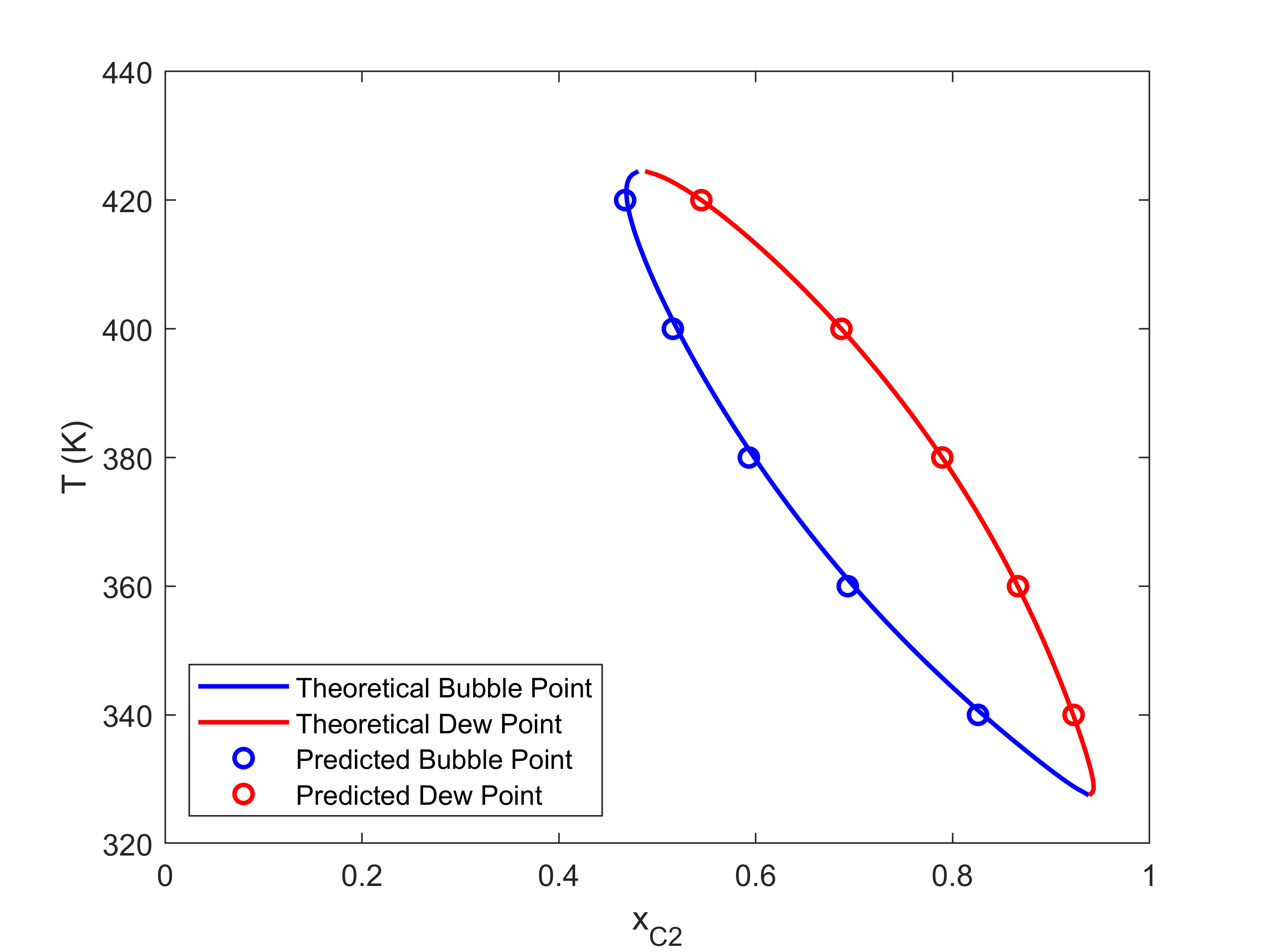}
            \label{figTxEnvelopesE}
        \end{subfigure}
        \hfill 
        \begin{subfigure}{0.48\textwidth}
            \centering
            \caption{}
            \includegraphics[width=\textwidth]{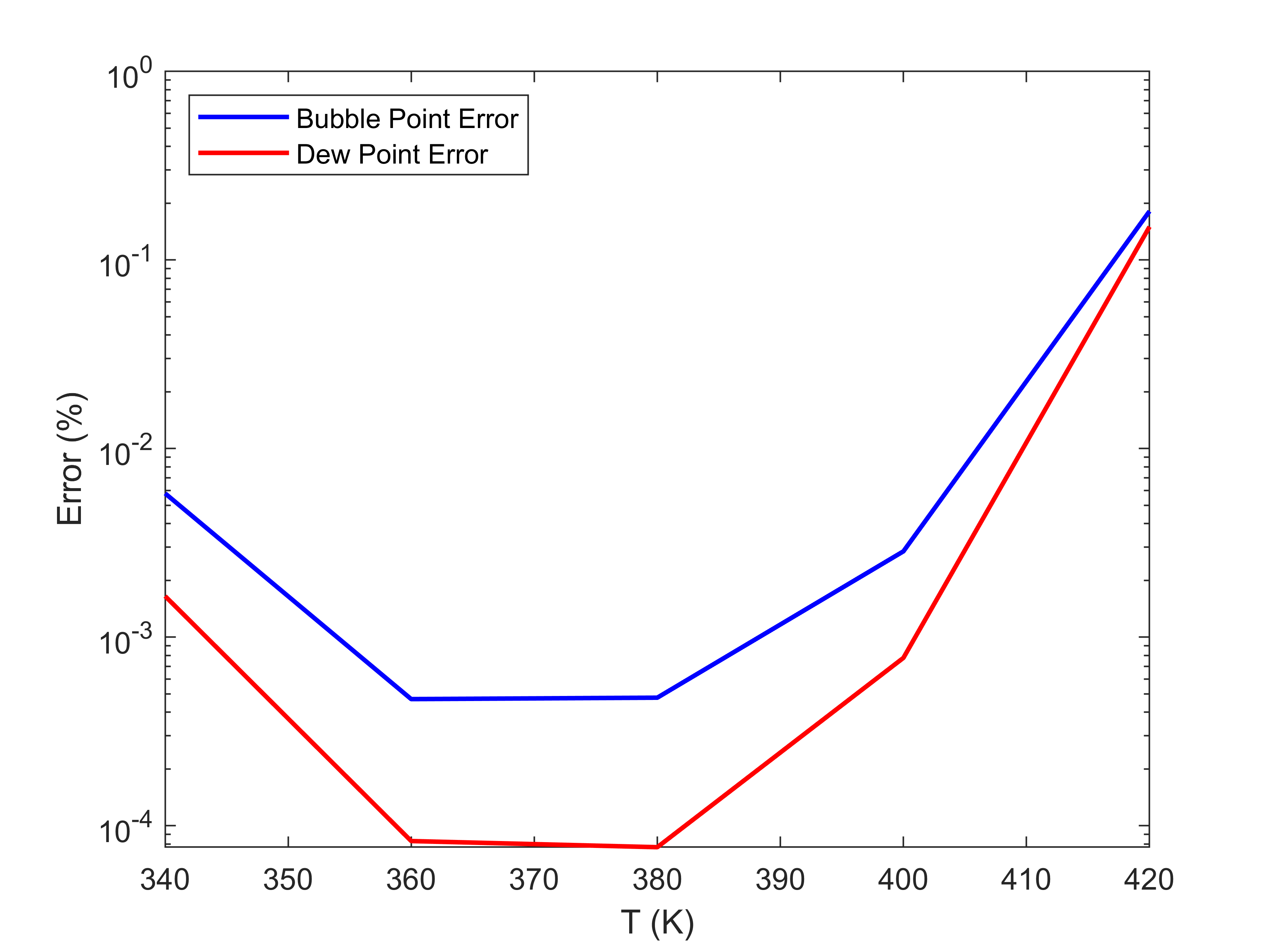}
            \label{figTxEnvelopesF}
        \end{subfigure}
        \hfill 

\caption{Left: The T-x envelopes at (a) P = 30 bar, (c) P = 45 bar, and (e) P = 58 bar, where the solid lines represent the theoretical results and the dots represent the results predicted by LBM. Right: The relative error (\%) in the predicted values at (b) P = 30 bar, (d) P = 45 bar, and (f) P = 58 bar.}
\label{figTxEnvelopes}
\end{figure}

Next, this process is repeated at constant pressure to create T-x envelopes. Figure \ref{figPTmainb} shows three relevant isobars $p_1=30\ bar$, $p_2=45\ bar$, and $p_3=58\ bar$ such that: $p_1<p_{c,C5}<p_{c,C2}$, $p_{c,C5}<p_2<p_{c,C2}$, and $p_{c,C5}<p_{c,C2}<p_3$. At $p_1$ the T-x envelope will begin from $x_{C2}=0$ and end at $x_{C2}=1$, at $p_2$ the T-x envelope will begin after $x_{C2}=0$ and end at $x_{C2}=1$, and at $p_3$ the T-x envelope will begin after $x_{C2}=0$ and end before $x_{C2}=1$. This behavior is replicated using the fugacity-based LBM using the same values of $n_x$, $n_y$, $\tau$, $\kappa_{C2}$, $\kappa_{C5}$, $W$, $S_V$ and the number of time steps as those used to generate the P-x envelopes. The temperature is varied at a given pressure, and $\rho_{i,V}$ and $\rho_{i,L}$ are initialized accordingly. The theoretical T-x envelope generated by performing flash calculations using the PR EOS and the T-x envelope generated by plotting the equilibrium values of temperature and phase composition from the fugacity-based LBM are shown in Figures \ref{figTxEnvelopesA}, \ref{figTxEnvelopesC}, and \ref{figTxEnvelopesE}. The deviation of the predicted values from the theoretical ones is slightly greater for the T-x envelope than for the P-x envelope. This is because for the T-x envelope, we are assuming that the pressure remains constant at its initial value (at $p_1$, $p_2$, or $p_3$) for each temperature condition in the envelope. However, the pressure deviates slightly from the initial value as the initialization is not done at the true equilibrium because of the ``tanh" interface approximation as discussed in Section \ref{seccase1}. When the errors at each temperature are calculated using the theoretical values of composition at the actual equilibrium pressure (and not the pressure of the chosen isobar), the error versus temperature graphs are shown in Figures \ref{figTxEnvelopesB}, \ref{figTxEnvelopesD}, and \ref{figTxEnvelopesF}. The errors range in orders of magnitude from $10^{-2}\ \%$ to $10^{-8}\ \%$, again showing excellent agreement between predicted and theoretical results.

For the reader's reference, the equivalent vapor and liquid phase density versus pressure plots for the simulations in Figure \ref{figPxEnvelope} and the equivalent vapor and liquid phase density versus temperature plots for the simulations in Figure \ref{figTxEnvelopes} are shown in Appendix \ref{appendixDenplot}.

\subsection{Three component mixtures}
\label{seccase3}
\begin{figure}[H]
    \centering
        \begin{subfigure}{0.48\textwidth}
            \centering
            \caption{}
            \includegraphics[width=\textwidth]{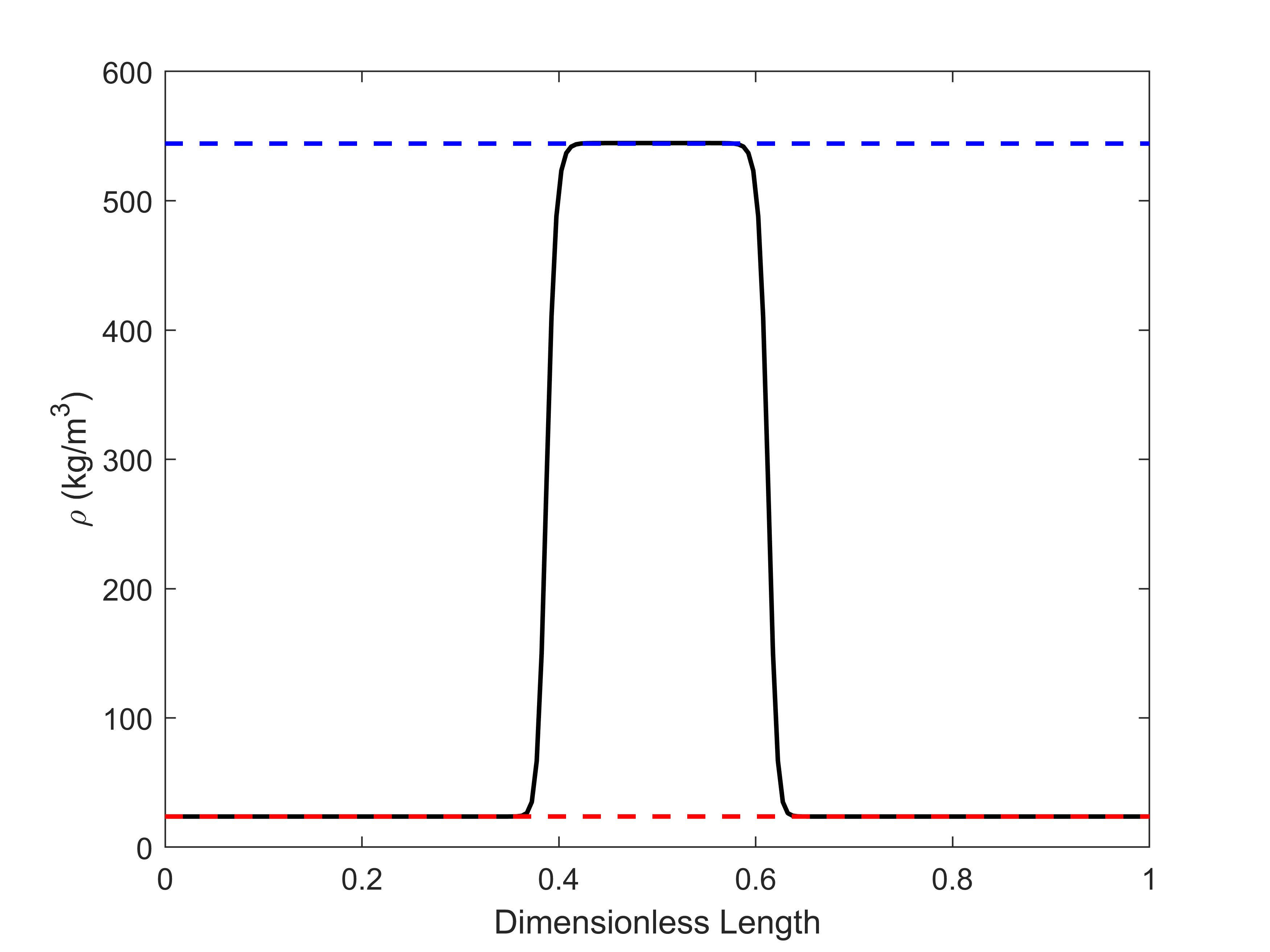}
        \end{subfigure}%
        \hfill
        \begin{subfigure}{0.48\textwidth}
            \centering
            \caption{}
            \includegraphics[width=\textwidth]{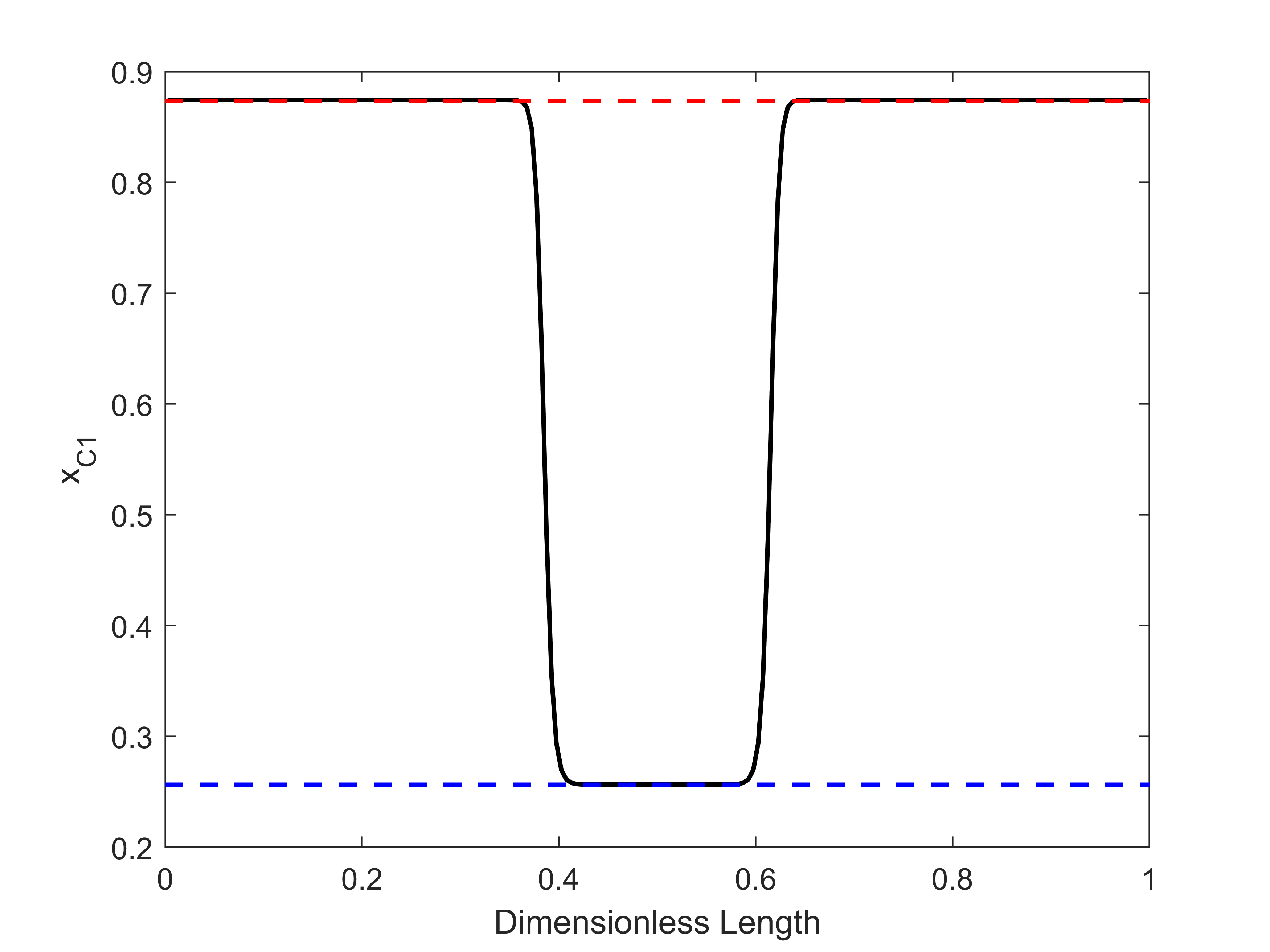}
        \end{subfigure}
        \hfill        
        \begin{subfigure}{0.48\textwidth}
            \centering
            \caption{}
            \includegraphics[width=\textwidth]{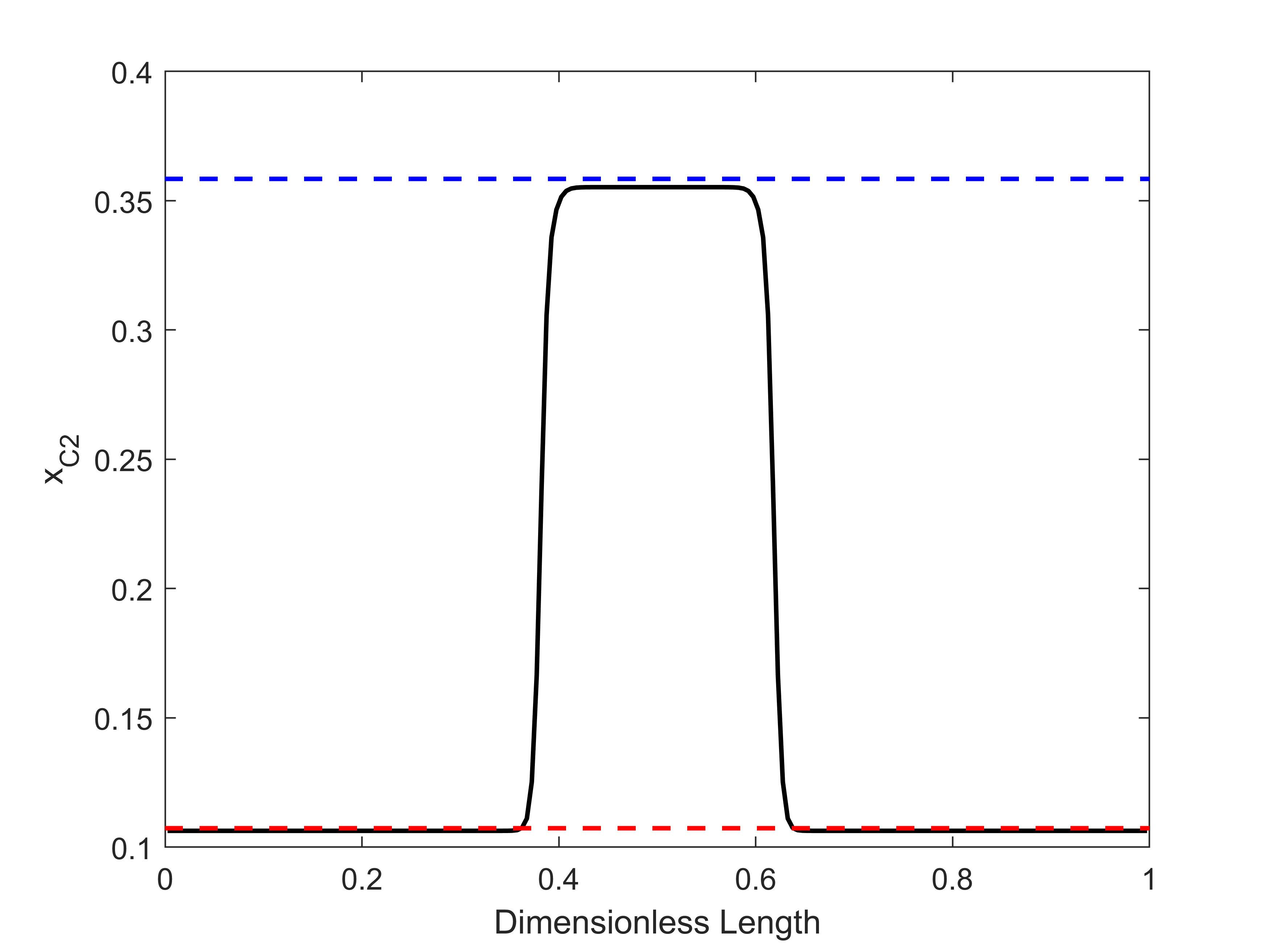}
        \end{subfigure}
        \hfill 
        \begin{subfigure}{0.48\textwidth}
            \centering
            \caption{}
            \includegraphics[width=\textwidth]{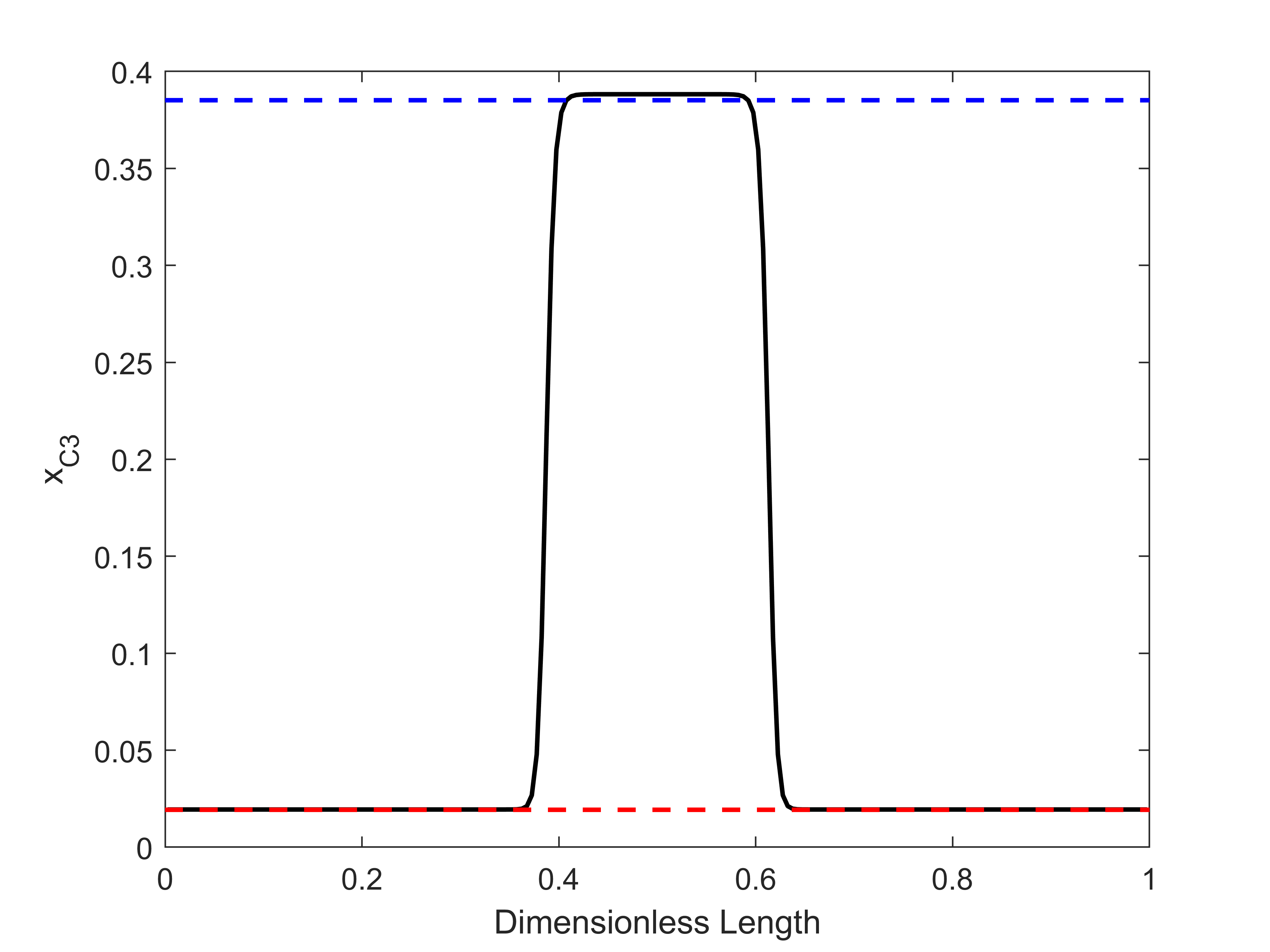}
        \end{subfigure}
        \hfill 

\caption{The equilibrium profiles for the (a) density, (b) composition of C1, (c) composition of C2, and (d) composition of C3. The solid black line shows the results from LBM and the dashed blue and red lines show the
theoretical values from a flash calculation for the liquid and vapor phase respectively.}
\label{figcase3}
\end{figure}

In this section, we show that our formulation is not restricted to binary mixtures by simulating the two-phase vapor-liquid system with a flat interface using three components. Methane (C1), Ethane (C2), and Propane (C3) with the overall mole fractions 0.4, 0.3, and 0.3, respectively, are chosen for this case. The PR EOS is used. The size of the computational domain is $200\times 2\ (n_x\times n_y)$, the relaxation time $\tau=1$, and the values of interfacial tension strength are: $\kappa_{C1}=0.05$, $\kappa_{C2}=0.10$ and $\kappa_{C3}=0.15$. The system is initialized at a temperature of 216.483 K and a pressure of 20.684 bar. The density is initialized using Equation \ref{eqInit}, with $W=4$ and $\rho_{i,V}$, $\rho_{i,L}$ and $S_V$ calculated by performing a flash calculation using the PR EOS at 216.483 K and 20.684 bar. The simulation is run for 1,000,000 time steps until equilibrium is reached. The results of density and composition versus dimensionless length ($x/n_x$) are shown in Figure \ref{figcase3}. The results from a flash calculation at the temperature and updated pressure at the equilibrium state are also included in Figure \ref{figcase3}.

The deviation of the results of the fugacity-based LBM from the results of the flash calculation is quantified with a relative error (Equation \ref{eqerror}) and tabulated in Table \ref{tabcase3}. The low error values show that the fugacity-based LBM also predicts the correct values for multicomponent mixtures. 

\begin{table}[h]
\caption{The relative error values in the density, C1 composition, C2 composition, and C3 composition in the different phases.}
\label{tabcase3}
\centering
\begin{tabular}{|c|c|c|}
\hline
\ Property\ \  & Liquid Region Error (\%) & Vapor Region Error (\%) \\
\hline
Density  & 0.0705                   & 0.0677                  \\
$x_{C1}$       & 0.0521                   & 0.0970                  \\
$x_{C2}$       & 0.8962                   & 0.9207                  \\
$x_{C3}$       & 0.7994                   & 0.7320                 \\
\hline
\end{tabular}
\end{table}

\subsection{Spinodal Decomposition}
\begin{figure}[H]
    \centering
        \begin{subfigure}{0.48\textwidth}
            \centering
            \caption{}
            \includegraphics[width=\textwidth]{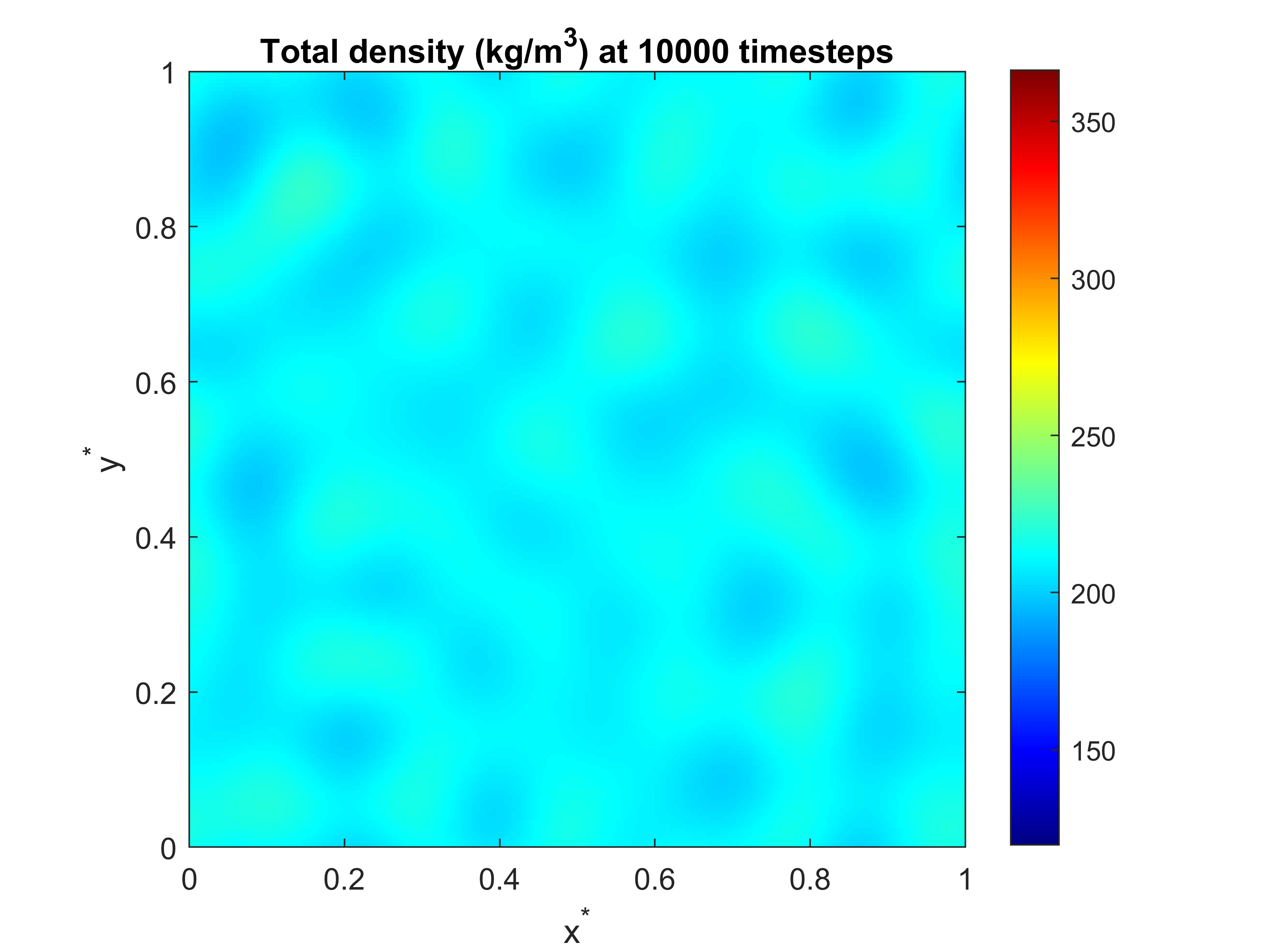}
            \label{figSpinodalA}
        \end{subfigure}%
        \hfill
        \begin{subfigure}{0.48\textwidth}
            \centering
            \caption{}
            \includegraphics[width=\textwidth]{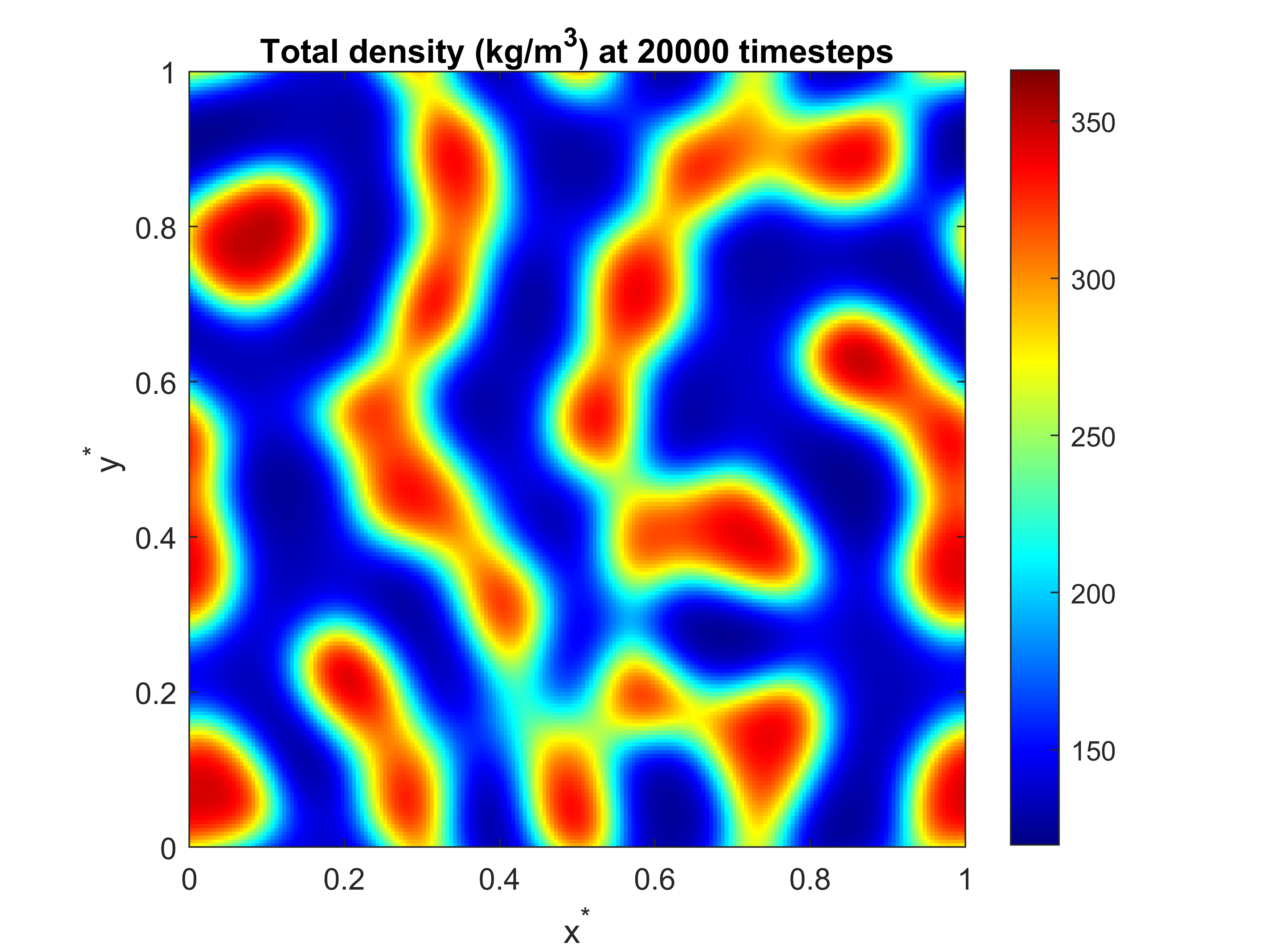}
            \label{figSpinodalB}
        \end{subfigure}
        \hfill        
        \begin{subfigure}{0.48\textwidth}
            \centering
            \caption{}
            \includegraphics[width=\textwidth]{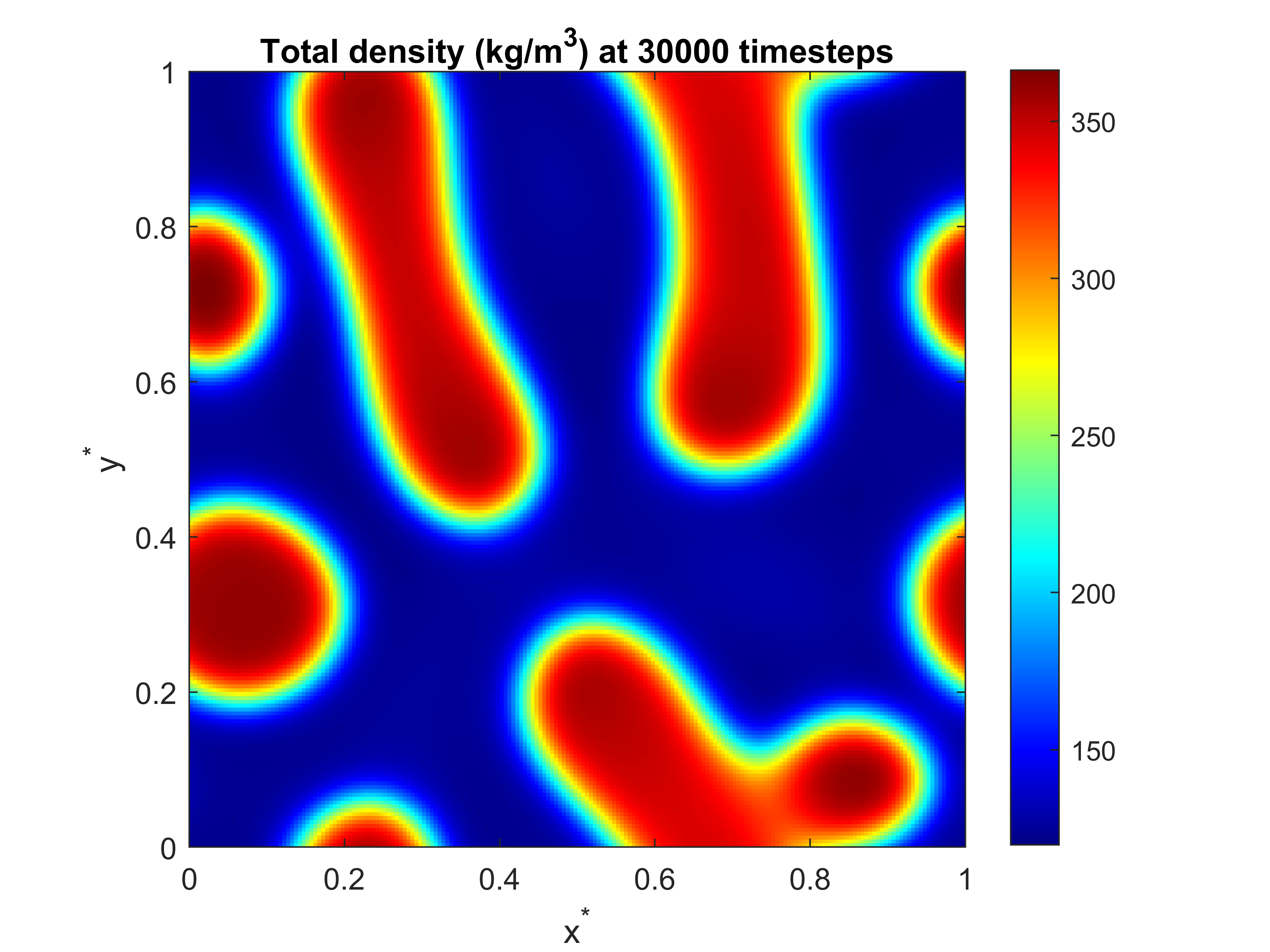}
            \label{figSpinodalC}
        \end{subfigure}
        \hfill 
        \begin{subfigure}{0.48\textwidth}
            \centering
            \caption{}
            \includegraphics[width=\textwidth]{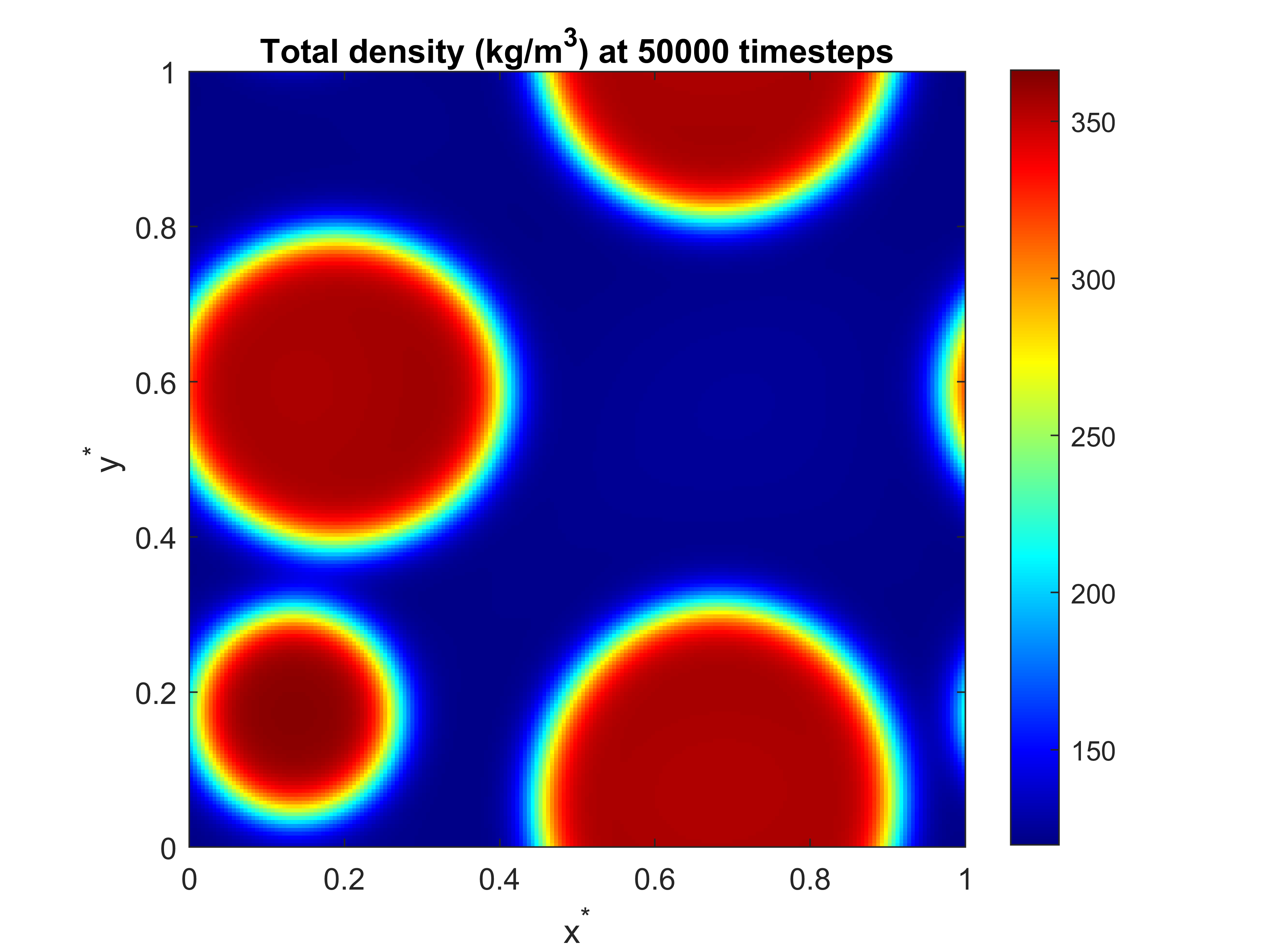}
            \label{figSpinodalD}
        \end{subfigure}
        \hfill 
        \begin{subfigure}{0.48\textwidth}
            \centering
            \caption{}
            \includegraphics[width=\textwidth]{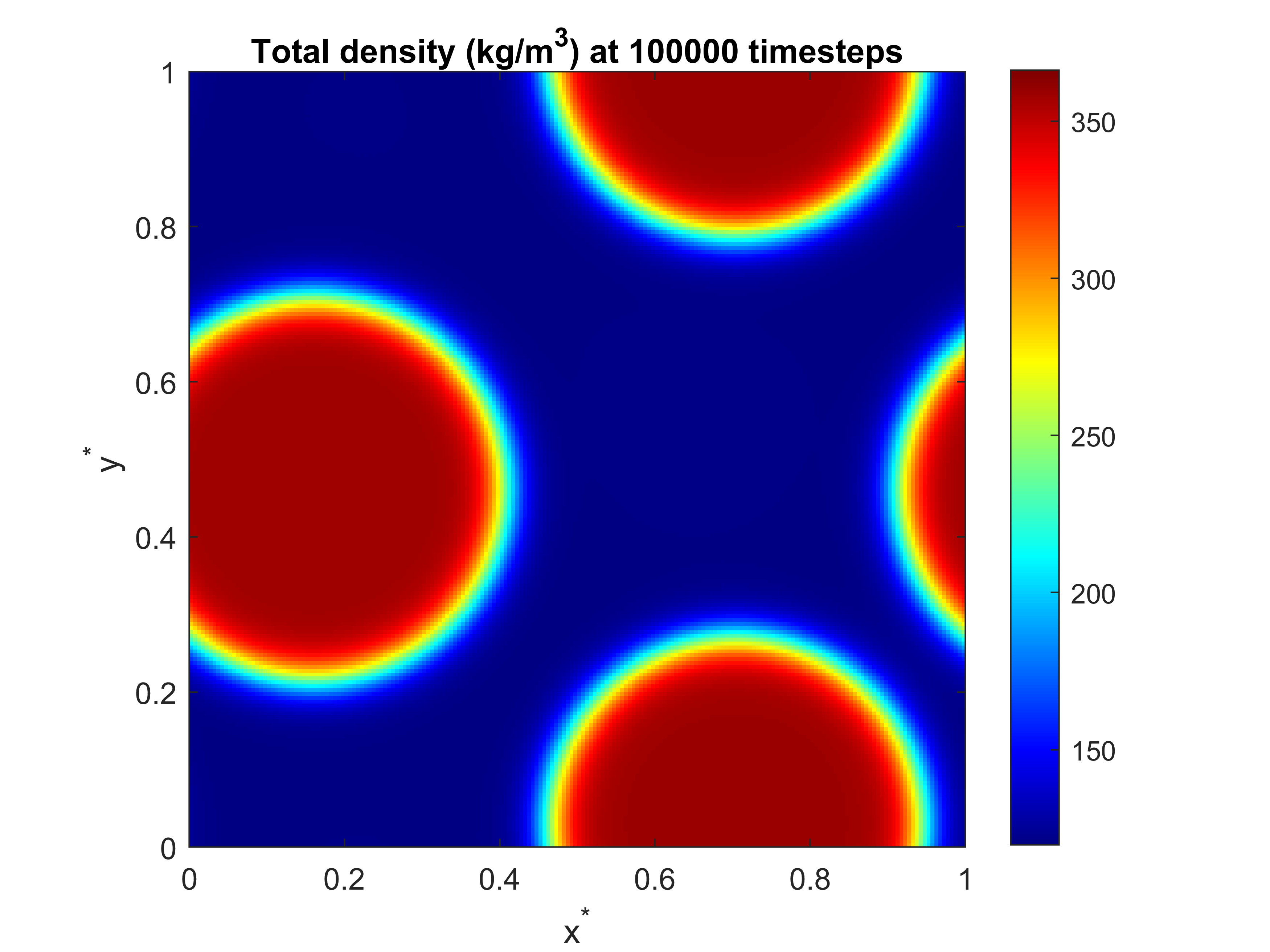}
            \label{figSpinodalE}
        \end{subfigure}
        \hfill 
        \begin{subfigure}{0.48\textwidth}
            \centering
            \caption{}
            \includegraphics[width=\textwidth]{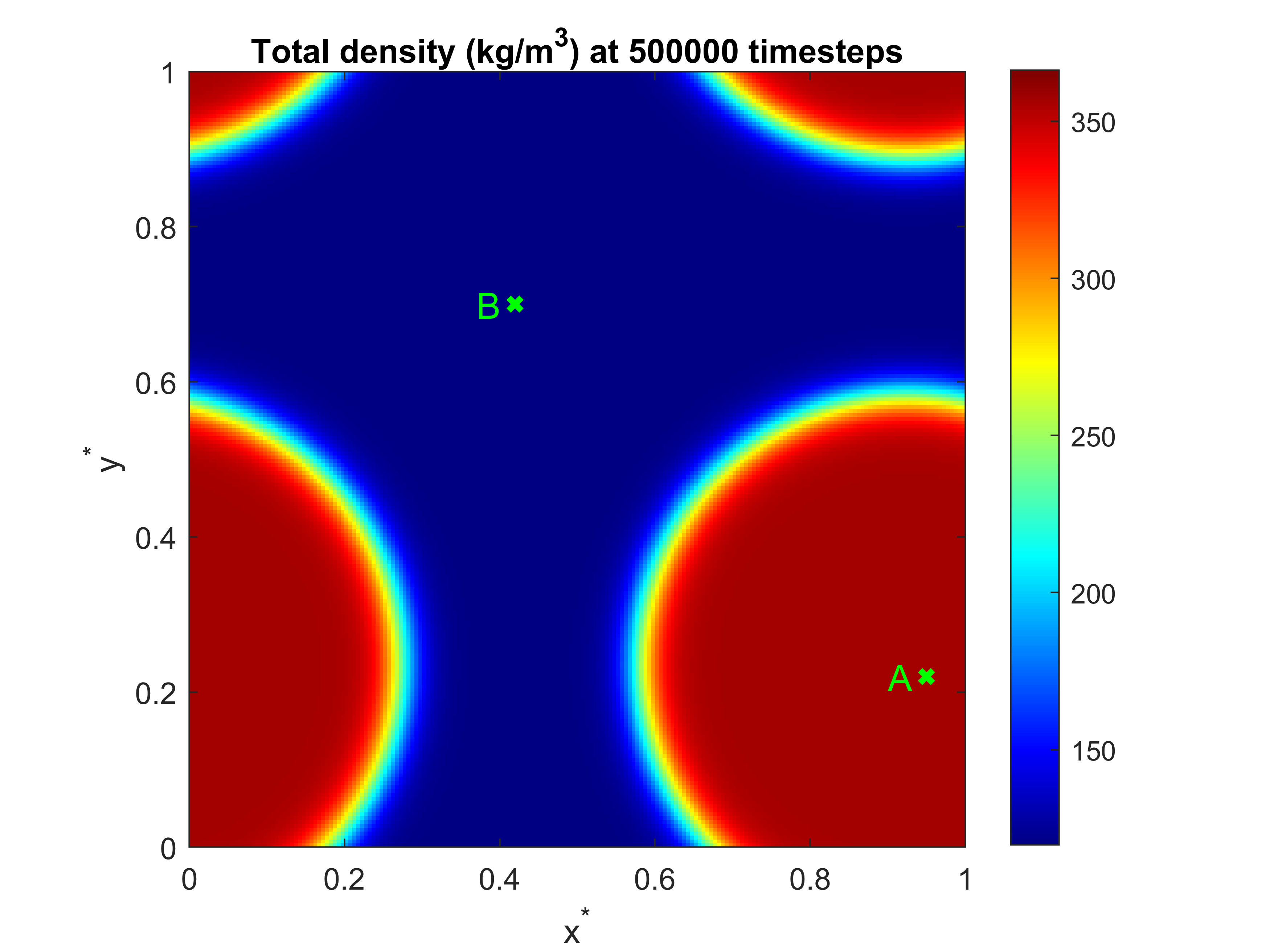}
            \label{figSpinodalF}
        \end{subfigure}
        \hfill
\caption{A uniform system separating into different phases shown at time (in lattice units) (a) 10,000, (b) 20,000, (c) 30,000, (d) 50,000, (e) 100,000, and (f) 500,000. The points marked A and B, in (f), represent the points in the liquid and vapor region, respectively, where relevant properties are measured.}
\label{figSpinodal}
\end{figure}

So far, we have shown results of our method initialized at equilibrium conditions, using Equation \ref{eqInit}. In this section, we show that our model allows for phase separation, in agreement with thermodynamics, when initialized far from equilibrium. To do this we simulate a case of spinodal decomposition of a system with Ethane (C2) and n-Pentane (C5), using the PR EOS. The size of the computational domain is $200 \times 200\ (n_x \times n_y)$, the relaxation time $\tau = 1.15$, and the values of interfacial tension strength are: $\kappa_{C2}=0.10$ and $\kappa_{C5}=0.15$. The temperature in the domain is $387.70\ K$. The initial composition in the domain is kept uniform at $x_{C2}=0.62$ and $x_{C5}=0.38$ and the density is also kept uniform and such the pressure in the domain is $50\ bar$. As can be seen in Figure \ref{figPxEnvelopeC}, at these conditions the domain should decompose into two separate phases. To trigger this decomposition, we introduce a random fluctuation in the total density of the system, and run the simulation for $500,000$ time steps. The evolution of the total density with time is shown in Figure \ref{figSpinodal}. In these plots, the domain is characterized in terms of dimensionless lengths $x^*=x/n_x$ and $y^*=y/n_y$.

Figure \ref{figSpinodal} confirms that our model allows for phase separation of a domain initialized far from equilibrium. Now we test whether the results are consistent with thermodynamics. It should be noted that in Sections \ref{seccase1}, \ref{seccase2}, and \ref{seccase3}, to test for consistency with thermodynamics, the results from LBM were compared with results from a flash calculation. However, a flash calculation assumes a flat interface, therefore, we cannot use the same approach to verify the results in this section. Instead we verify our results by testing the iso-fugacity criterion, which applies to both flat and curved interfaces and is the kernel of the flash calculation. The fugacity of each component in the liquid phase (obtained at point A marked in Figure \ref{figSpinodalF}) and the vapor phase (obtained at point B marked in Figure \ref{figSpinodalF}) is measured and tabulated in Table \ref{tabFug}. It can be seen that the fugacity of each component is the same in the liquid and vapor phase, which confirms that the iso-fugacity criterion holds and that the phase separation predicted by the fugacity-based LBM is consistent with thermodynamics.

\begin{table}[h]
\caption{The fugacity of each component measured in each phase.}
\label{tabFug}
\centering
\begin{tabular}{|c|c|c|}
\hline
\ Component\ \  & Fugacity in liquid phase (bar) & Fugacity in vapor phase (bar) \\
\hline
C2  & 35.1276     & 35.1262        \\
C5  & 4.4563      & 4.4567          \\
\hline
\end{tabular}
\end{table}

\subsection{Oscillating Droplet}
\begin{figure}[h]
    \centering
    \includegraphics[scale=0.5]{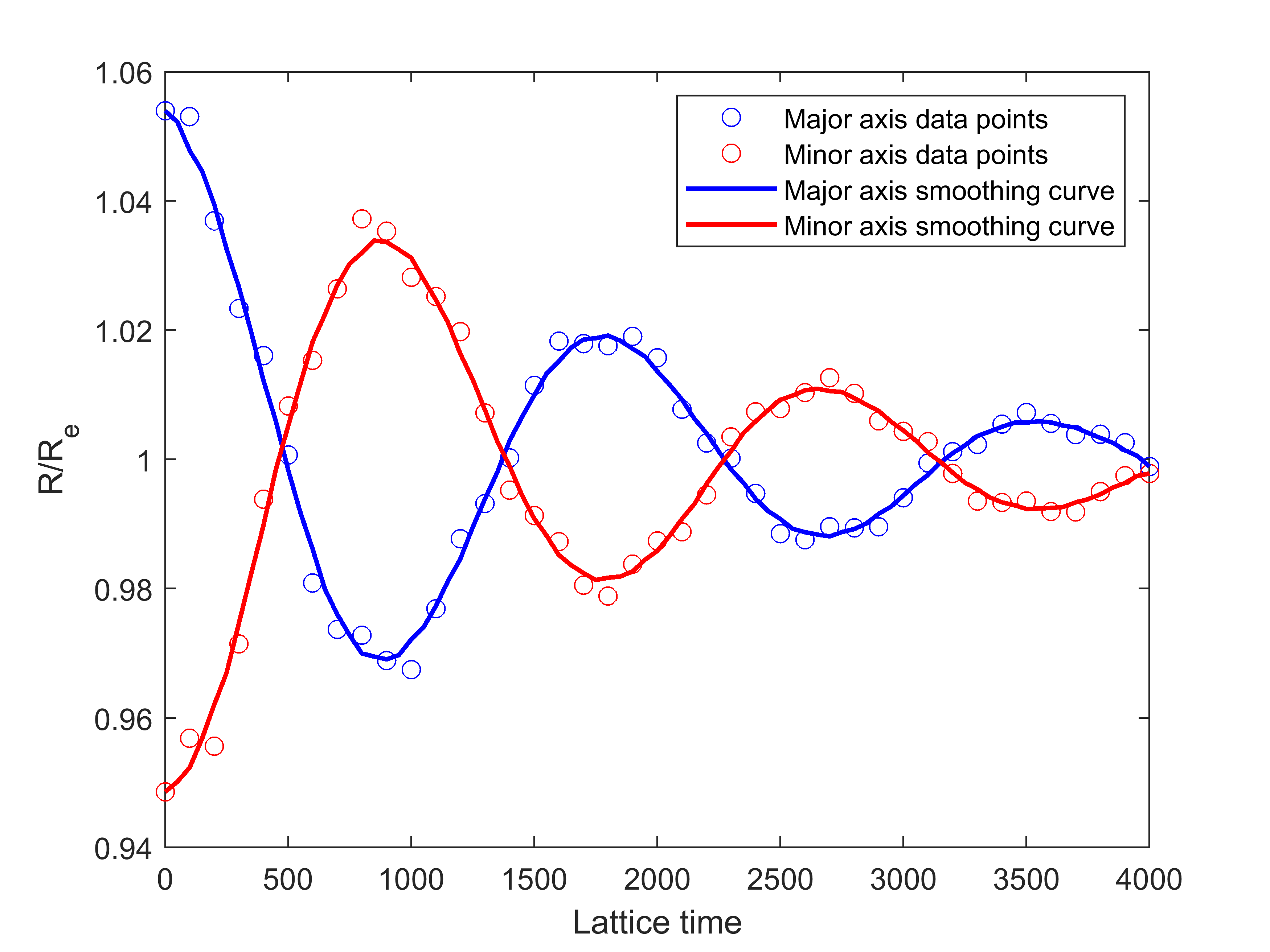}
    \vspace{-0.35cm}
    \caption{The ratio of the droplet radius to the equilibrium radius versus time (in lattice units). The dots represent the raw data obtained from the simulation and the solid lines are the smooth curves fit to this data.}
    \label{figOscillating}
\end{figure}

For all the tests done so far, we have verified results under static conditions and in this section, we will test our model at dynamic conditions. We simulate the case of an elliptical droplet oscillating due to the effects of surface tension, which has been used to benchmark single component \cite{Li2013} and multicomponent LBM models \cite{Peng2021}. For this case Propane (C3) and n-Pentane (C5) are simulated using the PR EOS. The size of the computational domain is $200\times 200\ (n_x\times n_y)$, the relaxation time $\tau=0.85$, and the values of interfacial tension strength are: $\kappa_{C3}=0.05$ and $\kappa_{C5}=0.10$. The temperature of the system is set to $340\ K$ and the density is initialized using Equation \ref{eqInit2D}.
\begin{equation}
    \label{eqInit2D}
    \rho_i(x,y,t=0)=\frac{\rho_{i,L}+\rho_{i,V}}{2}-\frac{\rho_{i,L}-\rho_{i,V}}{2}\tanh{\left(\frac{2 \left[\sqrt{\left(x-x_c\right)^2+\left(y-y_c\right)^2/h^2}-R_0\right]}{W}\right)}
\end{equation}
Here $x_c$ and $y_c$ are the $x$ and $y$ coordinates of the center of the droplet set to be $x_c=n_x/2$ and $y_c=n_y/2$. The parameters controlling the shape of the elliptical droplet are set to be: $R_0 =30$, $h=0.9$, and $W=4$. Additionally, $\rho_{i,V}$ and $\rho_{i,L}$ are calculated by performing a flash calculation using the PR EOS at $T=340\ K$ and $p=5\ bar$, which corresponds to a density ratio of $50.97$. With this initialization, the elliptical droplet will oscillate, and the time period of this oscillation can be determined theoretically, as shown in Equation \ref{eqPeriod} \cite{Lamb1932}.
\begin{equation}
    \label{eqPeriod}
    T_a=2\pi \left[n\left(n^2-1\right)\frac{\sigma}{\rho_LR_e^3}\right]^{-1/2}
\end{equation}
Here $T_a$ is the time period of the oscillations, $n$ is the mode of oscillation ($n=2$ for this case), $\sigma$ is the interfacial tension and $R_e$ is the equilibrium radius given by $R_e=\sqrt{R_{max}R_{min}}$, with $R_{max}=R_0$ and $R_{min}=R_0h$. The simulation is run for 4000 time steps, and the ratio of the droplet radius to the equilibrium radius is plotted with time, along both the major and minor axis. It should be noted that the radius does not have a clear definition due to the presence of a diffuse interface and it is calculated by defining the boundary of the droplet to be the location where $\rho=(\rho_V+\rho_L)/2$. The location of this boundary is determined by linearly interpolation of the density distribution in space. This uncertainty in the radius introduces a scatter in the data extracted from the simulation, and the scatter is further amplified since we are capturing very small variations in the radius. Therefore, a smooth curve is fit to the data extracted from the simulation. The raw data and the smoothing curves are shown in Figure \ref{figOscillating}. From Figure \ref{figOscillating}, the period of the oscillations can be determined to be 1787.5 time steps (in lattice units), which agrees closely with the theoretical period of 1717.5 time steps, determined through Equation \ref{eqPeriod}.

\section{Discussion and Conclusion}
\label{secConclusion}
In this paper, we introduced a model for the free energy LBM, which can not only handle MCMP simulations of partially miscible fluids but is also able to incorporate any multicomponent EOS, without the need for any empirically determined tuning parameter to achieve full thermodynamic consistency. This model calculates the force required in the free energy model for each component based on the fugacity of that component, a quantity readily available for any multicomponent EOS. Using this fugacity-based LBM model, we could show that accurate EOSs, like the PR and SRK equations, could easily be incorporated into LBM. Additionally, we were able to use the fugacity-based LBM to extensively analyze a binary system's pressure, temperature, and composition behavior by replicating various characteristic pressure-composition and temperature-composition envelopes. We further showed that the fugacity-based LBM model could be applied to any number of components. All results from the fugacity-based LBM showed excellent agreement with the results predicted by multicomponent thermodynamics, with deviations well below 1\%. We also tested the case of spinodal decomposition for a binary mixture and showed that a uniform system could decompose into two phases, in agreement with thermodynamics. Finally, we tested the model under dynamic conditions by simulating the case of an oscillating droplet and showing that the oscillation period obtained from the simulation agrees with the theoretical prediction.

A few things need to be stressed out about this procedure. Firstly, our model is not restricted to the use of the well-balanced LBM formulation. It can apply to the standard LBM formulation, although there will be slight deviations from thermodynamic predictions. A general form of our proposed force for the standard multicomponent LBM would be given by Equation \ref{eqForceGen}.
\begin{equation}
    \label{eqForceGen}
    \mathbf{F}_i=-\tilde{\rho}_i RT\nabla\ln{f_i}-\tilde{\rho}_i\nabla\mu_{I,i}+c_s^2\nabla\rho_i
\end{equation}
Secondly, the well-balanced LBM is unstable at low relaxation time values and high density ratios. This problem can be addressed with a more stable version of the Well-balanced LBM as shown in Ref. \cite{Zhang2022WB}. Thirdly, the interfacial tension in our model is set indirectly, through the parameter $\kappa_i$. While the interfacial tension cannot be obtained beforehand, it can be calculated for a curved interface through the Young-Laplace test or for a flat interface by integrating the difference between the normal and transversal components of the pressure tensor across the interface, resulting in Equation \ref{eqMCsigma} \cite{Carey1980a}. Both of these approaches would require knowing simulation results.
\begin{equation}
    \label{eqMCsigma}
    \sigma=\int_{-\infty}^{+\infty} \left[\sum_{i=1}^{N_c}\sum_{j=1}^{N_c} \kappa_{ij} \frac{d \tilde{\rho}_i}{d x} \frac{d \tilde{\rho}_j}{d x}\right]dx
\end{equation}
 Lastly, the expression for fugacity is generally written with dimensionless groups inside the natural logs as shown in Equations \ref{eqfugvdW}, \ref{eqfugSRK} and \ref{eqfugPR} in Appendix \ref{secAppendixEOS}. This way of calculating fugacity could potentially lead to instabilities as the pressure can take on a negative value in the interface region which can cause the term  $\ln{\left[\left(\tilde{v}-b_m\right)p/(RT)\right]}$ to turn into a log of a negative number. This term will appear in the fugacity expression for any EOS. To avoid this, we recommend calculating the natural log of the fugacity at each node as shown in Equation \ref{eqfugupdateVDW} for the vdW EOS, Equation \ref{eqfugupdateSRK} for the SRK EOS, or Equation \ref{eqfugupdatePR} for the PR EOS. Then Equation \ref{eqforcefinal} can be used with the natural log of the fugacity to calculate the component force. 
\begin{equation}
    \label{eqfugupdateVDW}
    \ln{f_i} = \ln{x_i}+ \frac{b_i}{\tilde{v}-b_{m}}-\ln{\left[\frac{\left(\tilde{v}-b_m\right)}{RT}\right]}-\frac{2}{RT\tilde{v}}\sum_{j=1}^{N_c}x_ja_{ij}
\end{equation}

\begin{equation}
    \label{eqfugupdateSRK}
    \ln{f_i} = \ln{x_i}+ \frac{b_i}{b_{m}}\left[\frac{p\tilde{v}}{RT}-1\right]-\ln{\left[\frac{\left(\tilde{v}-b_m\right)}{RT}\right]}+\frac{(a\alpha)_{m}}{b_mRT}\left[\frac{b_i}{b_m}-\frac{2}{(a\alpha)_m}\sum_{j=1}^{N_c}x_j(a\alpha)_{ij}\right]\ln{\left[1+\frac{b_m}{\tilde{v}}\right]}
\end{equation}

\begin{equation}
    \label{eqfugupdatePR}
         \ln{f_i} = \ln{x_i}+ \frac{b_i}{b_{m}}\left[\frac{p\tilde{v}}{RT}-1\right]-\ln{\left[\frac{\left(\tilde{v}-b_m\right)}{RT}\right]}+\frac{(a\alpha)_{m}}{2\sqrt{2}b_mRT}\left[\frac{b_i}{b_m}-\frac{2}{(a\alpha)_m}\sum_{j=1}^{N_c}x_j(a\alpha)_{ij}\right]\ln{\left[\frac{\tilde{v}+\left(1+\sqrt{2}\right)b_m}{\tilde{v}+\left(1-\sqrt{2}\right)b_m}\right]}
\end{equation}

\begin{acknowledgments}
Funding support from the William A. Fustos Family Professorship in Energy and Mineral Engineering at the Pennsylvania State University is gratefully acknowledged.
\end{acknowledgments}

%

\appendix
\section{Pitfalls of deriving free energy from an equation of state}
\label{appendixA}
In this study, we propose to create a multicomponent LBM model that does not rely on deriving the free energy functional from an equation of state (EOS), as is done for single component models. To see why deriving the free energy functional from an EOS would not work in a multicomponent case, we start by writing the total differential of the Helmholtz free energy given by Equation \ref{eqHelmholtz}. 
\begin{equation}
\label{eqHelmholtz}
    d\Psi=-SdT-pdV+\sum_{i=1}^{N_c}\mu_idn_i
\end{equation}
Here $S$ is the entropy. We can identify that the pressure is the negative partial derivative of the free energy with respect to volume as shown in Equation \ref{eqpresspartial}.
\begin{equation}
\label{eqpresspartial}
    p=-\left(\frac{\partial\Psi}{\partial V}\right)_{T,n_i}
\end{equation}
Since the free energy is mathematically homogeneous, we can derive an intensive free energy per unit mole, $\tilde{\psi}$, through Equation \ref{eqmolarFreeEn}. In the literature, the intensive free energy is usually defined per unit volume, however, it carries the same information as if were defined per unit mole and in this case, it would simplify the calculations.
\begin{equation}
    \label{eqmolarFreeEn}
    \tilde{\psi}\left(T,\frac{V}{n},\frac{n_i}{n}\right)=\tilde{\psi}\left(T,\tilde{v},x_i\right)=\frac{\Psi(T,V,n_i)}{n}
\end{equation}
Here $\tilde{v}$ is the volume per unit mole (molar volume). The pressure can then be written as shown in Equation \ref{eqpresspartial2}
\begin{equation}
\label{eqpresspartial2}
    p=-\left(\frac{\partial\tilde{\psi}}{\partial \tilde{v}}\right)_{T,x_i}
\end{equation}
Integrating Equation \ref{eqpresspartial2} at constant temperature and composition the molar free energy will be given by Equation \ref{eqmolarfreeEnInt} and the extensive free energy will be given by Equation \ref{eqfreeEnInt}.
\begin{subequations}
\begin{equation}
\label{eqmolarfreeEnInt}
    \tilde{\psi}=-\int pd\tilde{v}+C(T,x_i)
\end{equation}
\begin{equation}
\label{eqfreeEnInt}
    \Psi=-n\int pd\tilde{v}+nC(T,x_i)
\end{equation}
\end{subequations}
Here C is a constant of integration which is a function of temperature and composition. The chemical potential of component `$i$' is the partial derivative of the free energy with respect to the moles of component `$i$'; therefore, it will be given by Equation \ref{eqChemPotAp}.
\begin{equation}
\label{eqChemPotAp}
    \mu_i = \frac{\partial}{\partial n_i}\left[ -n\int pd\tilde{v} \right]_{T,V,n_{j\ne i}}+\frac{\partial}{\partial n_i}\left[nC(T,x_i) \right]_{T,V,n_{j\ne i}}
\end{equation}
The first term on the left-hand side of Equation \ref{eqChemPotAp} can be evaluated with an EOS, whereas the second term is unknown. For a single component, isothermal simulation, this constant of integration can be ignored as we are interested in the gradient of chemical potential, which will be calculated as follows:
\begin{equation*}
        \mu = \frac{\partial}{\partial n}\left[ -n\int pd\tilde{v} \right]_{T,V}+\frac{\partial}{\partial n}\left[nC(T) \right]_{T,V}
\end{equation*}
\begin{equation*}
        \mu = \frac{\partial}{\partial n}\left[ -n\int pd\tilde{v} \right]_{T,V}+C(T)
\end{equation*}
\begin{equation*}
        \nabla\mu =\nabla \left[\frac{\partial}{\partial n}\left( -n\int pd\tilde{v} \right)_{T,V}\right]
\end{equation*}
However, this cannot be done for a multicomponent case since the constant of integration is a function of temperature and composition, and therefore it cannot be ignored.

\section{Pitfalls of deriving free energy from the canonical partition function}
\label{appendixB}
The van der Waals EOS has a firm footing in statistical mechanics. In this appendix, we will show how a free energy functional can be derived for a van der Waals fluid from the canonical partition function. We will also show how this can not be extended to other EOSs, hence the need for our proposed model.
The canonical partition function, Q, for a system is given by Equation \ref{eqcanonical} \cite{Hill2012}.
\begin{equation}
    \label{eqcanonical}
    Q=\frac{Q_{int}^N}{N!\Lambda^{3N}}\int_{\mathbf{r}^N}\exp{\left(-\frac{U}{k_BT}\right)}d\mathbf{r}^N
\end{equation}
Here $N$ is the number of molecules in the system, $k_B$ is the Boltzmann constant, $Q_{int}$ is the contribution of the partition function due to the internal structure of the molecules, $U$ is the total potential energy of the system, and $\Lambda=\sqrt{h^2/(2\pi mk_BT)}$ is the thermal de Broglie wavelength, with $h$ being the Planck's constant and $m$ being the mass of a molecule. $\mathbf{r}^N$ is the collection of the position vector of each molecule, $\mathbf{r}_i$, in the system such that:
\begin{equation*}
    d\mathbf{r}^N=d\mathbf{r}_1d\mathbf{r}_2 ... d\mathbf{r}_N
\end{equation*}
The potential energy of the system is dependent of the position of every single molecule in the system; however, it can be approximated as a position independent mean potential energy, $\bar{U}$. The canonical partition function can then be written as shown in Equation \ref{eqQ2}.
\begin{equation}
\label{eqQ2}
    Q=\frac{Q_{int}^N}{N!\Lambda^{3N}} \exp{\left(-\frac{\bar{U}}{k_BT}\right)} \int_{\mathbf{r}^N}d\mathbf{r}^N
\end{equation}
The integral in Equation \ref{eqQ2} is simply a volume integral evaluated N times. If the molecules are assumed as hard spheres, the volume available for the molecules to move will have to account for the excluded volume due to the volume of the molecules. The available volume is $(V-N\bar{b})$, where $\bar{b}$ is the excluded volume per molecule and therefore, Equation \ref{eqQ2} becomes Equation \ref{eqQ3}.
\begin{equation}
    \label{eqQ3}
    Q=\frac{Q_{int}^N}{N!\Lambda^{3N}} \left(V-N\bar{b}\right)^N \exp{\left(-\frac{\bar{U}}{k_BT}\right)} 
\end{equation}
The free energy is given by: $\Psi=-k_BT\ln{Q}$. Therefore, the free energy functional can be written as shown in Equation \ref{eqQ4}.
\begin{equation}
    \label{eqQ4}
    \Psi=-Nk_BT\ln{\left(V-N\bar{b}\right)}+\bar{U}+Nk_BT\left(\ln{N}-1\right)+3Nk_BT\ln{\Lambda}-Nk_BT\ln{Q_{int}}
\end{equation}
Using an intermolecular pair potential based on hard sphere repulsion and Lennard-Jones type attraction, the mean potential energy can be approximated as shown in Equation \ref{eqMeanPot} \cite{Hill2012}.
\begin{equation}
    \label{eqMeanPot}
    \bar{U}=-\bar{a}\frac{N^2}{V}
\end{equation}
Here, $\bar{a}$ is the attraction term in an EOS when written for the number of molecules and not the number of moles. The free energy functional will then be given by Equation \ref{eqQ5}.
\begin{equation}
    \label{eqQ5}
    \Psi=-Nk_BT\ln{\left(V-N\bar{b}\right)}-\bar{a}\frac{N^2}{V}+Nk_BT\left(\ln{N}-1\right)+3Nk_BT\ln{\Lambda}-Nk_BT\ln{Q_{int}}
\end{equation}
The pressure can be computed from the partial derivative of the free energy as follows:
\begin{equation*}
    p=-\left(\frac{\partial\Psi}{\partial V}\right)_{T,N}
\end{equation*}
Keeping in mind that $Q_{int}$ and $\Lambda$ are not functions of volume, the pressure is given by Equation \ref{eqvdwB}.
\begin{equation}
\label{eqvdwB}
    p=\frac{Nk_BT}{V-N\bar{b}}-\bar{a}\frac{N^2}{V^2}
\end{equation}
When written in terms of moles, using the $a$ and $b$ for the attraction parameter and co-volume respectively ($a=\bar{a}N_A^2$ and $b=\bar{b}N_A$, where $N_A$ is the Avogadro constant), Equation \ref{eqvdwB} is the van der Waals EOS (Equation \ref{eqvdw2B}).
\begin{equation}
\label{eqvdw2B}
    p=\frac{RT}{\tilde{v}-b}-\frac{a}{\tilde{v}^2}
\end{equation}
The free energy given by Equation \ref{eqQ5} is the free energy of a van der Waals fluid. More recent cubic EOSs are empirical modifications to the van der Waals EOS and have the same form. They can be generalized as shown in Equation \ref{eqform}.
\begin{equation}
    \label{eqform}
    p=\frac{RT}{\tilde{v}-b}+p_{att}
\end{equation}
Here $p_{att}$ is the contribution of the attractive interactions between molecules to the pressure. For the van der Waals EOS $p_{att}=-a/\tilde{v}^2$, for the Peng-Robinson (PR) EOS $p_{att}=-a\alpha/(\tilde{v}^2+2b\tilde{v}-b^2)$ and for the Soave-Redlich-Kwong (SRK) EOS $p_{att}=-a\alpha/(\tilde{v}^2+b\tilde{v})$. These EOSs will be discussed in detail in Appendix \ref{secAppendixEOS}. It can be seen that $p_{att}$ emerges from the mean potential energy of the system (Equation \ref{eqMeanPot}). The relationship between $p_{att}$ and $\bar{U}$ is given by Equation \ref{eqpatt}.
\begin{equation}
    \label{eqpatt}
    p_{att}=-\left(\frac{\partial\bar{U}}{\partial V}\right)_{T,n}
\end{equation}
The free energy functional can be written for any EOS by integrating Equation \ref{eqpatt} and substituting into Equation \ref{eqQ4}. However, integrating Equation \ref{eqpatt} would introduce a constant of integration which is a function of the temperature and moles, $C(T,n)$, as shown in Equation \ref{eqQgen}.
\begin{equation}
    \label{eqQgen}
    \Psi=-Nk_BT\ln{\left(V-N\bar{b}\right)}-\int p_{att} dV+C(T,n)+Nk_BT\left(\ln{N}-1\right)+3Nk_BT\ln{\Lambda}-Nk_BT\ln{Q_{int}}
\end{equation}
For the Free Energy LBM, we require an expression for the chemical potential for which we need to take the partial derivative of the free energy functional with respect to the moles. The constant of integration would affect the results when taking this partial derivative and hence a unique chemical potential cannot be obtained. This constant of integration cannot be designed for a particular EOS as any choice of $C(T,n)$ will retrieve that EOS (the EOS is retrieved by taking a partial derivative of the free energy functional with respect to volume and the constant of integration is not a function of volume). However, there is no way of gauging the effect of the designed $C(T,n)$ on the chemical potential. Therefore, EOSs like the PR and SRK can not be incorporated into our model the same way as the van der Waals EOS.

\section{Common equations of state and their respective fugacity expressions}
\label{secAppendixEOS}
\subsection{Equations of state}
A popular class of equations of state (EOSs) is cubic EOSs. In this Appendix, we will go through some important cubic EOSs and their extensions to multicomponent mixtures. For a pure component `$i$', a cubic EOS has a parameter accounting for molecular attraction, $a_i$ or $(a\alpha)_i$, and a parameter accounting for the volume occupied by the molecules, $b_i$ (also known as the co-volume). For the case of a multicomponent mixture, these pure component parameters are replaced with mixture parameters: $a_m$ or $(a\alpha)_m$, and $b_m$. The mixture parameters are calculated using van der Waals mixing rules by considering attractions between like and unlike component pairs and an average molecular volume \cite{Kwak1986}. The first cubic EOS was the van der Waals (vdW) EOS \cite{vdw1873}. Two extensions of the vdW EOS known for their accuracy are the Soave-Redlich-Kwong (SRK) \cite{Soave1972} and Peng-Robinson (PR) \cite{Peng1976}  EOS. The multicomponent vdW, SRK, and PR EOS are summarized in the following sections.

\subsubsection{vdW equation of state}
The vdW EOS is given by Equation \ref{eqvdwEOS}.
\begin{equation}
    \label{eqvdwEOS}
    p=\frac{RT}{\tilde{v}-b_m}-\frac{a_m}{\tilde{v}^2}
\end{equation}
Here $p$ is the pressure, $R$ is the universal gas constant, $T$ is the temperature, and $\tilde{v}$ is the molar volume. For a mixture with $N_c$ components, the mixing rules for the attraction term and co-volume term are given by Equation \ref{eqmixingrulesavdW} and Equation \ref{eqmixingrulesbm} respectively.

\begin{subequations}
\label{eqmixingrulesavdW}
\begin{equation}
    a_m=\sum_{i=1}^{N_c}\sum_{j=1}^{N_c}x_ix_ja_{ij}
\end{equation}
\begin{equation}
\label{eqmixingrulesavdW2}
    a_{ij}=\sqrt{a_ia_j}\left(1-\delta_{ij} \right)
\end{equation}
\end{subequations}

\begin{equation}
\label{eqmixingrulesbm}
    b_m=\sum_{i=1}^{N_c}x_ib_i
\end{equation}
Here $x_i$ is the mole fraction of component `$i$' and $\delta_{ij}$ is the binary interaction parameter between component `$i$' and component `$j$'. The attraction and co-volume terms for component `$i$' are given by $a_i=\frac{27}{64}\frac{R^2T_{c,i}^2}{P_{c,i}}$ and $b_i=\frac{1}{8}\frac{RT_{c,i}}{P_{c,i}}$ and $T_{c,i}$ and $P_{c,i}$ are the critical temperature and critical pressure of component `$i$' respectively.


\subsubsection{SRK equation of state}
The SRK EOS is given by Equation \ref{eqSRKEOS}.
\begin{equation}
    \label{eqSRKEOS}
    p=\frac{RT}{\tilde{v}-b_m}-\frac{(a\alpha)_m}{\tilde{v}\left(\tilde{v}+b_m\right)}
\end{equation}
For a mixture with $N_c$ components, the mixing rules for the attraction term and co-volume term are given by Equation \ref{eqmixingrulesaSRK} and Equation \ref{eqmixingrulesbm} respectively.
\begin{subequations}
\label{eqmixingrulesaSRK}
\begin{equation}
    (a\alpha)_m=\sum_{i=1}^{N_c}\sum_{j=1}^{N_c}x_ix_j\left(a\alpha\right)_{ij}
\end{equation}
\begin{equation}
\label{eqmixingrulesaSRK2}
    \left(a\alpha\right)_{ij}=\sqrt{(a\alpha)_i(a\alpha)_j}\left(1-\delta_{ij} \right)
\end{equation}
\end{subequations}
Here $a_i=0.4274802\frac{R^2T_{c,i}^2}{P_{c,i}}$, $b_i=0.08664035\frac{RT_{c,i}}{P_{c,i}}$, and:

\begin{equation*}
    \alpha_i=\left[1+\left(0.48+1.574\omega_i-0.176\omega_i^2 \right) \left(1-T_{r,i}^{0.5} \right) \right]^2 
\end{equation*}
where $\omega_i$ is the accentric factor of component `$i$' and $T_{r,i}=\frac{T}{T_{c,i}}$.

\subsubsection{PR equation of state}
The PR EOS is given by Equation \ref{eqPREOS}.
\begin{equation}
    \label{eqPREOS}
    p=\frac{RT}{\tilde{v}-b_m}-\frac{(a\alpha)_m}{\tilde{v}^2+2b_m\tilde{v}-b_m^2}
\end{equation}
For a mixture with $N_c$ components, the mixing rules for the attraction term and co-volume term are given by Equation \ref{eqmixingrulesaSRK} and Equation \ref{eqmixingrulesbm} respectively, with $a_i=0.457235529\frac{R^2T_{c,i}^2}{P_{c,i}}$, $ b_i=0.077796074\frac{RT_{c,i}}{P_{c,i}}$, and:

\begin{equation*}
    \alpha_i=
    \begin{cases}
    \left[1+\left(0.374640+1.54226\omega_i-0.26992\omega_i^2 \right) \left(1-T_{r,i}^{0.5} \right) \right]^2 & \text{if $\omega_i\le0.49$ }\\
    \left[1+\left(0.379642+1.48503\omega_i-0.164423\omega_i^2+0.016666\omega_i^3 \right) \left(1-T_{r,i}^{0.5} \right) \right]^2 & \text{if $\omega_i>0.49$}\\
    \end{cases}
\end{equation*}

\subsection{Fugacity expressions from relevant equations of state}
\label{secAppendixFug}
In this appendix, we will show how to derive an expression for the fugacity using its definition given by Equation \ref{eqfugdef}. Then we will show how this expression can be evaluated using the vdW, SRK, and PR EOS. To obtain an expression for the fugacity of component `$i$' we start by integrating Equation \ref{eqfugdef1} by varying the pressure while keeping temperature and composition constant to obtain Equation \ref{eqfugexpress}
\begin{equation}
    \label{eqfugexpress}
    \ln{\left[f_i\left(T,p,x_i\right)\right]}-\ln{\left[f_i\left(T,p^{ref},x_i\right)\right]}=\frac{\mu_{B,i}(T,p,x_i)-\mu_{B,i}(T,p^{ref},x_i)}{RT}
\end{equation}
Here $x_i$ is the mole fraction for component `$i$' and $p^{ref}$ is the reference pressure for integration. Taking the reference pressure to be approaching 0, $p^{ref}=p_0$, we can use the reference state given by Equation \ref{eqfugdef2} ($f_i(T,p_0,x_i)=x_ip_0$). Additionally, at low pressure, the mixture behaves like an ideal gas mixture (denoted by superscript IGM). Therefore, we get Equation \ref{eqfugexpress2}.
\begin{equation}
    \label{eqfugexpress2}
    \ln{\left[f_i\left(T,p,x_i\right)\right]}-\ln{\left[x_ip_0\right]}=\frac{\mu_{B,i}(T,p,x_i)-\mu_{B,i}^{IGM}(T,p_0,x_i)}{RT}
\end{equation}
For an ideal gas mixture, Equation \ref{dchemIGM} holds \cite{sandler2006}.
\begin{equation}
\label{dchemIGM}
    \mu_{B,i}^{IGM} (T,p,x_i)-\mu_{B,i}^{IGM} (T,p_0,x_i) = RT \ln{\frac{p}{p_0}}
\end{equation}
Combining Equations \ref{eqfugexpress2} and \ref{dchemIGM} we get:
\begin{equation}
    \label{eqfugexpress3}
    \ln{\left[\frac{f_i(T,p,x_i)}{x_ip}\right]}=\mu_{B,i}(T,p,x_i)-\mu_{B,i}^{IGM}(T,p,x_i)
\end{equation}
The right hand side of Equation \ref{eqfugexpress3} is the residual chemical potential which can be evaluated to give Equation \ref{eqfug} \cite{sandler2006}.
\begin{equation}
    \label{eqfug}
           \ln{\left[\frac{f_i(T,p,x_i)}{x_ip}\right]} =  \frac{1}{RT} \int_{0}^p \left[\left(\frac{\partial V}{\partial n_i}\right)_{T,p,n_{j\ne i}}-\frac{RT}{p}\right] dp 
\end{equation}
Here $n_i$ is the moles of component `i' and $V$ is the volume. Applying the cyclic rule for the properties p, V, and $n_i$ holding T and $n_{j\ne i}$ constant, we get:
\begin{equation}
    \label{eqcyclic}
    \left(\frac{\partial V}{\partial n_i}\right)_{T,p,n_{j\ne i}} \left(\frac{\partial p}{\partial V}\right)_{T,n_{i}} \left(\frac{\partial n_i}{\partial p}\right)_{T,V,n_{j\ne i}}=-1
\end{equation}
The integral in Equation \ref{eqfug} is at constant temperature and composition, therefore the term $\left(\frac{\partial p}{\partial V}\right)_{T,n_{i}}$ in Equation \ref{eqcyclic} will turn into a total derivative giving Equation \ref{eqcyclic2} \cite{Koretsky2013}.
\begin{equation}
\label{eqcyclic2}
    \left(\frac{\partial V}{\partial n_i}\right)_{T,p,n_{j\ne i}}dp=- \left(\frac{\partial p}{\partial n_i}\right)_{T,V,n_{j\ne i}}dV
\end{equation}
Substituting into Equation \ref{eqfug}:
\begin{equation}
    \label{eqfug2}
       \ln{\left[\frac{f_i(T,p,x_i)}{x_ip}\right]} = -\frac{1}{RT} \int_{\infty}^V \left[\left(\frac{\partial p}{\partial n_i}\right)_{T,V,n_{j\ne i}}\right] dV -\frac{1}{RT} \int_{0}^p \left[\frac{RT}{p}\right] dp 
\end{equation}
The partial derivative $\left(\frac{\partial p}{\partial n_i}\right)_{T,V,n_{j\ne i}}$ can be evaluated using a pressure explicit EOS, like Equation \ref{eqvdwEOS}, \ref{eqSRKEOS}, or \ref{eqPREOS}. While evaluating this derivative, the composition in the EOS needs to be expressed as $x_i=\frac{n_i}{\sum_i n_i}$ and molar volume needs to be expressed as $\tilde{v}=\frac{V}{\sum_i n_i}$. Then Equation \ref{eqfug2} can be solved to find the fugacity of component `i'. The fugacity expression for the vdW, SRK and PR EOS are given by Equations \ref{eqfugvdW}, \ref{eqfugSRK} and \ref{eqfugPR} respectively \cite{Koretsky2013,Tosun2013}. 

\begin{equation}
    \label{eqfugvdW}
    \ln{\left[\frac{f_i}{x_ip}\right]} = \frac{b_i}{\tilde{v}-b_{m}}-\ln{\left[\frac{\left(\tilde{v}-b_m\right)p}{RT}\right]}-\frac{2}{RT\tilde{v}}\sum_{j=1}^{N_c}x_ja_{ij}
\end{equation}

\begin{equation}
    \label{eqfugSRK}
    \ln{\left[\frac{f_i}{x_ip}\right]} = \frac{b_i}{b_{m}}\left[\frac{p\tilde{v}}{RT}-1\right]-\ln{\left[\frac{\left(\tilde{v}-b_m\right)p}{RT}\right]}+\frac{(a\alpha)_{m}}{b_mRT}\left[\frac{b_i}{b_m}-\frac{2}{(a\alpha)_m}\sum_{j=1}^{N_c}x_j(a\alpha)_{ij}\right]\ln{\left[1+\frac{b_m}{\tilde{v}}\right]}
\end{equation}

\begin{equation}
    \label{eqfugPR}
    \ln{\left[\frac{f_i}{x_ip}\right]} = \frac{b_i}{b_{m}}\left[\frac{p\tilde{v}}{RT}-1\right]-\ln{\left[\frac{\left(\tilde{v}-b_m\right)p}{RT}\right]}+\frac{(a\alpha)_{m}}{2\sqrt{2}b_mRT}\left[\frac{b_i}{b_m}-\frac{2}{(a\alpha)_m}\sum_{j=1}^{N_c}x_j(a\alpha)_{ij}\right]\ln{\left[\frac{\tilde{v}+\left(1+\sqrt{2}\right)b_m}{\tilde{v}+\left(1-\sqrt{2}\right)b_m}\right]}
\end{equation}

\section{Density versus pressure and density vs temperature plots from Section \ref{seccase2}}
\label{appendixDenplot}
\begin{figure}[H]
    \centering
        \begin{subfigure}{0.4\textwidth}
            \centering
            \caption{}
            \includegraphics[width=\textwidth]{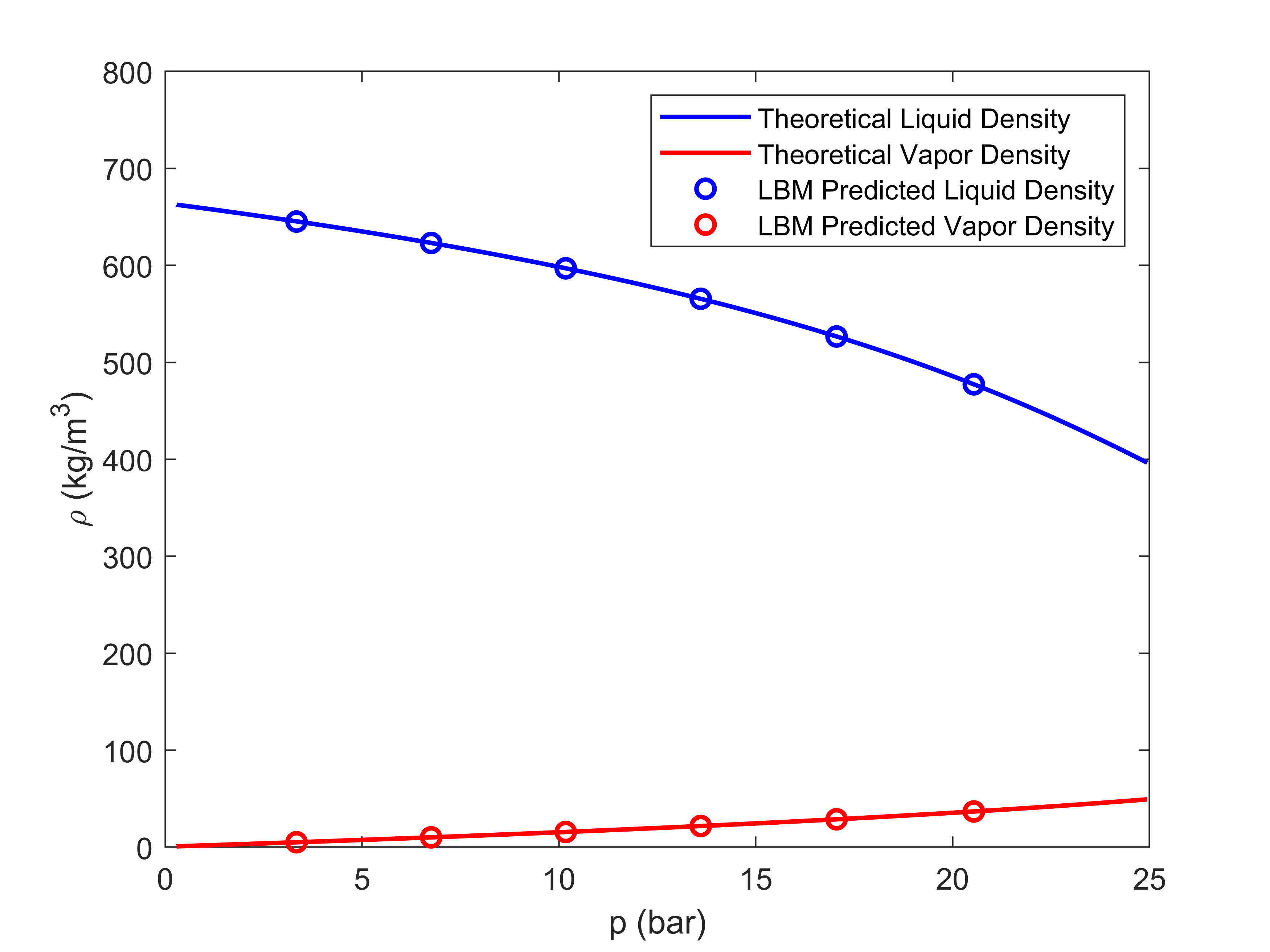}
        \end{subfigure}%
        \hfill
        \begin{subfigure}{0.4\textwidth}
            \centering
            \caption{}
            \includegraphics[width=\textwidth]{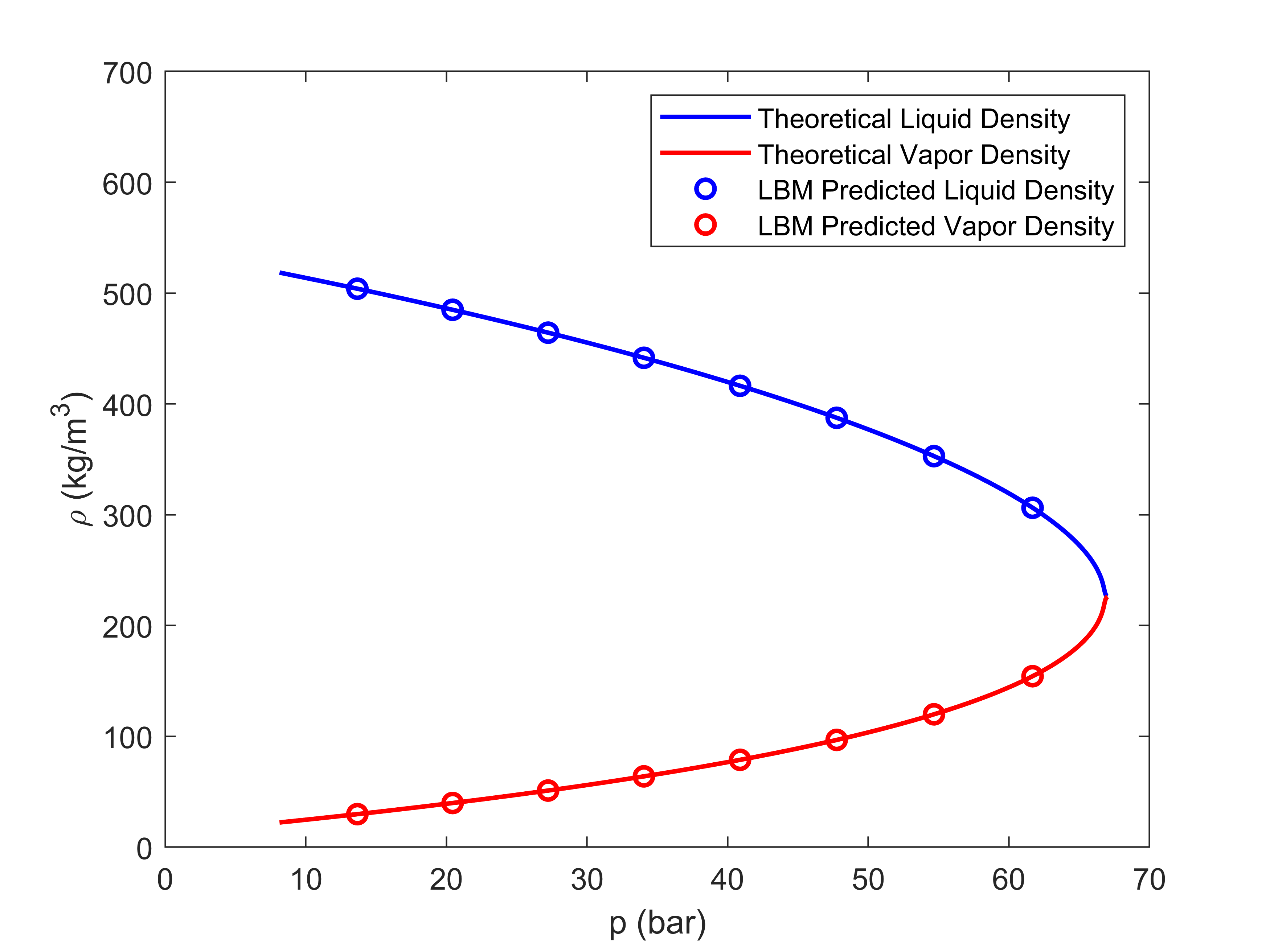}
        \end{subfigure}
        \hfill        

\caption{The liquid density vs pressure (blue) and vapor density vs pressure (red) at (a) T = 274.96 K and (b) T = 387.70 K. The solid lines represent the theoretical results whereas the dots represent the results predicted by LBM.}
\label{figrhoP}
\end{figure}

Several vapor-liquid equilibrium cases were simulated in Section \ref{seccase2} to obtain the p-x and T-x envelopes. The density of the vapor phase and the density of the liquid phase at each of the pressures the simulations were run at to generate the p-x envelope are plotted in Figure \ref{figrhoP}. The theoretical vapor and liquid phase densities predicted by a flash calculation are also shown.

\begin{figure}[H]
    \centering
        \begin{subfigure}{0.33\textwidth}
            \centering
            \caption{}
            \includegraphics[width=\textwidth]{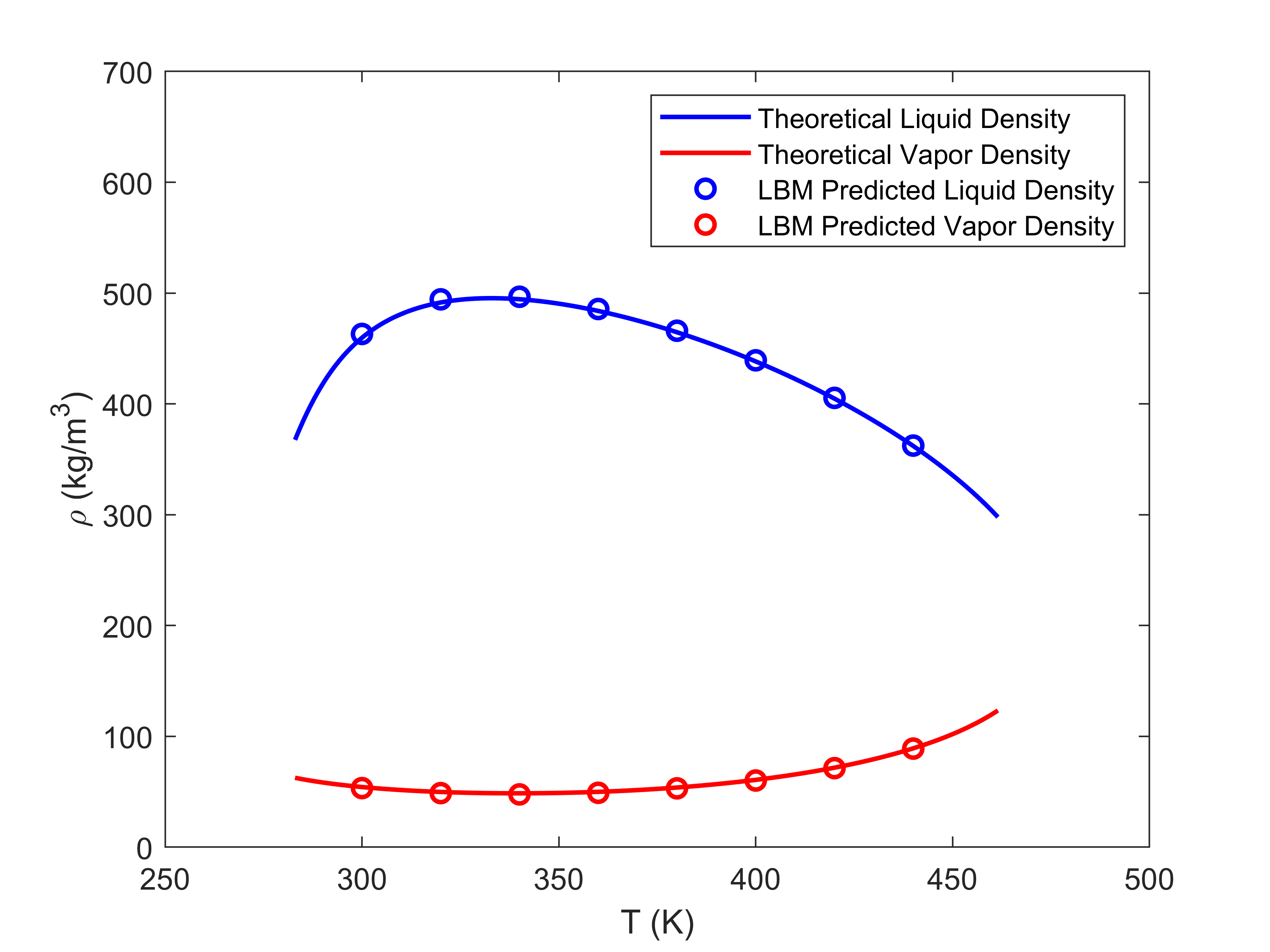}
        \end{subfigure}%
        \hfill
        \begin{subfigure}{0.33\textwidth}
            \centering
            \caption{}
            \includegraphics[width=\textwidth]{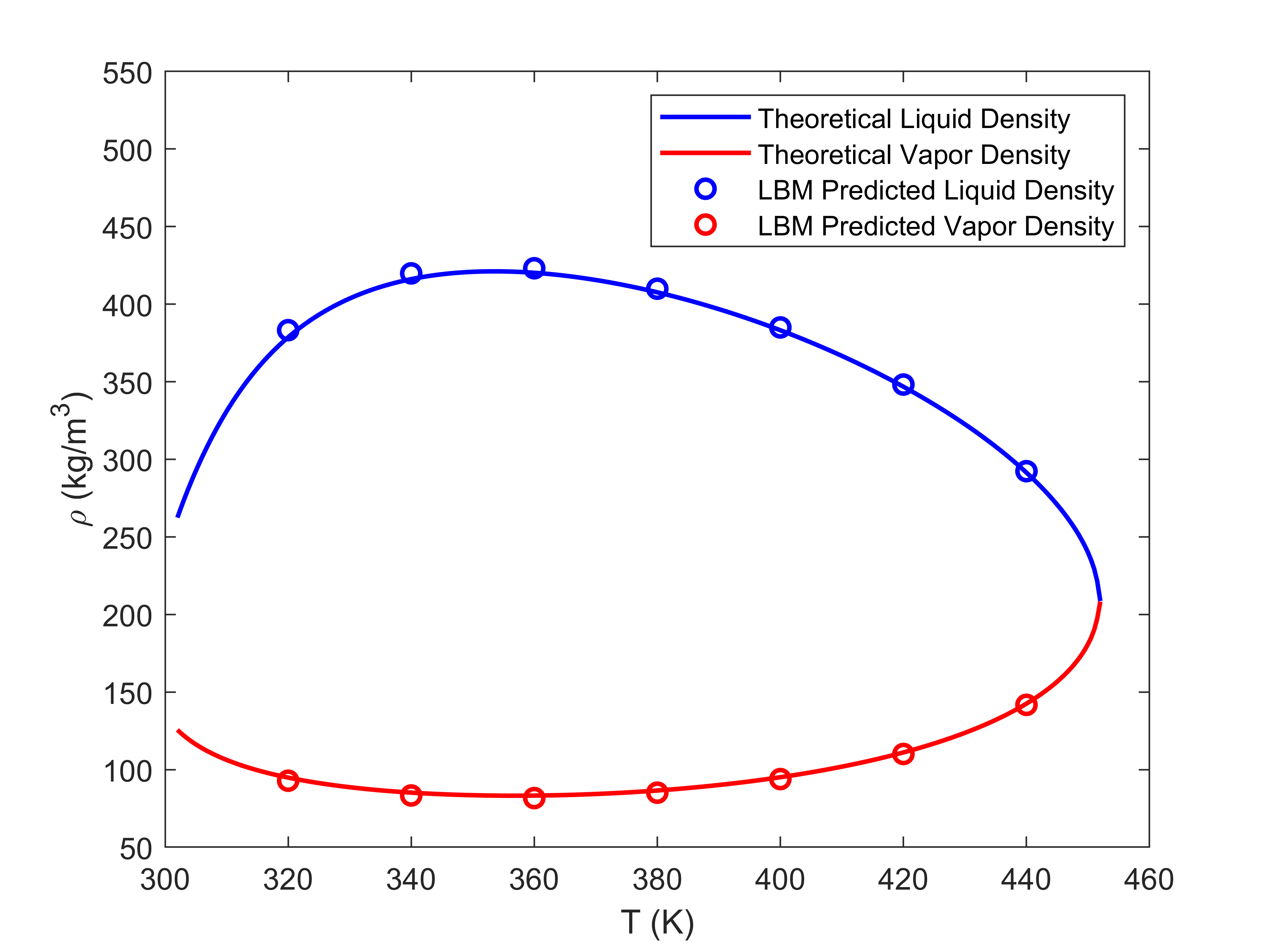}
        \end{subfigure}
        \hfill        
        \begin{subfigure}{0.33\textwidth}
            \centering
            \caption{}
            \includegraphics[width=\textwidth]{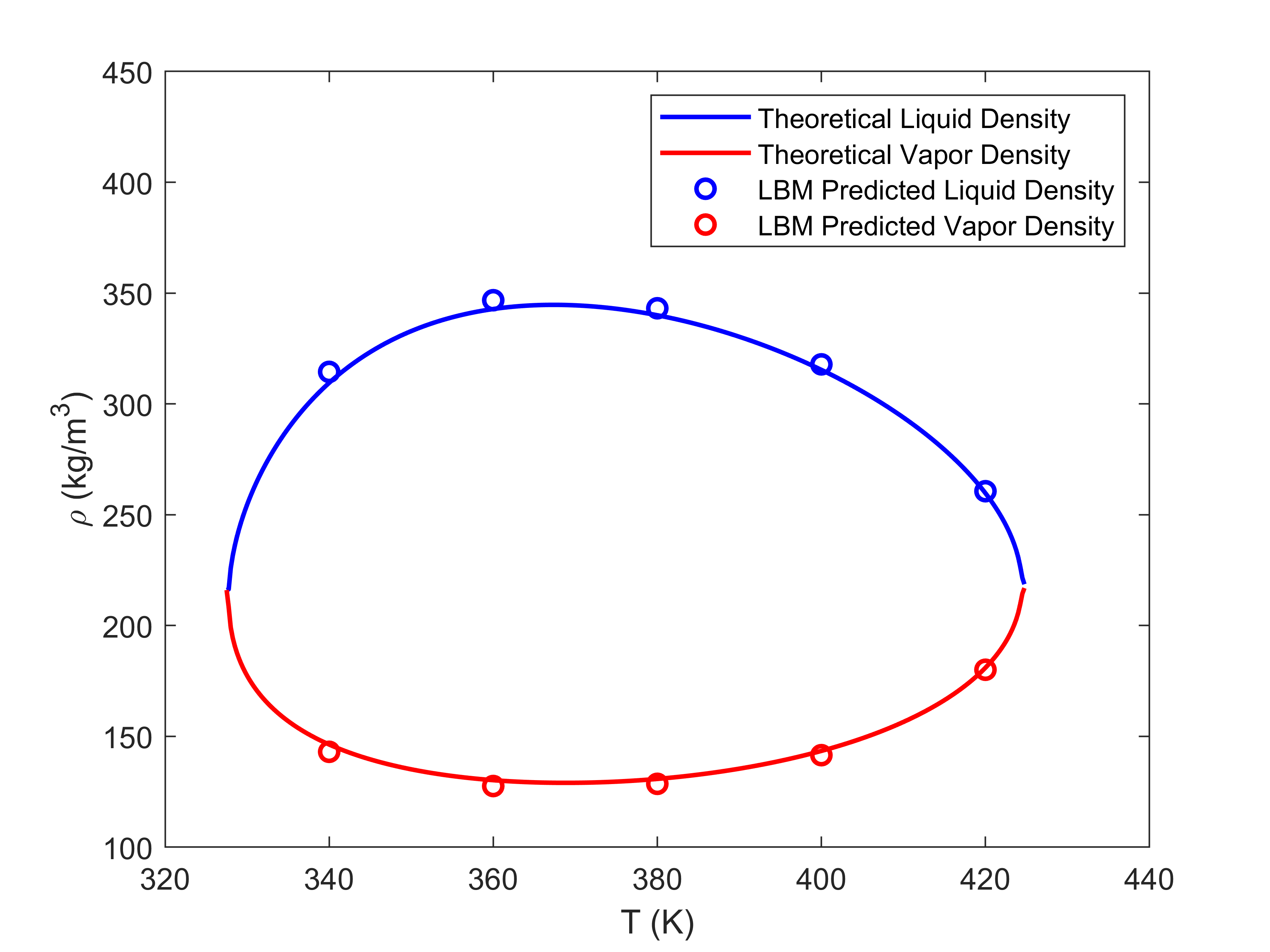}
        \end{subfigure}
        \hfill 

\caption{The liquid density vs temperature (blue) and vapor density vs temperature (red) at (a) P = 30 bar, (b) P = 45 bar, and (c) P = 58 bar. The solid lines represent the theoretical results whereas the dots represent the results predicted by LBM.}
\label{figrhoT}
\end{figure}

 The density of the vapor phase and the density of the liquid phase at each of the temperatures the simulations were run at to generate the T-x envelope are plotted in Figure \ref{figrhoT}. The theoretical vapor and liquid phase densities predicted by a flash calculation are also shown. In this figure, we see a greater deviation of the LBM results from the theoretical results and this is because the theoretical results are generated at a constant pressure where as the equilibrium pressure in LBM slightly varies, as discussed in Section \ref{seccase2} in reference to the T-x envelopes from Figure \ref{figTxEnvelopes}.

%

\end{document}